\documentclass[12pt]{article}
\usepackage{jheppub}
\usepackage[utf8]{inputenc}
\usepackage{amsmath}
\usepackage{dsfont}
\usepackage{amssymb}
\usepackage{amsthm}
\usepackage{mathrsfs}
\usepackage{appendix}
\usepackage{bbold}
\usepackage{color}
\usepackage{pgfplots}
\usepackage{pgf}
\usepackage{slashed}
\usepackage{booktabs}
\usepackage{array}
\usepackage{subfigure}
\usepackage{comment}

\newcommand{\Ssus}{16}                 
\newcommand{\xiParam}{2.0}             
\newcommand{\PhiZeroOverMstar}{1200.0} 

\usepackage[dvipsnames]{xcolor}

\numberwithin{equation}{section}      

\begin{document}
\title{\vspace{-6mm}Geometric Baryogenesis with Chiral-Time Equivalence\vspace{-1mm}}

\author[1,2]{Sameer Ahmad Mir}
\emailAdd{sameerphst@gmail.com}
\affiliation[1]{Canadian Quantum Research Center, 460 Doyle Ave 106, Kelowna, BC V1Y 0C2, Canada}
\affiliation[2]{Department of Computer Sciences, Asian School of Business, Noida, Uttar Pradesh, 201303, India}
\author[1]{Arshid Shabir}
\emailAdd{aslone186@gmail.com}
\author[3]{Swatantra Kumar Tiwari}
\emailAdd{swatantra@allduniv.ac.in}
\affiliation[3]{Department of Physics, University of Allahabad, Prayagraj 211002, India}
\author[1,4,5,6]{Mir Faizal}
\emailAdd{mirfaizalmir@googlemail.com}
\affiliation[4]{Irving K. Barber School of Arts and Sciences, University of British Columbia Okanagan, Kelowna, BC V1V 1V7, Canada}
\affiliation[5]{Department of Mathematical Sciences, Durham University, Upper Mountjoy, Stockton Road, Durham DH1 3LE, UK}
\affiliation[6]{Computational Mathematics Group, Hasselt University, Agoralaan Gebouw D, Diepenbeek, 3590 Belgium}

\abstract{The asymmetry between matter and antimatter demands a cause as simple as it is profound. Here we show that a single geometric principle Chiral-Time Equivalence (CTE)-suffices to generate and correlate the required CP violation with the time orientation of the cosmos. Promoting the Immirzi parameter to a pseudoscalar Nambu-Goldstone field $\Phi$, CTE fixes the leading operators: a shift-symmetric derivative portal $((\partial_\mu\Phi)J^\mu_{B-L}/M_*)$ that acts as a dynamical chemical potential in FRW, and a topological term $(\Phi\,R\tilde R)$ that imprints parity on tensor modes. In thermal equilibrium this structure produces gravity-assisted leptogenesis, whose magnitude is set at the decoupling temperature by susceptibilities rather than by tuned departures from equilibrium. A fully flavored Boltzmann network with curvature sources captures flavor transfer and washout, while slow-roll and resonant regimes are established via thermodynamic and Kubo formulas. Consistency is secured by an EFT analysis (stability, perturbative unitarity, and BBN safety), and by explicit elimination of EC torsion and control of dCS birefringence in the small-coupling domain. The most striking prediction is a sign locking among $\eta_B$, tensor chirality $\chi_T$, and the drift of $\Phi$, together with a tri-observable relation that ties $\eta_B$ to cosmic birefringence $\Delta\alpha$ and $\chi_T$. Thus a single, symmetry-protected geometric origin renders the baryon excess testable by TB/EB correlations and stochastic-wave chirality, and calculable within a minimal, ultraviolet-anchored effective theory.
}

\maketitle

\section{Introduction}\label{sec1}

The observed cosmic baryon asymmetry, commonly quoted as the baryon-to-photon ratio \(\eta_B \equiv n_B/n_\gamma \simeq 6\times 10^{-10}\) with percent-level \textit{Planck} precision \cite{Planck2018}, requires an early-universe dynamical origin. Any successful explanation must satisfy (or effectively implement) Sakharov’s conditions: (i) baryon number violation, (ii) C and CP violation, and (iii) departure from thermal equilibrium \cite{Sakharov1967}. The Standard Model (SM) provides electroweak sphalerons violating \((B+L)\) while conserving \((B-L)\), and CP violation through the CKM phase; however, the electroweak transition is not first order for the observed Higgs mass and CKM CP violation is far too small to account for \(\eta_B\) \cite{FarrarShaposhnikov1993,RiottoTrodden1999,MorrisseyRamseyMusolf2012}. This motivates controlled extensions of the SM and/or a reappraisal of gravitational ingredients inevitably present in the early universe.

Conventional proposals illustrate both promise and limitations. Electroweak baryogenesis relies on a first-order electroweak phase transition, with bubble-wall CP sources biasing sphalerons; it is testable but typically requires extra scalars and tuned couplings while evading EDM bounds \cite{MorrisseyRamseyMusolf2012,RiottoTrodden1999}. Leptogenesis links the asymmetry to neutrino mass generation: out-of-equilibrium decays of heavy Majorana neutrinos generate a lepton asymmetry later reprocessed by sphalerons \cite{FukugitaYanagida1986,BuchmullerPecceiYanagida2005,DavidsonNardiNir2008}, but often points to very high scales and can be sensitive to phases and flavor dynamics \cite{DavidsonNardiNir2008}. Gravitational baryogenesis exploits the fact that an expanding background breaks Lorentz invariance, allowing curvature-dependent chemical potentials, e.g.\ \((\partial_\mu R) J^\mu_B/M^2\) \cite{Davoudiasl2004}; yet in radiation-dominated FRW one has \(R=0\) at leading order, so realistic implementations typically require departures from radiation domination, model-dependent gravity modifications, or loop-suppressed sources \cite{Davoudiasl2004,RiottoTrodden1999}. These considerations suggest seeking a framework where CP-odd gravitational data are calculable, the chemical potential is symmetry-protected, and the mechanism is robust to the thermal history.

In this work we develop \emph{Geometric Baryogenesis with Chiral-Time Equivalence} (CTE): a symmetry-based, EFT-controlled mechanism in which CP violation is tied to geometric parity-odd invariants and to the single slow hydrodynamic charge surviving in the hot plasma. The key observation is that cosmology selects a preferred clock (breaking time reparametrizations down to the comoving flow), while axial rotations are anomalous in curved spacetime. We postulate a diagonal transformation-CTE-that mixes an infinitesimal time diffeomorphism along the cosmological four-velocity with an axial \(U(1)_A\) rotation, and we promote the Barbero-Immirzi parameter of Einstein-Cartan-Holst gravity to a pseudoscalar Nambu-Goldstone field \(\Phi\) realizing this symmetry nonlinearly \cite{Holst1996,RovelliThiemann1998,Shapiro2002,NiehYan1982,TaverasYunes2008,CalcagniMercuri2009,Mercuri2009}. Anomaly matching then forces the parity-odd coupling \(\Phi\,R\tilde R\), i.e.\ dynamical Chern-Simons (dCS) gravity \cite{JackiwPi2003,AlexanderYunes2009}. In the hot SM plasma, fast Yukawas, strong sphalerons, and hypercharge neutrality reduce the charge sector to a single approximately conserved slow mode \((B-L)\) \cite{HarveyTurner1990}. Symmetry and power counting therefore single out the unique leading portal \((\partial_\mu \Phi)J_{B-L}^\mu/M_*\), so that in FRW \(\mu_{B-L} \equiv \dot{\Phi}/M_*\) acts as a dynamical chemical potential. Unlike \((\partial_\mu R)J^\mu\), this bias is protected by the shift symmetry of \(\Phi\) and persists in radiation domination; meanwhile \(R\tilde R\) vanishes on the homogeneous background but induces calculable parity violation for gravitational waves \cite{JackiwPi2003,AlexanderYunes2009}.

CTE has two key geometric implications. First, in first-order (vielbein-connection) variables, torsion is nonpropagating and algebraically tied to axial fermion currents; eliminating it yields controlled contact interactions and a clean pseudoscalar sector associated with the Holst/Nieh-Yan structure \cite{Shapiro2002,NiehYan1982,CalcagniMercuri2009}. Second, the dCS term \(\Phi R\tilde R\) is topological, sourcing a Cotton tensor and producing helicity-dependent tensor friction while remaining well posed in the small-coupling regime \cite{AlexanderYunes2009,JackiwPi2003}. These features correlate a late-time observable-tensor chirality in a stochastic gravitational-wave background (SGWB-with the early-time sign of \(\dot\Phi\), and hence with the sign of the baryon asymmetry. Parity violation is testable via TB/EB CMB correlations and via interferometric probes of the SGWB Stokes-\(V\) parameter \cite{LueWangKamionkowski1999,SetoTaruya2007,GluscevicKamionkowski2010,Qiao2023}.

Conceptually, CTE recasts spontaneous baryogenesis \cite{CohenKaplan1988} as the low-energy manifestation of a geometric symmetry and addresses three persistent issues. First, the CP-violating source is not \emph{ad hoc}: it is the unique leading operator compatible with the CTE shift symmetry and the hydrodynamic slow mode, \((\partial\Phi)\!\cdot\!J_{B-L}/M_*\). Second, the mechanism reduces reliance on special out-of-equilibrium dynamics: an asymmetry is generated while \(B{-}L\)-violating processes remain in kinetic equilibrium, and the final yield is set at their decoupling temperature \(T_D\), analogous to chemical freeze-out \cite{MorrisseyRamseyMusolf2012,RiottoTrodden1999}. Third, the same \(\dot\Phi\) that biases \((B-L)\) fixes tensor chirality through dCS birefringence, yielding “sign locking’’ and accompanying amplitude relations that correlate \(\eta_B\) with TB/EB spectra or SGWB circular polarization-a lever arm absent in standard leptogenesis and curvature-gradient baryogenesis \cite{FukugitaYanagida1986,DavidsonNardiNir2008,Davoudiasl2004}.

Technically, the EFT is controlled by two small parameters: \(H/M_*\ll 1\) for the derivative portal and \(\varepsilon_{\rm CS}\sim \mu_g \dot\Phi/M_{\rm Pl}^2 H\ll 1\) for dCS backreaction, ensuring unitarity and tensor stability in the relevant epoch \cite{AlexanderYunes2009,JackiwPi2003}. The operator basis is minimal: \(R\tilde R=0\) in homogeneous FRW ensures no double counting between \(\Phi R\tilde R\) and the portal as a background \((B-L)\) source, while on tensor backgrounds the same coupling cleanly tracks parity violation without contaminating scalar dynamics. The ultraviolet anchor is provided by the axial ABJ anomaly and its gravitational/torsional extensions \cite{Adler1969,AlvarezGaumeWitten1984,ChandiaZanelli1997}. In this way, CTE links a symmetry-protected chemical potential to geometric parity violation and to a unique hydrodynamic slow mode, yielding a predictive baryogenesis framework testable with current and near-future CMB and gravitational-wave measurements \cite{LueWangKamionkowski1999,SetoTaruya2007,GluscevicKamionkowski2010,Qiao2023,Planck2018}.

\section{Chiral-Time Equivalence (CTE): Core Principle}\label{sec2}

\begin{figure}[htb]
    \centering
    \includegraphics[width=\linewidth, height=0.9\linewidth]{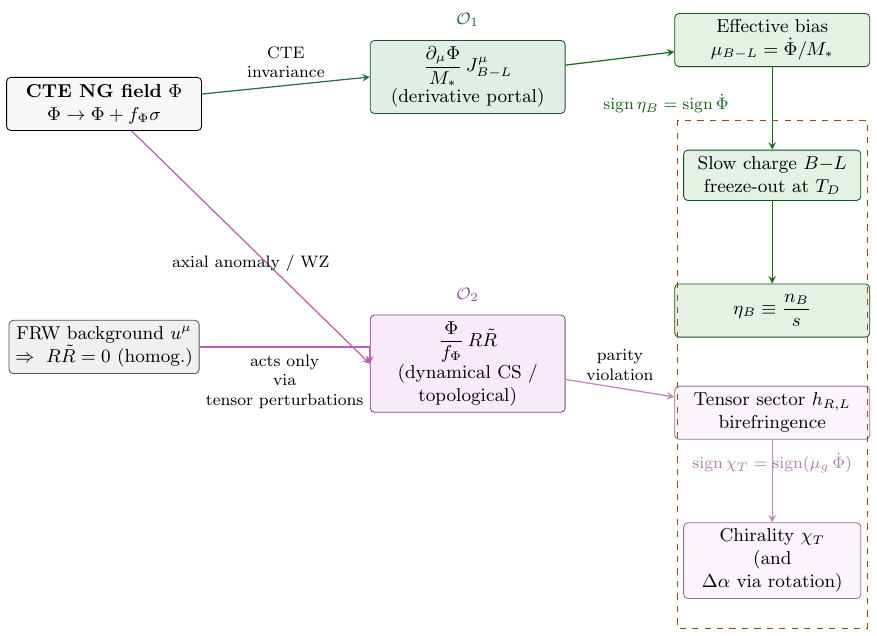}
     \caption{\textbf{CTE-locked geometric baryogenesis.}
  The unique shift-symmetric portal \(\mathcal{O}_1=(\partial_\mu\Phi/M_*)J^\mu_{B-L}\) yields \(\mu_{B-L}=\dot\Phi/M_*\) and fixes the \(B{-}L\) asymmetry at decoupling \(T_D\). The parity-odd coupling \(\mathcal{O}_2=(\Phi/f_\Phi)R\tilde R\) vanishes on homogeneous FRW but induces tensor birefringence and chirality \(\chi_T\) for perturbations, implying the sign relation
  \(\mathrm{sign}\,\eta_B=\mathrm{sign}\,\chi_T=\mathrm{sign}\,\dot\Phi\) (up to the convention for \(\mu_g\)).}
  \label{fig:cte-schematic}
\end{figure}

Chiral-Time Equivalence (CTE) ties the reversal of the time orientation selected by a cosmological background to the axial (chiral) phase of fermions. In a hot, expanding Universe a future-directed timelike vector field $u^\mu$ is selected by the radiation rest frame, so standard time reversal is not a symmetry of the \emph{state}. CTE restores a discrete invariance of the \emph{action} by compensating time-orientation reversal with an axial rotation and a shift of a pseudoscalar St\"uckelberg field $\Phi$. When realized nonlinearly, CTE yields a conserved Noether current in the chiral limit and sharply constrains CP-odd operators: at lowest dimension, the only local, shift-symmetric, parity-odd couplings consistent with CTE are the derivative portal $(\partial_\mu \Phi)J_{B-L}^\mu/M_*$ and the topological densities $(\Phi R\tilde R)$ and $(\Phi \mathcal{N}\mathcal{Y})$, where $R\tilde R$ is the Pontryagin density and $\mathcal{N}\mathcal{Y}$ is the Nieh-Yan invariant \cite{Sakharov1967,tHooft:1976PRD,Fujikawa:1979PRL,Weinberg:2008cosmo,JackiwPi2003,AlexanderYunes2009,Davoudiasl2004}.
We work on a four-dimensional Lorentzian manifold $(\mathcal{M},g_{\mu\nu})$ with signature $(-,+,+,+)$ equipped with a tetrad $e^a{}_\mu$ and a metric-compatible spin connection $\omega_\mu{}^{ab}$. Chiral fermions $\psi$ minimally coupled to gauge fields $A_\mu$ and torsion have Dirac action
\begin{equation}
S_{\rm D}=\int d^4x\, e\, \bar{\psi}\big(i\gamma^\mu \mathcal{D}_\mu-m\big)\psi,
\qquad
\gamma^\mu \equiv e^\mu{}_a \gamma^a,
\qquad
\mathcal{D}_\mu \equiv \partial_\mu + \frac{1}{4}\omega_{\mu ab}\gamma^{ab}-i A_\mu,
\label{eq:Dirac}
\end{equation}
where $e\equiv \det(e^a{}_\mu)$, $\gamma^{ab}\equiv \tfrac{1}{2}[\gamma^a,\gamma^b]$, and $\gamma^5\equiv i \gamma^0\gamma^1\gamma^2\gamma^3$. Decomposing $\omega_{\mu ab}=\tilde{\omega}_{\mu ab}+K_{\mu ab}$ into Levi-Civita and contorsion parts, the torsion is
\begin{equation}
T^a{}_{\mu\nu} = 2\partial_{[\mu}e^a{}_{\nu]} + 2 \omega^a{}_{b[\mu} e^b{}_{\nu]},
\end{equation}
with totally antisymmetric axial vector $S_\mu \propto \epsilon_{\mu\nu\rho\sigma} T^{\nu\rho\sigma}$ coupling to the axial current $J_5^\mu \equiv \bar{\psi}\gamma^\mu\gamma^5\psi$ \cite{Hehl:1976RMP,Shapiro2002,NiehYan1982}. We augment the theory by a shift-symmetric pseudoscalar $\Phi\to\Phi+\mathrm{const}$ that compensates anomalous phases and encodes the Nambu-Goldstone (NG) mode of spontaneously broken CTE:
\begin{equation}
S_\Phi = \int d^4x \, e \, \Bigg[ -\frac{1}{2}(\nabla \Phi)^2-V(\Phi)
+ \frac{\partial_\mu \Phi}{M_*} J_{B-L}^\mu
+ \frac{c_A}{f_\Phi} \Phi F\tilde{F}
+ \frac{c_g}{f_\Phi} \Phi R\tilde{R}
+ \frac{c_{\rm NY}}{f_\Phi} \Phi \mathcal{N}\mathcal{Y} \Bigg],
\label{eq:PhiAction}
\end{equation}
where $F\tilde{F} \equiv \tfrac{1}{2}\epsilon^{\mu\nu\rho\sigma} F_{\mu\nu} F_{\rho\sigma}$, $R\tilde{R} \equiv \tfrac{1}{2} \epsilon^{\mu\nu\rho\sigma} R^\alpha{}_{\beta\mu\nu} R^\beta{}_{\alpha\rho\sigma}$, and $\mathcal{N}\mathcal{Y}\equiv d(e^a \wedge T_a)=T^a \wedge T_a-e^a \wedge e^b \wedge R_{ab}$ \cite{NiehYan1982,ChandiaZanelli1997}. Here $M_*$ and $f_\Phi$ are the portal and decay scales, and $c_A,c_g,c_{\rm NY}$ are fixed by anomaly matching.
CTE acts as the diagonal transformation
\begin{equation}
\mathcal{T}_\chi(\sigma):\quad
x^\mu \mapsto x^\mu-\sigma(x) u^\mu,
\qquad
\psi \mapsto e^{i \sigma(x) \gamma^5} \mathcal{T}\psi,
\qquad
\Phi \mapsto \Phi + f_\Phi \sigma(x),
\label{eq:CTEdef}
\end{equation}
where $\mathcal{T}$ is microscopic (antiunitary) time reversal on spinors and $\sigma(x)$ is local. The diffeomorphism generated by $\sigma u^\mu$ contributes the Noether current $T^{\mu\nu}u_\nu$, the chiral rotation contributes $J_5^\mu$, and the shift of $\Phi$ compensates the anomalous Jacobian. Using Fujikawa’s method \cite{Fujikawa:1979PRL}, the axial anomaly in a torsional background is
\begin{equation}
\nabla_\mu J_5^\mu = 2i m \bar{\psi}\gamma^5 \psi + \frac{g^2}{16\pi^2} F\tilde{F}
+\frac{1}{384\pi^2} R\tilde{R} + c_{\rm NY} \mathcal{N}\mathcal{Y},
\label{eq:axialAnomaly}
\end{equation}
where the regulator-dependent Nieh-Yan coefficient is absorbed by the $(\Phi,\mathcal{N}\mathcal{Y})$ counterterm \cite{ChandiaZanelli1997,Shapiro2002}. The Wess-Zumino variation in \eqref{eq:PhiAction} cancels the topological densities,
\begin{equation}
\delta_\sigma S_{\rm WZ} =
-\int d^4x \, e \, \sigma(x)\Bigg( \frac{g^2}{16\pi^2} F\tilde{F}
+\frac{1}{384\pi^2} R\tilde{R} + c_{\rm NY} \mathcal{N}\mathcal{Y} \Bigg),
\end{equation}
so the CTE Ward identity becomes (up to explicit chiral breaking by $m$)
\begin{equation}
\nabla_\mu \mathcal{J}^\mu_{\rm CTE} = 2 i m \bar{\psi} \gamma^5 \psi,
\qquad
\mathcal{J}^\mu_{\rm CTE} \equiv T^{\mu\nu} u_\nu + J_5^\mu-\frac{1}{f_\Phi}
\Big( c_A K^\mu_{\rm CS}[A] + c_g K^\mu_{\rm CS}[\omega] + c_{\rm NY} Y^\mu \Big),
\label{eq:CTEcurrent}
\end{equation}
with $\nabla_\mu K^\mu_{\rm CS}[A]=F\tilde{F}$, $\nabla_\mu K^\mu_{\rm CS}[\omega]=R\tilde{R}$, and $\nabla_\mu Y^\mu=\mathcal{N}\mathcal{Y}$ \cite{JackiwPi2003,AlexanderYunes2009}. Thus, in the chiral limit the time-translation current along $u^\mu$ and the axial current admit a Chern-Simons improvement that renders the diagonal CTE symmetry exact.
The baryon asymmetry enters through the grand-canonical functional coupling the NG mode to the slow conserved charge $(B-L)$,
\begin{equation}
\mathcal{F}_{B-L}[\Phi,g] = \int d^4x \, e \, \frac{\partial_\mu \Phi}{M_*} J_{B-L}^\mu,
\label{eq:BLfunctional}
\end{equation}
which is CTE invariant because $\partial_\mu\Phi$ is invariant and $J_{B-L}^\mu$ is anomaly-free in the SM. This defines the spacetime-dependent chemical potential $\mu_{B-L}\equiv u^\mu\partial_\mu\Phi/M_*$ that biases baryon- and lepton-number violating reactions in (near) equilibrium, while $(\Phi R\tilde{R})$ and $(\Phi\mathcal{N}\mathcal{Y})$ are geometric CP-odd sources activated when parity-violating dynamics yields $\langle R\tilde{R}\rangle\neq 0$ \cite{Davoudiasl2004,LueWangKamionkowski1999,Alexander:2006PRL}.
In spatially flat FRW, $ds^2=-dt^2+a^2(t)d\vec{x}^2$, with comoving $u^\mu=(1,\vec 0)$ (so $u_\mu u^\mu=-1$), the homogeneous gradient is timelike,
\begin{equation}
\big\langle \nabla_\mu \Phi \big\rangle = \dot\Phi_0(t)\,\delta^0_{\mu} = -\,\dot\Phi_0(t)\,u_\mu,
\qquad
\big\langle \nabla^\mu \Phi \big\rangle = -\,\dot\Phi_0(t)\,u^\mu,
\end{equation}
and spontaneously breaks the diagonal generator in \eqref{eq:CTEdef}. {We decompose the CTE field into a homogeneous background plus fluctuations, $\Phi(x)=\Phi_0(t)+\delta\Phi(x)$, and in the FRW background $ds^2$ we assume spatial homogeneity for $\Phi_0$, i.e. $\partial_i\Phi_0=0$. Consequently the background gradient is purely timelike and aligned with the comoving four-velocity.} After integrating out torsion and fast modes, the low-energy Lagrangian is
\begin{equation}
\mathcal{L}_{\rm eff} = -\frac{1}{2} (\partial \Phi)^2-V(\Phi)
+ \frac{\partial_\mu \Phi}{M_*} J_{B-L}^\mu
+ \frac{c_g}{f_\Phi} \Phi R\tilde{R}
+ \frac{c_{\rm NY}}{f_\Phi} \Phi \mathcal{N}\mathcal{Y} + \ldots,
\label{eq:Leff}
\end{equation}
and the $\Phi$ equation of motion reads
\begin{equation}
\nabla_\mu \nabla^\mu \Phi-V'(\Phi) =
-\frac{1}{M_*} \nabla_\mu J_{B-L}^\mu
-\frac{c_g}{f_\Phi} R\tilde{R}
-\frac{c_{\rm NY}}{f_\Phi} \mathcal{N}\mathcal{Y}.
\label{eq:PhiEOM}
\end{equation}
This identifies three sources for the NG background: hydrodynamic drift via $(B-L)$, a Pontryagin source, and a Nieh-Yan source. In radiation domination $\langle R\tilde{R}\rangle=0$ at the homogeneous level, but parity-violating tensor fluctuations in CS gravity generate a chiral GW ensemble with $\langle R\tilde{R}\rangle\neq 0$, correlating the sign of $\dot\Phi_0$ with tensor helicity \cite{LueWangKamionkowski1999,JackiwPi2003,AlexanderYunes2009}.
The functional \eqref{eq:BLfunctional} reduces homogeneously to
\begin{equation}
\mathcal{F}_{B-L} = \int dt \, a^3(t) \, \frac{\dot{\Phi}_0(t)}{M_*} n_{B-L}(t),
\label{eq:BLhom}
\end{equation}
where $n_{B-L}\equiv u_\mu J_{B-L}^\mu$. Near equilibrium, $n_{B-L}=\chi_{B-L}\mu_{B-L}+\mathcal{O}(\mu_{B-L}^3)$ with $\chi_{B-L}\sim T^2$ \cite{Weinberg:2008cosmo}. When fast SM reactions set all other chemical potentials to zero, the slow mode $(B-L)$ freezes at $T_D$, yielding the comoving excess
\begin{equation}
Y_{B-L} \equiv \frac{n_{B-L}}{s} \simeq
c_{\rm eq}\,\frac{\dot{\Phi}_0}{M_* T}\Big|_{T=T_D}
+ c_g \frac{\Phi_0}{f_\Phi}\frac{\langle R\tilde{R} \rangle}{s H}\Big|_{T=T_D}
+ c_{\rm NY} \frac{\Phi_0}{f_\Phi}\frac{\langle \mathcal{N}\mathcal{Y} \rangle}{s H}\Big|_{T=T_D},
\label{eq:YBL}
\end{equation}
with $s$ the entropy density, $H$ the Hubble rate, and $c_{\rm eq}$ a calculable susceptibility factor. Equation \eqref{eq:YBL} is the macroscopic imprint of the CTE Ward identity \eqref{eq:CTEcurrent}: a time-asymmetric cosmological state is compatible with an exact microscopic symmetry once the time flow is compensated by a chiral rotation and a shift of $\Phi$, leaving precisely the anomaly topological densities as geometric sources.
CTE also fixes the CP-odd operator basis. Invariance under $\Phi\to\Phi+f_\Phi\sigma$ and $\psi\to e^{i\sigma\gamma^5}\psi$ (up to total derivatives) implies that at mass dimension five the only independent geometric operators are
\begin{equation}
\mathcal{O}_1 = \frac{\partial_\mu \Phi}{M_*} J_{B-L}^\mu,
\qquad
\mathcal{O}_2 = \frac{\Phi}{f_\Phi} R\tilde{R},
\label{eq:Ops}
\end{equation}
with optional torsional companion $(\Phi/f_\Phi)\mathcal{N}\mathcal{Y}$. Figure~\ref{fig:cte-schematic} summarizes the FRW bias $\mu_{B-L}=\dot\Phi/M_*$ from $\mathcal{O}_1$ and tensor birefringence/chirality from $\mathcal{O}_2$.

The operator $\mathcal{O}_1$ is the unique derivative portal to an anomaly-free slow charge and furnishes $\mu_{B-L}=\dot{\Phi}_0/M_*$. The operator $\mathcal{O}_2$ is topological and P-odd: it modifies tensor propagation through dCS gravity and generates a helicity asymmetry $\Pi \propto \mu_g \dot{\Phi}_0/M_{\rm Pl}^2$, with $\mu_g\equiv c_g$ up to normalization \cite{JackiwPi2003,AlexanderYunes2009}. The Nieh-Yan coupling becomes physical in the presence of torsion; in Einstein-Cartan theory torsion is nondynamical and induces a finite axial contact interaction and a rapidly redshifting ($\propto a^{-6}$) spin density \cite{Hehl:1976RMP,Shapiro2002}.
The same operators feed into transport. Covariantly, the baryon-number current obeys
\begin{multline}
\nabla_\mu J_B^\mu = \mathcal{C}_{\rm EW}[J] + \mathcal{C}_{\Delta B, \Delta L}[J]
+ \frac{1}{M_*}\nabla_\mu\!\left(\partial^\mu \Phi \,\frac{\partial \mathcal{P}}{\partial \mu_{B-L}}\right)
+ \frac{c_g}{f_\Phi} \mathcal{S}_{\rm CS}[g,\Phi]
+ \frac{c_{\rm NY}}{f_\Phi} \mathcal{S}_{\rm NY}[e,\omega,\Phi],
\label{eq:Transport}
\end{multline}
where $\mathcal{C}$ are collision integrals for baryon/lepton violation and $\mathcal{P}$ is the pressure; the last terms arise from varying $\int e\,\Phi R\tilde{R}$ and $\int e\,\Phi\mathcal{N}\mathcal{Y}$. In exactly homogeneous FLRW with $\Phi=\Phi(t)$ one has $R\tilde R=0$ at the background level, so $\dot\Phi_0/M_*$ acts as a homogeneous chemical potential and only parity-odd tensor perturbations can generate a coarse-grained $\langle R\tilde R\rangle\neq 0$. Consequently, the sign of the final asymmetry is locked to $\mathrm{sign}(\dot\Phi_0)$ and to tensor handedness, yielding an observationally falsifiable link to TB/EB correlations and SGWB helicity \cite{LueWangKamionkowski1999,Alexander:2006PRL,AlexanderYunes2009}.
In summary, CTE promotes a diagonal combination of time flow and chiral phase to an exact symmetry in the chiral limit, realized by a shift-symmetric pseudoscalar $\Phi$. The Ward identity \eqref{eq:CTEcurrent} fixes the anomaly-canceling improvement terms, while the operator analysis isolates a unique hydrodynamic portal and topological parity-odd couplings. This provides the backbone for the subsequent quantitative analysis of yields and parity observables \cite{Weinberg:2008cosmo,AlexanderYunes2009}.

\section{Einstein-Cartan-Immirzi-Chern-Simons (ECICS)}\label{sec3}

We formulate the gravitational sector in first-order (tetrad) variables on an oriented four-dimensional Lorentzian manifold $(\mathcal{M},g)$ with signature $(-,+,+,+)$. The fundamental fields are the coframe $e^a=e^a{}_\mu dx^\mu$ and a metric-compatible $\mathfrak{so}(1,3)$ connection $\omega^{ab}=\omega^{ab}{}_\mu dx^\mu$ ($\omega^{ab}=-\omega^{ba}$), with $\eta_{ab}=\mathrm{diag}(-,+,+,+)$ and $g_{\mu\nu}=\eta_{ab}e^a{}_\mu e^b{}_\nu$, $e\equiv \det(e^a{}_\mu)=\sqrt{-g}$. The curvature and torsion two-forms are
\begin{equation}
R^{ab}=d\omega^{ab}+\omega^a{}_c\wedge\omega^{cb},
\qquad
T^a=De^a=de^a+\omega^a{}_b\wedge e^b,
\label{eq:curvaturetorsion}
\end{equation}
and we employ the internal Lorentz dual $\star$ on algebra indices, $(\star X)^{ab}\equiv \tfrac{1}{2}\epsilon^{ab}{}_{cd}X^{cd}$, and the spacetime Hodge dual $*$ on forms. We adopt $\epsilon_{0123}=+1$ (local frame), so $\epsilon_{\mu\nu\rho\sigma}=e\,\varepsilon_{\mu\nu\rho\sigma}$ with $\varepsilon_{0123}=+1$.
The Einstein-Cartan (EC) action is the Palatini action in differential forms,
\begin{equation}
S_{\mathrm{EC}}[e,\omega]=\frac{1}{2\kappa}\int_{\mathcal{M}} e^a\wedge e^b\wedge \star R_{ab},
\qquad
\kappa\equiv 8\pi G,
\label{eq:EC}
\end{equation}
reducing to Einstein-Hilbert when torsion vanishes. The parity-odd Holst deformation introduces the real Immirzi parameter $\gamma\in\mathbb{R}\setminus\{0\}$ \cite{Holst1996,Ashtekar:1986PRL},
\begin{equation}
S_{\mathcal{H}}[e,\omega;\gamma]=\frac{1}{2\kappa\gamma}\int_{\mathcal{M}} e^a\wedge e^b\wedge R_{ab},
\label{eq:Holst}
\end{equation}
so that
\begin{equation}
S_{\mathrm{EC+H}}[e,\omega;\gamma]=\frac{1}{2\kappa}\int_{\mathcal{M}} e^a\wedge e^b\wedge\Big(\star R_{ab}+\frac{1}{\gamma}R_{ab}\Big).
\label{eq:ECH}
\end{equation}
By the Nieh-Yan identity \cite{NiehYan1982,ChandiaZanelli1997}
\begin{equation}
d\!\left(e^a\wedge T_a\right)=T^a\wedge T_a-e^a\wedge e^b\wedge R_{ab},
\label{eq:NYidentity}
\end{equation}
the Holst density differs from a total derivative by a torsion-squared term. Hence $S_{\mathcal{H}}$ is inert in the torsionless sector but affects the algebraic torsion induced by spin, controlling parity properties of fermion-induced contact interactions \cite{Hehl:1976RMP,Shapiro2002,FreidelMinicTakeuchi:2005PRD,PerezRovelli:2006PRD,Mercuri:2006PRD}.
We also include the gravitational Chern-Simons (CS) coupling, where a pseudoscalar $\theta(x)$ multiplies the Pontryagin density \cite{JackiwPi2003,AlexanderYunes2009}:
\begin{equation}
S_{\mathrm{CS}}[\theta,\omega]=\frac{\alpha}{4}\int_{\mathcal{M}} \theta\,R^{ab}\wedge R_{ab}
=\frac{\alpha}{4}\int_{\mathcal{M}} \theta\, dQ_{\mathrm{CS}}(\omega),
\label{eq:SCS}
\end{equation}
with $Q_{\mathrm{CS}}(\omega)=\omega^{ab}\wedge d\omega_{ab}+\tfrac{2}{3}\,\omega^a{}_c\wedge\omega^c{}_d\wedge\omega^{db}$. In components,
\begin{multline}
{}^{\star}RR \equiv \frac{1}{2}\,\epsilon^{\mu\nu\rho\sigma}\,R^\alpha{}_{\beta\mu\nu}R^\beta{}_{\alpha\rho\sigma}
=\frac{1}{e}\,\partial_\mu K^\mu(\Gamma), \\
K^\mu=\epsilon^{\mu\alpha\beta\gamma}\left(\Gamma^\rho{}_{\alpha\sigma}\partial_\beta \Gamma^\sigma{}_{\gamma\rho}
+\frac{2}{3}\Gamma^\rho{}_{\alpha\sigma}\Gamma^\sigma{}_{\beta\eta}\Gamma^\eta{}_{\gamma\rho}\right),
\label{eq:Pontryagin}
\end{multline}
where $\Gamma^\rho{}_{\mu\nu}$ is the affine connection built from $(e^a{}_\mu,\omega^{ab}{}_\mu)$ after eliminating nonmetricity. We endow $\theta$ with standard dynamics,
\begin{equation}
S_{\theta}[\theta,e]=-\frac{\beta}{2}\int_{\mathcal{M}} d\theta\wedge *d\theta-\int_{\mathcal{M}} e\,V(\theta),
\label{eq:Stheta}
\end{equation}
with $\alpha$ controlling parity violation in the curvature sector and $\beta>0$ the kinetic normalization.
Spin-$\tfrac{1}{2}$ matter couples minimally to $(e^a,\omega^{ab})$ via \cite{Hehl:1976RMP,Shapiro2002}
\begin{equation}
S_{\psi}[e,\omega,\psi]=\int d^4x\,e\left[\frac{i}{2}\left(\bar{\psi}\gamma^\mu D_\mu \psi-D_\mu \bar{\psi}\,\gamma^\mu \psi\right)-m\,\bar{\psi}\psi\right],
\qquad
D_\mu\psi\equiv \partial_\mu\psi+\frac{1}{4}\omega_{\mu ab}\gamma^{ab}\psi,
\label{eq:DiracAction}
\end{equation}
with $\gamma^\mu=e^\mu{}_a\gamma^a$, $\gamma^{ab}\equiv \tfrac{1}{2}[\gamma^a,\gamma^b]$, and $\gamma^5\equiv i\gamma^0\gamma^1\gamma^2\gamma^3$. The total ECICS action is
\begin{equation}
S_{\mathrm{ECICS}}[e,\omega,\theta;\psi]
= S_{\mathrm{EC+H}}[e,\omega;\gamma] + S_{\mathrm{CS}}[\theta,\omega] + S_{\theta}[\theta,e] + S_{\psi}[e,\omega,\psi].
\label{eq:SECICS}
\end{equation} 
Varying \eqref{eq:SECICS} independently with respect to $\omega^{ab}$ (using $\delta R^{ab}=D(\delta\omega^{ab})$ and integrating by parts) yields the Cartan equation with Holst and CS deformations,
\begin{multline}
\tau^{ab}
=\frac{1}{\kappa}\,e^{[a}\wedge\left(\star+\frac{1}{\gamma}\right)R^{b]}{}_{c}\wedge e^c
-\frac{1}{\kappa}\,D\!\left[e^{[a}\wedge\left(\star+\frac{1}{\gamma}\right)e^{b]}\right]
+\frac{\alpha}{2}\,d\theta\wedge R^{ab},
\label{eq:CartanGeneral}
\end{multline}
where $\tau^{ab}\equiv \delta S_{\psi}/\delta\omega_{ab}$ is the spin three-form. Using $De^a=T^a$ and $D(e^a\wedge e^b)=T^a\wedge e^b-e^a\wedge T^b$, the purely gravitational part reduces to a linear map on $e^{[a}\wedge T^{b]}$. Defining the internal projector $P\equiv \mathbb{1}+\gamma^{-1}\star$ with inverse $P^{-1}=\tfrac{1}{1+\gamma^2}\left(\mathbb{1}-\gamma^{-1}\star\right)$ and neglecting the $(d\theta)$ term for the moment,\footnote{The CS piece $d\theta\wedge R^{ab}$ is suppressed when $d\theta$ is small on curvature scales \cite{AlexanderYunes2009}; including it does not change the algebraic (nondynamical) character of torsion.} one obtains
\begin{equation}
P^{ab}{}_{cd}\,e^{[c}\wedge T^{d]}=\kappa\,\tau^{ab}.
\label{eq:CartanCompact}
\end{equation}
For a minimally coupled Dirac field, $\tau^{ab}$ is proportional to the axial current one-form $J_5\equiv J_5^a e_a$ with $J_5^a\equiv \bar{\psi}\gamma^a\gamma^5\psi$ \cite{Hehl:1976RMP,Shapiro2002}. Solving \eqref{eq:CartanCompact} gives $\omega^{ab}=\tilde{\omega}^{ab}+K^{ab}$, with contorsion (components) \cite{FreidelMinicTakeuchi:2005PRD,PerezRovelli:2006PRD,Mercuri:2006PRD}
\begin{align}
K_{abc}&=\frac{\kappa}{4}\,\frac{1}{1+\gamma^2}\left(\epsilon_{abcd}\,J_5^d-\frac{1}{\gamma}\,(\eta_{ac}J_{5b}-\eta_{bc}J_{5a})\right),
\nonumber\\
T^a&=K^a{}_b\wedge e^b.
\label{eq:contorsion}
\end{align}
Thus torsion is nondynamical and fully sourced by $J_5^\mu$; substituting \eqref{eq:contorsion} back into the action generates local four-fermion operators. In the EC limit $\gamma\to\infty$ one recovers the axial-axial Hehl-Datta term, while finite $\gamma$ induces a parity-violating admixture controlled by $1/(1+\gamma^2)$ \cite{Hehl:1976RMP,FreidelMinicTakeuchi:2005PRD,Mercuri:2006PRD}. 
Varying with respect to the coframe and eliminating torsion via \eqref{eq:contorsion} yields the modified Einstein equation
\begin{equation}
G_{\mu\nu}(\tilde{g})+\frac{\alpha}{\kappa}\,C_{\mu\nu}(\theta,\tilde{g})
=\kappa\!\left(T^{(\psi)}_{\mu\nu}+T^{(4\psi)}_{\mu\nu}\right)+T^{(\theta)}_{\mu\nu},
\label{eq:EinsteinCS}
\end{equation}
where $T^{(4\psi)}_{\mu\nu}$ encodes the torsion-induced contact terms, $T^{(\theta)}_{\mu\nu}=\beta\left(\nabla_\mu\theta\nabla_\nu\theta-\tfrac{1}{2}g_{\mu\nu}(\nabla\theta)^2\right)-g_{\mu\nu}V(\theta)$, and the parity-odd Cotton-like tensor is \cite{JackiwPi2003,AlexanderYunes2009,YunesPretorius:2009PRD}
\begin{multline}
C^{\mu\nu}(\theta,\tilde{g})
=-(\nabla_\sigma \theta)\,\epsilon^{\sigma\alpha\beta(\mu}\,\nabla_\alpha R^{\nu)}{}_{\beta}
-\frac{1}{2}\,(\nabla_\sigma\nabla_\tau \theta)\,{}^{\star}R^{\tau(\mu\nu)\sigma}, \\
{}^{\star}R^{\mu\nu}{}_{\rho\sigma}\equiv \tfrac{1}{2}\,\epsilon^{\mu\nu\alpha\beta}R_{\alpha\beta\rho\sigma}.
\label{eq:Cotton}
\end{multline}
Equation \eqref{eq:EinsteinCS} shows two conceptually distinct parity-odd channels: dCS gravity through $C_{\mu\nu}$ (affecting tensor/vector propagation and TB/EB-type signatures), and Holst-induced contact effects through $T^{(4\psi)}_{\mu\nu}$ when $\langle J_5^\mu\rangle\neq 0$ (short-range and Planck suppressed) \cite{AlexanderYunes2009,Shapiro2002}. Finally, varying with respect to $\theta$ gives
\begin{equation}
\beta\,\Box\theta - V'(\theta) = -\,\frac{\alpha}{4}\,{}^{\star}RR,
\label{eq:thetaEOM}
\end{equation}
so curvature backgrounds with nonzero Pontryagin density source $\theta$. 
For completeness, in Euclidean signature one may define the Pontryagin index \cite{Holst1996,ChandiaZanelli1997}
\begin{equation}
P\equiv \frac{1}{32\pi^2}\int_{\mathcal{M}} R^{ab}\wedge R_{ab}\in\mathbb{Z},
\end{equation}
so for constant $\theta$ the CS term contributes the topological weight $S_{\rm CS}=8\pi^2\,\alpha\,\theta\,P$, while for spacetime-dependent $\theta$ one may equivalently use the boundary-current form $\int d\theta\wedge Q_{\rm CS}$ implied by \eqref{eq:SCS}-\eqref{eq:Pontryagin}. The Immirzi parameter weights the torsional completion of an otherwise total derivative via \eqref{eq:NYidentity}, while the CS coupling endows $\theta$ with a topological current whose divergence is ${}^{\star}RR$ \cite{JackiwPi2003,AlexanderYunes2009}.
Collecting the ingredients, the parity-odd gravitational-topological sector underpinning our framework is
\begin{multline}
S_{\mathrm{ECICS}}[e,\omega,\theta;\psi]
=\frac{1}{2\kappa}\int_{\mathcal{M}} e^a\wedge e^b\wedge\Big(\star R_{ab}+\frac{1}{\gamma}R_{ab}\Big)
+\frac{\alpha}{4}\int_{\mathcal{M}} \theta\,R^{ab}\wedge R_{ab} \\
-\frac{\beta}{2}\int_{\mathcal{M}} d\theta\wedge *d\theta
-\int_{\mathcal{M}} e\,V(\theta)
+S_{\psi},
\label{eq:FinalAction}
\end{multline}
to be supplemented by the algebraic torsion solution \eqref{eq:contorsion}, the modified Einstein equation \eqref{eq:EinsteinCS}, and the scalar equation \eqref{eq:thetaEOM}. The ECICS system is therefore a parity-violating extension of Einstein-Cartan gravity with two CP-odd structures: the Holst density (weighted by $\gamma$) controlling the torsional response to spin, and the CS density (sourced by $\theta$) modifying the metric equations via $C_{\mu\nu}$.

\section{Stuckelberg \texorpdfstring{$U(1)_{B-L}$}{U(1)_{B-L}} Portal}\label{sec4}
A minimal gauge-invariant way to give a mass to a vector associated with baryon minus lepton number is the St\"uckelberg mechanism, which realizes a mass term without a Higgs field charged under the Abelian factor and without explicit breaking of the gauge symmetry \cite{Stueckelberg:1938,Ruegg:2003IMPA}. We gauge $U(1)_{B-L}$ with an Abelian field $B_\mu$ and coupling $g_{BL}$, and introduce a pseudoscalar St\"uckelberg field $\sigma$ providing the longitudinal mode. Integrating out the heavy vector generates a derivative coupling to the conserved $(B\!-\!L)$ current, and (when present) UV-induced topological counterterms furnish a gauge-invariant interface to curvature Chern-Simons structures \cite{Kors:2004PLB,Kors:2004JHEP,GreenSchwarz:1984PLB,Preskill:1991AnnPhys}.
On a curved background $(g_{\mu\nu},\nabla_\mu)$, with $F_{\mu\nu}\equiv \nabla_\mu B_\nu-\nabla_\nu B_\mu$ and conserved current $\nabla_\mu J^\mu_{B-L}=0$, the St\"uckelberg Lagrangian is
\begin{equation}
\mathcal{L}_{\text{Stk}}
=
-\frac{1}{4}F_{\mu\nu}F^{\mu\nu}
+\frac{1}{2}\big(m_B B_\mu-\nabla_\mu\sigma\big)\big(m_B B^\mu-\nabla^\mu\sigma\big)
+g_{BL}B_\mu J^\mu_{B-L},
\label{eq:LStueckelberg}
\end{equation}
invariant under local $U(1)_{B-L}$ transformations
\begin{equation}
\delta_\alpha B_\mu=\nabla_\mu\alpha,
\qquad
\delta_\alpha \sigma=m_B\alpha,
\qquad
\delta_\alpha J^\mu_{B-L}=0.
\label{eq:U1transform}
\end{equation}
It is convenient to define the St\"uckelberg scale $f_\sigma$ by $m_B\equiv g_{BL}f_\sigma$. Varying \eqref{eq:LStueckelberg} yields
\begin{equation}
\nabla_\nu F^{\nu\mu}+m_B^2 B^\mu-m_B\nabla^\mu\sigma+g_{BL}J^\mu_{B-L}=0,
\qquad
\square\sigma-m_B\nabla_\mu B^\mu=0,
\label{eq:EOM_basic}
\end{equation}
with $\square\equiv \nabla_\mu\nabla^\mu$. No gauge fixing is required; imposing Lorenz gauge $\nabla_\mu B^\mu=0$ is only an intermediate simplification for the heavy-mass expansion, and the resulting operator basis is manifestly gauge invariant \cite{Ruegg:2003IMPA}. {In particular, the EFT obtained after integrating out the heavy vector is expressed entirely in terms of gauge-invariant operators (conserved currents and $F_{\mu\nu}$), so any intermediate gauge choice cannot affect the resulting local operator basis.}
To parameterize possible UV-induced curvature couplings in a gauge-covariant way, introduce the gravitational Chern-Simons current $K^\mu_{\mathrm{CS}}$ satisfying $\nabla_\mu K^\mu_{\mathrm{CS}}=\tfrac{1}{2}{}^{\star}RR$, where ${}^{\star}RR\equiv \tfrac{1}{2}\epsilon^{\mu\nu\rho\sigma}R^\alpha{}_{\beta\mu\nu}R^\beta{}_{\alpha\rho\sigma}$. The gauge-invariant pair of couplings
\begin{equation}
\mathcal{L}_{\text{top}}=\xi\,B_\mu K^\mu_{\mathrm{CS}}+\frac{\xi}{2m_B}\sigma\,{}^{\star}RR
\label{eq:LtopGaugeInvariant}
\end{equation}
is the gravitational analogue of Green--Schwarz inflow: under \eqref{eq:U1transform} the variation of $B_\mu K^\mu_{\mathrm{CS}}$ is canceled by the shift of $\sigma$ \cite{GreenSchwarz:1984PLB,Preskill:1991AnnPhys,JackiwPi2003,AlexanderYunes2009}. In an anomaly-free $U(1)_{B-L}$ realization (e.g.\ with three $\nu_R$) the net $\xi$ can vanish at one loop, but \eqref{eq:LtopGaugeInvariant} remains the appropriate covariant parameterization of any residual UV-induced topological coupling.
In the massive-vector regime $m_B^2\gg\{\square,R,g_{BL}\sqrt{J^2}\}$, one can solve \eqref{eq:EOM_basic} locally as an expansion in $m_B^{-1}$. In Lorenz gauge, the leading solution is
\begin{equation}
B^\mu
=
\frac{1}{m_B}\nabla^\mu\sigma
-\frac{g_{BL}}{m_B^2}J^\mu_{B-L}
-\frac{\xi}{m_B^2}K^\mu_{\mathrm{CS}}
+\mathcal{O}\!\left(\frac{\nabla^2}{m_B^3},\,\frac{R}{m_B^3}\right).
\label{eq:Bsolution}
\end{equation}
Substituting \eqref{eq:Bsolution} into $\mathcal{L}_{\text{Stk}}+\mathcal{L}_{\text{top}}$ and completing the square yields the low-energy effective Lagrangian
\begin{multline}
\mathcal{L}_{\text{eff}}
=
\frac{g_{BL}}{m_B}\nabla_\mu\sigma\,J^\mu_{B-L}
+\frac{\xi}{m_B}\nabla_\mu\sigma\,K^\mu_{\mathrm{CS}}
-\frac{g_{BL}^2}{2m_B^2}J_{B-L,\mu}J^\mu_{B-L}
-\frac{g_{BL}\xi}{m_B^2}J_{B-L,\mu}K^\mu_{\mathrm{CS}}
-\frac{\xi^2}{2m_B^2}K_{\mathrm{CS},\mu}K^\mu_{\mathrm{CS}} \\
+\frac{\xi}{2m_B}\sigma\,{}^{\star}RR
+\cdots,
\label{eq:LeffFull}
\end{multline}
where the ellipsis denotes terms suppressed by additional powers of $m_B^{-1}$ and by derivatives/curvature. The $\nabla\sigma\cdot K_{\rm CS}$ and $\sigma\,{}^{\star}RR$ terms combine into a total derivative using $\nabla_\mu K^\mu_{\mathrm{CS}}=\tfrac{1}{2}{}^{\star}RR$:
\begin{equation}
\frac{\xi}{m_B}\nabla_\mu\sigma\,K^\mu_{\mathrm{CS}}+\frac{\xi}{2m_B}\sigma\,{}^{\star}RR
=
\frac{\xi}{m_B}\nabla_\mu\!\left(\sigma\,K^\mu_{\mathrm{CS}}\right),
\label{eq:boundary}
\end{equation}
and thus do not affect local bulk dynamics on globally hyperbolic spacetimes with standard falloff. The universal infrared bulk interaction is therefore the derivative portal
\begin{equation}
\mathcal{L}_{\text{portal}}
=
\frac{1}{M_*}\nabla_\mu\sigma\,J^\mu_{B-L}
-\frac{1}{2M_*^2}J_{B-L,\mu}J^\mu_{B-L}
+\mathcal{O}\!\left(\frac{\nabla^2}{m_B^3},\,\frac{R}{m_B^3}\right),
\qquad
M_*\equiv \frac{m_B}{g_{BL}}=f_\sigma,
\label{eq:PortalFinal}
\end{equation}
together with contact current-current interactions required by decoupling \cite{AppelquistCarazzone:1975PRD}. The operator $(\nabla_\mu\sigma\,J^\mu_{B-L}/M_*)$ is gauge invariant under \eqref{eq:U1transform}, covariant on curved backgrounds, and is the unique dimension-five coupling of a St\"uckelberg mode to a conserved hydrodynamic current.
For completeness, including $\mathcal{L}_{\text{top}}$ the equations of motion may be written as
\begin{equation}
\nabla_\nu F^{\nu\mu}+m_B^2 B^\mu-m_B\nabla^\mu\sigma+g_{BL}J^\mu_{B-L}+\xi K^\mu_{\mathrm{CS}}=0,
\label{eq:EOM_B_full}
\end{equation}
\begin{equation}
\square\sigma-m_B\nabla_\mu B^\mu+\frac{\xi}{2m_B}\,{}^{\star}RR=0,
\label{eq:EOM_sigma_full}
\end{equation}
whose compatibility follows from the Bianchi identity and $\nabla_\mu J^\mu_{B-L}=0$. At energies well below $m_B$, the dynamics reduces to \eqref{eq:PortalFinal} (plus higher-derivative/curvature corrections), providing the gauge-invariant and EFT-controlled portal that underlies the $(B\!-\!L)$ chemical bias used in the subsequent cosmological analysis.

\section{Cosmological Dynamics of \texorpdfstring{$\Phi$}{Phi}}
\label{sec5}
The pseudoscalar $\Phi$ provides the homogeneous, CP-odd background that selects a time orientation in the hot early Universe and biases otherwise CP-symmetric reaction networks through its derivative couplings. We treat $\Phi$ as spatially coherent on a spatially flat FLRW background with Hubble rate
\begin{equation}
H(t)\equiv \frac{\dot a(t)}{a(t)}\,.
\label{eq:FLRWmetric}
\end{equation}
At energies well below heavy thresholds, the homogeneous dynamics is captured by the covariant effective action
\begin{equation}
S_\Phi = \int d^4x\,\sqrt{-g}\,\bigg[-\frac{1}{2}\nabla_\mu\Phi\,\nabla^\mu\Phi-V(\Phi)
+ \frac{\alpha_g}{f_\Phi}\,\Phi\,R\tilde R
+ \frac{\alpha_T}{\Lambda^3}\,\Phi\,\mathcal{I}(g,\mathcal{T})\bigg],
\label{eq:Seff}
\end{equation}
where $R\tilde R \equiv \tfrac{1}{2}\epsilon^{\mu\nu\rho\sigma}R^\alpha{}_{\beta\mu\nu}R^\beta{}_{\alpha\rho\sigma}$ and $\mathcal{I}(g,\mathcal{T})$ is a covariant scalar built from the metric and the plasma stress tensor $\mathcal{T}_{\mu\nu}$ (e.g.\ $\mathcal{I}=G_{\mu\nu}\mathcal{T}^{\mu\nu}$). Varying \eqref{eq:Seff} yields
\begin{equation}
\square \Phi-V'(\Phi) = -\,\frac{\alpha_g}{f_\Phi}\,R\tilde R-\frac{\alpha_T}{\Lambda^3}\,\mathcal{I}(g,\mathcal{T}),
\label{eq:PhiCovEOM}
\end{equation}
with $\square\equiv \nabla_\mu\nabla^\mu$. For $\Phi=\Phi(t)$ on FLRW,
\begin{equation}
\ddot{\Phi} + 3H\dot{\Phi} + V'(\Phi) = S(t),
\qquad
S(t)\equiv -\,\frac{\alpha_g}{f_\Phi}\,\langle R\tilde R\rangle-\frac{\alpha_T}{\Lambda^3}\,\langle \mathcal{I}(g,\mathcal{T})\rangle,
\label{eq:backgroundEOM}
\end{equation}
where $\langle\cdots\rangle$ denotes a coarse-grained average over subhorizon plasma/metric fluctuations. In exactly homogeneous FLRW one has $R\tilde R=0$, but parity-violating tensor ensembles can generate $\langle R\tilde R\rangle\neq 0$ stochastically; $\langle\mathcal{I}\rangle$ is a scalar functional of the cosmological fluid. Equation \eqref{eq:backgroundEOM} isolates Hubble friction, the conservative force $V'(\Phi)$, and an external driving $S(t)$.
The stress tensor $T^{(\Phi)}_{\mu\nu}\equiv -\tfrac{2}{\sqrt{-g}}\,\delta S_\Phi/\delta g^{\mu\nu}$ takes the form
\begin{equation}
T^{(\Phi)}_{\mu\nu}= \nabla_\mu\Phi\,\nabla_\nu\Phi
-g_{\mu\nu}\Big(\tfrac{1}{2}\nabla_\alpha\Phi\,\nabla^\alpha\Phi + V(\Phi)\Big)
+\Pi^{(\mathrm{curv})}_{\mu\nu},
\label{eq:Tmunu}
\end{equation}
where $\Pi^{(\mathrm{curv})}_{\mu\nu}$ collects contributions from the curvature operators in \eqref{eq:Seff}. For homogeneous $\Phi(t)$, the canonical energy density and pressure are
\begin{equation}
\rho_\Phi = \frac{1}{2}\dot{\Phi}^2 + V(\Phi),
\qquad
p_\Phi = \frac{1}{2}\dot{\Phi}^2-V(\Phi),
\label{eq:rhopp}
\end{equation}
and $\Pi^{(\mathrm{curv})}_{\mu\nu}$ is either a boundary term (for $\Phi R\tilde R$ after isotropic averaging) or suppressed by $\Lambda^{-3}$ and background curvatures. The background expansion satisfies
\begin{equation}
\begin{aligned}
3M_{\mathrm {Pl}}^2 H^2 &= \rho_r + \rho_m + \rho_\Phi + \delta\rho_{\mathrm{curv}},\\
-2M_{\mathrm {Pl}}^2 \dot H &= \frac{4}{3}\rho_r + \rho_m + \dot{\Phi}^2 + \delta p_{\mathrm{curv}},
\end{aligned}
\label{eq:Friedmann2}
\end{equation}
with $(\delta\rho_{\mathrm{curv}},\delta p_{\mathrm{curv}})$ denoting any residual homogeneous contributions from the irrelevant operators in \eqref{eq:Seff}. Combining \eqref{eq:backgroundEOM} and \eqref{eq:rhopp} gives the exact energy-transfer relation
\begin{equation}
\dot{\rho}_\Phi + 3H(\rho_\Phi + p_\Phi) = \dot{\Phi}\,S(t),
\label{eq:continuity}
\end{equation}
so the driving $S(t)$ performs work on the $\Phi$ fluid with power density $\dot\Phi\,S$. 
If $|\ddot{\Phi}|\ll 3H|\dot{\Phi}|$, Eq.~\eqref{eq:backgroundEOM} reduces to
\begin{equation}
3H\dot{\Phi} + V'(\Phi) \simeq S(t),
\label{eq:slowroll}
\end{equation}
so that
\begin{align}
\dot{\Phi}(t) &\simeq \frac{S(t)-V'(\Phi)}{3H(t)},\nonumber
\\
\epsilon_\Phi &\equiv \frac{1}{2M_{\mathrm{Pl}}^2}\frac{\dot{\Phi}^2}{H^2}
\simeq \frac{\big(S(t)-V'(\Phi)\big)^2}{18\,M_{\mathrm{Pl}}^2 H^4}.
\label{eq:slowrollSolution}
\end{align}
When $S(t)$ varies slowly over a Hubble time and $V'(\Phi)$ is subdominant, $\dot\Phi/H$ is approximately constant and the sign of $\dot\Phi$ is set by $\mathrm{sign}[S(t)-V'(\Phi)]$, providing a sign-definite CP-odd background without requiring departures from local thermal equilibrium \cite{Linde:1983PLB,Mukhanov:2005book,Weinberg:2008cosmo,CohenKaplan1988,JoyceShaposhnikov:1997PRL,DineKusenko:2003RMP}. 
If $V'(\Phi)\approx 0$ and $S(t)\approx 0$ on Hubble timescales, then $\dot\Phi\propto a^{-3}$ and
\begin{equation}
\rho_\Phi = \frac{1}{2}\dot{\Phi}^2 \propto a^{-6},
\qquad
p_\Phi = \rho_\Phi,
\label{eq:stiff}
\end{equation}
so $\Phi$ behaves as a stiff fluid ($w_\Phi=+1$). This rapid dilution makes early kinetic misalignment harmless by BBN provided $\rho_\Phi/\rho_r\ll 1$ at the end of reheating \cite{Turner:1983PRD,KolbTurner:1990book}. 
Near a minimum with $m_\Phi^2\equiv V''(\Phi_\star)\gg H^2$, writing $\Phi=\Phi_\star+\delta\Phi$ and neglecting $S(t)$ on sub-Hubble timescales gives
\begin{equation}
\ddot{\delta\Phi}+3H\dot{\delta\Phi}+m_\Phi^2\,\delta\Phi \simeq 0,
\label{eq:oscillator}
\end{equation}
with WKB solution
\begin{equation}
\delta\Phi(t) \simeq \frac{\mathcal{A}}{a^{3/2}(t)}\,\cos\!\big(m_\Phi t+\varphi\big),
\label{eq:WKB}
\end{equation}
and oscillation-averaged equation of state
\begin{equation}
\langle \rho_\Phi \rangle \simeq \frac{m_\Phi^2\,\mathcal{A}^2}{2\,a^3(t)},
\qquad
\langle p_\Phi \rangle \simeq 0,
\label{eq:matterlike}
\end{equation}
so the condensate behaves as nonrelativistic matter \cite{Turner:1983PRD,Marsh:2016PhysRep}. A slowly varying $S(t)$ yields a particular solution $\delta\Phi_{\mathrm{part}}\approx S/m_\Phi^2$ that adiabatically tracks the driving.
When $V'(\Phi)\approx 0$, Eq.~\eqref{eq:backgroundEOM} admits the first integral
\begin{equation}
\dot{\Phi}(t) = a^{-3}(t)\left[\int^t dt'\,a^3(t')\,S(t') + C\right],
\label{eq:firstIntegral}
\end{equation}
with constant $C$ set by initial conditions. For $a\propto t^p$ (radiation: $p=\tfrac{1}{2}$; matter: $p=\tfrac{2}{3}$) and constant $S(t)=S_0$, one obtains
\begin{equation}
\dot{\Phi}(t)=\frac{S_0}{3p+1}\,t + C\,t^{-3p},
\label{eq:constS}
\end{equation}
where the decaying mode is erased by Hubble friction. 
The curvature operators in \eqref{eq:Seff} are irrelevant and remain perturbative provided
\begin{equation}
\varepsilon_g \equiv \left|\frac{\alpha_g}{f_\Phi}\,\frac{\langle R\tilde R\rangle}{3H\dot{\Phi}}\right|\ll 1,
\qquad
\varepsilon_T \equiv \left|\frac{\alpha_T}{\Lambda^3}\,\frac{\langle \mathcal{I}\rangle}{3H\dot{\Phi}}\right|\ll 1,
\label{eq:smallcoupling}
\end{equation}
and backreaction is negligible for
\begin{equation}
\Omega_\Phi \equiv \frac{\rho_\Phi}{3M_{\mathrm{Pl}}^2 H^2}\ll 1,
\qquad
\left|\frac{\delta\rho_{\mathrm{curv}}}{3M_{\mathrm{Pl}}^2 H^2}\right|\ll 1.
\label{eq:backreaction}
\end{equation}
BBN constraints are conveniently expressed as an ``extra radiation'' bound. Defining
\begin{equation}
\Delta N_{\rm eff}(T_{\rm BBN})
\equiv \frac{\rho_\Phi(T_{\rm BBN})}{\rho_{\nu,1}(T_{\rm BBN})},
\qquad
\rho_{\nu,1}(T)=\frac{7}{8}\frac{\pi^2}{30}\,T_\nu^4,
\end{equation}
one has
\begin{equation}
\frac{\rho_\Phi(T_{\rm BBN})}{\rho_\gamma(T_{\rm BBN})}
=\frac{7}{8}\left(\frac{T_\nu}{T}\right)^4 \Delta N_{\rm eff}(T_{\rm BBN})
\ \lesssim\ 0.07\text{--}0.13,
\label{eq:BBN_DNeff_bound}
\end{equation}
using representative $\Delta N_{\rm eff}(T_{\rm BBN})\lesssim 0.3$ and $(T_\nu/T)^4\in[(4/11)^{4/3},1]$ across the BBN era \cite{KolbTurner:1990book,Weinberg:2008cosmo}. In the kinetic/stiff regime \eqref{eq:stiff}, during radiation domination $T\propto a^{-1}$ implies
\begin{equation}
\left.\frac{\rho_\Phi}{\rho_r}\right|_{T_{\rm BBN}}
=
\left.\frac{\rho_\Phi}{\rho_r}\right|_{T_D}
\left(\frac{T_{\rm BBN}}{T_D}\right)^2,
\qquad (w_\Phi\simeq 1),
\label{eq:BBN_scaling_stiff}
\end{equation}
so for $T_D\gg T_{\rm BBN}$ the BBN contribution is parametrically negligible. If instead $\Phi$ is oscillatory by BBN, $\langle\rho_\Phi\rangle\propto a^{-3}$ and
\begin{equation}
\left.\frac{\rho_\Phi}{\rho_r}\right|_{T_{\rm BBN}}
=
\left.\frac{\rho_\Phi}{\rho_r}\right|_{T_{\rm osc}}
\left(\frac{T_{\rm osc}}{T_{\rm BBN}}\right),
\qquad (w_\Phi\simeq 0),
\label{eq:BBN_scaling_osc}
\end{equation}
so BBN requires $\left.\rho_\Phi/\rho_r\right|_{T_{\rm osc}}\lesssim 0.1\,(T_{\rm BBN}/T_{\rm osc})\ll 1$ for $T_{\rm osc}\gg T_{\rm BBN}$; the onset of oscillations at $3H\simeq m_\Phi$ then bounds the oscillation amplitude (see Appendix~\eqref{apph}).
In all regimes, $\dot\Phi(t)$ supplies a homogeneous CP-odd, time-dependent background whose sign is dynamically fixed by the competition between $S(t)$ and $V'(\Phi)$, while its energy density remains safely subdominant in the spectator domain. The conversion of this background into a net comoving asymmetry is treated in the next section via the kinetic/thermodynamic response of the plasma \cite{CohenKaplan1988,JoyceShaposhnikov:1997PRL,DineKusenko:2003RMP,Davoudiasl2004,AlexanderYunes2009}.

\section{Equilibrium Leptogenesis from Geometry}\label{sec6}

Geometric parity violation in the early Universe can imprint a chiral, CP-odd background on the thermal plasma that biases the populations of left- and right-handed fermions even in strict thermal equilibrium. In such a setting, the microscopic time-reversal asymmetry is encoded by a pseudoscalar gravitational functional that couples to fermionic currents, thereby generating an effective axial or lepton-number chemical potential without invoking out-of-equilibrium decays. When electroweak sphalerons are active, this lepton bias is partially converted into a baryon asymmetry according to the standard equilibrium relations among chemical potentials in the electroweak-symmetric phase \cite{Sakharov1967,Lyth:1999PhysRept,DavidsonNardiNir2008}.
The long-wavelength, isotropic geometric background is captured at leading order by a homogeneous, timelike pseudoscalar gradient $b_\mu$ that couples to the axial current $J_5^\mu\equiv \bar{\psi}\gamma^\mu\gamma^5\psi$, and by a topological coupling to the gravitational Pontryagin density,
\begin{equation}
\mathcal{L}_{\mathrm{eff}} = \sum_f \bar{\psi}_f\, i\slashed{\nabla}\, \psi_f + b_\mu J_5^\mu + \frac{\alpha_g}{f}\, \Theta\, R\tilde{R} + \cdots,
\label{eq:LeffBias}
\end{equation}
where $f$ is a heavy mass scale, $\alpha_g$ is dimensionless, and $\Theta$ is a slowly varying pseudoscalar functional of the background geometry whose explicit form is not needed here. The ellipsis denotes higher-dimension operators suppressed by the same scale. Spatial isotropy selects $b_\mu=(b_0(t),\vec{0})$ in the comoving frame, so $b_0$ plays the role of an {axial} chemical potential $\mu_5$ for each Dirac species with a model-dependent weight. The topological term in \eqref{eq:LeffBias} is responsible for a curvature-induced axial anomaly $(\nabla_\mu J_5^\mu \supset \tfrac{1}{384\pi^2}R\tilde{R})$ \cite{Fujikawa:1979PRL}, and can be viewed as a source that feeds the homogeneous $b_0$ in the coarse-grained hydrodynamics of the chiral plasma \cite{AlexanderYunes2009}. Crucially, the term $b_0 J_5^0$ violates CPT spontaneously by selecting a preferred time direction and violates CP by distinguishing chiralities. Therefore, even in full kinetic and chemical equilibrium, left- and right-handed states acquire different single-particle spectra and occupation numbers.
To leading order in the parity-odd background, the Dirac equation for a massless fermion in \eqref{eq:LeffBias} yields
\begin{equation}
(\slashed{p}+ \slashed{b}\gamma^5)u(p)=0, \qquad p^\mu=(E,\vec{p}),  b^\mu=(b_0,\vec{0}),
\label{eq:Dirac1}
\end{equation}
whose dispersion relations follow from $(\slashed{p}+ \slashed{b}\gamma^5)^2=0$, i.e.
\begin{equation}
E_{R,L}(\vec{p}) = |\vec{p}| \mp b_0 + \mathcal{O}(b_0^2),
\label{eq:dispersion}
\end{equation}
with the upper (lower) sign for right- (left-) handed helicity. Thus, for vanishing vector chemical potentials, the equilibrium Fermi-Dirac distributions are shifted as
\begin{equation}
f_{R,L}(\vec{p}) = \frac{1}{e^{(|\vec{p}|\mp b_0)/T}+1}.
\label{eq:distributions}
\end{equation}
The lepton chiral asymmetry density for a single Dirac species is obtained by thermal averaging,
\begin{equation}
n_R-n_L = \int \!\frac{d^3p}{(2\pi)^3}\,\big[f_R(\vec{p})-f_L(\vec{p})\big]
 = \frac{T^2}{3}\,b_0 + \frac{1}{3\pi^2}\,b_0^3 + \mathcal{O}(b_0^5),
\label{eq:nRminusnL}
\end{equation}
where the linear term follows from standard susceptibilities of a relativistic fermion gas, $(\partial n/\partial \mu)_{\mu=0}=T^2/3$ for a Dirac species, and higher odd powers reflect the analytic expansion in $b_0/T$ \cite{KapustaGale:2006}. Summing over all chiral fermions with model-dependent weights $c_f$ that encode their axial coupling to the geometric background, one may write the equilibrium axial density as
\begin{equation}
n_5 \equiv \sum_f (n_{R,f}-n_{L,f}) = \chi_5\,\mu_5 + \mathcal{O}(\mu_5^3),   
\mu_5\equiv b_0,\qquad 
\chi_5 \equiv \sum_f \frac{T^2}{3}\,c_f,
\label{eq:axialsus}
\end{equation}
where $\chi_5$ is the axial susceptibility of the multi-flavor plasma. Equation \eqref{eq:axialsus} makes explicit that a purely geometric, parity-odd background produces a chiral population imbalance in exact thermal equilibrium.
The relevant conserved and anomalous currents in the symmetric electroweak phase include the lepton number $J_L^\mu$, the baryon number $J_B^\mu$, and the axial $J_5^\mu$. In the Standard Model, $B-L$ is anomaly-free whereas $B+L$ is violated by electroweak sphalerons \cite{tHooft:1976PRD,KhlebnikovShaposhnikov:1988NPB}. In a curved background with the effective interaction \eqref{eq:LeffBias}, the covariant divergence of the left-lepton current takes the schematic form
\begin{equation}
\nabla_\mu J_L^\mu = -\,\Gamma_{\Delta L}\,\frac{\mu_L}{T}\,T^3 + \frac{N_f}{32\pi^2}\,g^2\,W\tilde{W} + c_{\mathrm{grav}}\,\frac{1}{384\pi^2}\,R\tilde{R},
\label{eq:JLdiv}
\end{equation}
where $\Gamma_{\Delta L}$ represents the rate of fast $(\Delta L\neq 0)$ interactions that are in thermal equilibrium (for instance, those induced by a dimension-five Weinberg operator), $N_f=3$ is the number of families, $g$ is the $SU(2)_L$ gauge coupling, and $c_{\mathrm{grav}}$ is a model-dependent coefficient of the gravitational contribution to the chiral anomaly \cite{Fujikawa:1979PRL,DavidsonNardiNir2008}. In the adiabatic regime where all rapid interactions (gauge, Yukawa, and $\Delta L$ processes) are equilibrated and the background $b_0$ varies slowly compared to microscopic timescales, the right-hand side of \eqref{eq:JLdiv} vanishes upon thermal averaging, and the equilibrium lepton densities are determined by minimizing the free energy in the presence of the effective axial source $\mu_5=b_0$, subject to the usual neutrality and reaction constraints. Formally, writing the free-energy density $\Omega(T,\{\mu_i\};\mu_5)$ for all light species $i$ and linearizing in $\mu_5$, one finds
\begin{equation}
n_a = -\,\frac{\partial \Omega}{\partial \mu_a} = \sum_b \chi_{ab}(T)\,\mu_b^{\mathrm{eff}},  
\mu_b^{\mathrm{eff}} = \mu_b + \sum_f \mathsf{Q}_{bf}\,c_f\,\mu_5,
\label{eq:susmatrix}
\end{equation}
where $a,b$ label conserved or approximately conserved charges (e.g.\ $B-L$, hypercharge $Y$, individual flavor lepton numbers), $\chi_{ab}$ is the susceptibility matrix, and $\mathsf{Q}_{bf}$ are the corresponding charge assignments. Solving the linear constraints imposed by fast reactions and overall hypercharge neutrality yields a unique equilibrium value for the slow charge density $n_{B-L}$ proportional to $\mu_5$. It is convenient to parametrize the result as
\begin{equation}
n_{B-L}^{\mathrm{eq}} = \chi_{B-L}(T)\,\mu_{B-L}^{\mathrm{eff}},  
\chi_{B-L}(T) = \frac{T^2}{6}\,S,\qquad 
\mu_{B-L}^{\mathrm{eff}} = \mathcal{C}\,\mu_5,
\label{eq:nBLsus}
\end{equation}
where $S$ is a dimensionless susceptibility factor that depends on the particle content in equilibrium and $\mathcal{C}$ is a computable coefficient encoding how the axial source projects onto $B-L$ after all fast constraints are imposed \cite{HarveyTurner1990,Nardi:2006JHEP,Weinberg:2008cosmo}. Equation \eqref{eq:nBLsus} is the equilibrium, gravity-induced lepton bias that seeds baryogenesis.
Electroweak sphalerons violate $B+L$ and preserve $B-L$, thereby redistributing any preexisting $B-L$ asymmetry among baryons and leptons according to fixed algebraic relations among chemical potentials \cite{KhlebnikovShaposhnikov:1988NPB,Lyth:1999PhysRept}. In the unbroken phase with $N_f$ fermion families and $N_H$ Higgs doublets, one finds the linear conversion
\begin{equation}
B = c_s\,(B-L),\qquad 
c_s = \frac{8N_f + 4N_H}{22N_f + 13N_H}.
\label{eq:csph}
\end{equation}
For the Standard Model with $N_f=3$ and $N_H=1$, this gives $c_s=28/79$ \cite{HarveyTurner1990}. Combining \eqref{eq:nBLsus} with \eqref{eq:csph} and dividing by the entropy density $s=\tfrac{2\pi^2}{45}g_* T^3$ yields the equilibrium baryon yield,
\begin{equation}
\frac{n_B}{s} = c_s\,\frac{\chi_{B-L}(T)}{s}\,\mu_{B-L}^{\mathrm{eff}}
 = c_s\,\frac{15}{4\pi^2}\,\frac{S}{g_*}\,\frac{\mu_{B-L}^{\mathrm{eff}}}{T}
+ \mathcal{O}\!\left(\frac{\mu_{B-L}^{\mathrm{eff}}}{T}\right)^{\!3}.
\label{eq:etaB}
\end{equation}
The linear term in $\mu_{B-L}^{\mathrm{eff}}$ embodies the essence of gravity-assisted leptogenesis: a parity-odd geometric background plays the role of a {spurionic} chemical potential that biases lepton number in exact equilibrium, and electroweak sphalerons reprocess the resulting $B-L$ into a baryon excess with a fixed, model-independent coefficient $c_s$. The cubic correction in $(\mu_{B-L}^{\mathrm{eff}}/T)$ is numerically subleading whenever the geometric bias is perturbative, and higher-order effects can be systematically incorporated by extending the susceptibility expansion.

\begin{figure}[htb]
    \centering
    \includegraphics[width=0.9\linewidth, height=0.75\linewidth]{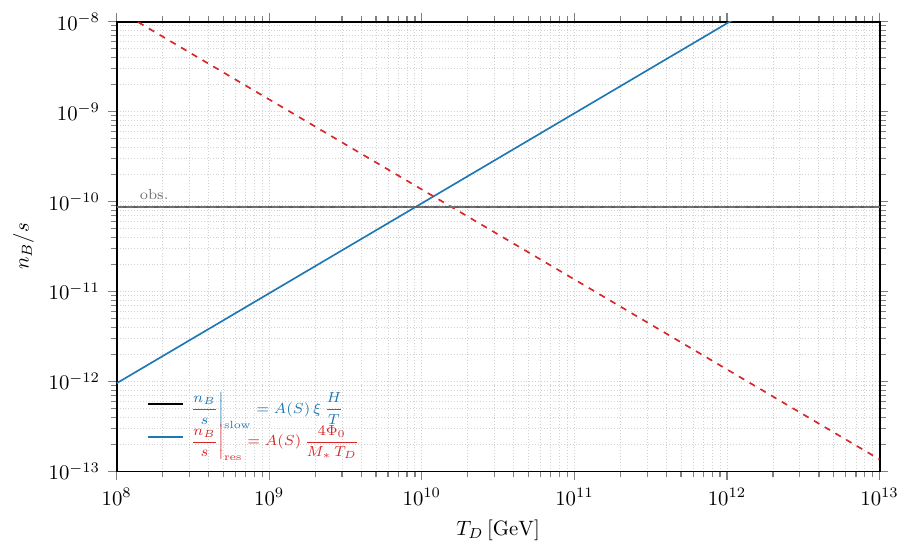}
     \caption{Slow-roll vs. resonant baryon yield as a function of the decoupling temperature \(T_D\).
  Defaults: \(g_*=106.75\), \(S=\Ssus\) (SM\(+\)3\(\nu_R\)), \(\xi=\xiParam\),
  \(M_{\rm Pl}=1.22\times10^{19}\,\mathrm{GeV}\) (unreduced), and \(\Phi_0/M_*=\PhiZeroOverMstar\).
  The gray band shows the observed yield \(n_B/s \simeq (8.7\pm0.1)\times10^{-11}\)
  (using \(\eta_B\simeq 6.1\times10^{-10}\) and \(n_B/s\approx \eta_B/7.04\)).
  }
  \label{fig:yields}
\end{figure}

The mechanism developed here realizes baryogenesis within a strictly equilibrium statistical ensemble by exploiting a geometric source of CPT and CP breaking. The parity-odd background $b_0$ modifies the dispersion relations of chiral fermions, producing a calculable lepton chiral asymmetry proportional to the axial susceptibility of the plasma. Rapid Standard-Model processes project this bias onto the single slow charge $B-L$, while electroweak sphalerons convert a fixed fraction $c_s$ of $B-L$ into baryon number. The final expression \eqref{eq:etaB} depends only on thermodynamic susceptibilities and on the geometric bias $\mu_{B-L}^{\mathrm{eff}}$, and is therefore universal within the class of gravity-assisted leptogenesis models. It furnishes an equilibrium pathway to the baryon asymmetry of the Universe consistent with a broad set of curvature-coupled frameworks and distinct from the traditional, decay-driven out-of-equilibrium scenarios \cite{Sakharov1967,Lyth:1999PhysRept,DavidsonNardiNir2008,AlexanderYunes2009}. 
{Figure~\ref{fig:yields} compares the analytic slow-roll and resonant baryon yields as functions of the decoupling temperature $T_D$, with the observed band shown for reference.
}

\section{Flavored Boltzmann Dynamics}\label{sec7}

Once a parity-odd geometric background establishes a chiral bias in the plasma, the final baryon asymmetry is set by flavor-resolved nonequilibrium transport in an expanding Universe. We work in spatially flat FLRW with $H\equiv \dot a/a$ and use comoving yields $Y_{L_\alpha}\equiv n_{L_\alpha}/s$, where $s$ is the entropy density. It is convenient to track
$Y_{\Delta_\alpha}\equiv Y_{B/3-L_\alpha}$ (conserved by electroweak sphalerons and fast Yukawas in the absence of explicit $(B\!-\!L)$ violation) and to collect them into
$\vec{Y}_\Delta\equiv (Y_{\Delta_e},Y_{\Delta_\mu},Y_{\Delta_\tau})^T$ \cite{Buchmuller:2005Ann,DavidsonNardiNir2008,Blanchet:2007JCAP,Drewes:2013IJMPE,AlexanderYunes2009}.
The flavored network is driven by a gravity-induced CP-odd source $\vec{S}^{\,\rm grav}$ and damped by a washout matrix $\mathsf{W}$,
\begin{equation}
\frac{d\vec{Y}_\Delta}{dt}=\frac{\vec{S}^{\,\rm grav}}{s}-\mathsf{W}\,\vec{Y}_\Delta,
\qquad
\mathsf{W}\equiv \frac{1}{s}\,\big(\Gamma_{\alpha\beta}\big)_{\alpha,\beta=e,\mu,\tau},
\label{eq:BoltzmannVector}
\end{equation}
where $\Gamma_{\alpha\beta}$ are thermal reaction densities governing relaxation, washout, and flavor transfer \cite{Buchmuller:2005Ann,DavidsonNardiNir2008,Blanchet:2007JCAP}. Using $z\equiv \Lambda_\star/T$ with $Hz\,d/dz=-\,d/dt$ (for a reference scale $\Lambda_\star$), Eq.~\eqref{eq:BoltzmannVector} becomes
\begin{equation}
\frac{d\vec{Y}_\Delta}{dz}
=
-\,\frac{1}{Hzs}\,\mathsf{W}(z)\,\vec{Y}_\Delta
+\frac{1}{Hzs}\,\vec{S}^{\,\rm grav}(z).
\label{eq:Boltzmannz}
\end{equation}
As the charged-lepton Yukawas successively equilibrate, $\mathsf{W}(z)$ interpolates between one-, two-, and three-flavor regimes, implemented by the standard flavor projectors (e.g.\ $\mathsf{W}\to \mathsf{P}^{(\tau)}\mathsf{W}\mathsf{P}^{(\tau)}+\mathsf{P}^{(\perp)}\mathsf{W}\mathsf{P}^{(\perp)}$ in the two-flavor regime) \cite{Abada:2006NPB,Nardi:2006JHEP,Blanchet:2007JCAP}. 
The source follows from the covariant divergence of the flavor currents $J^\mu_{L_\alpha}$. Schematically,
\begin{equation}
\big\langle \nabla_\mu J^\mu_{L_\alpha}\big\rangle
=
\mathcal{C}^{\rm coll}_\alpha+\mathcal{C}^{\rm grav}_\alpha,
\qquad
\mathcal{C}^{\rm grav}_\alpha
=
\frac{c_\alpha}{384\pi^2}\,\big\langle R\tilde R\big\rangle
+\Xi_\alpha(T)\,\partial_t\Upsilon(t),
\label{eq:divergence}
\end{equation}
where $\mathcal{C}^{\rm coll}_\alpha$ are the standard collision integrals, the first term is the topological Pontryagin source, and the second is a homogeneous CP-odd bias $\Upsilon(t)$ with Kubo coefficient $\Xi_\alpha(T)$ computable from thermal correlators in curved spacetime \cite{KapustaGale:2006,Beneke:2011NPB,Drewes:2013IJMPE}. Using entropy conservation in radiation domination, $ds/dt=-3Hs$, the contribution entering \eqref{eq:BoltzmannVector} is
\begin{equation}
\frac{S^{\,\rm grav}_\alpha}{s}
=
\frac{1}{s}\big\langle \nabla_\mu J^\mu_{L_\alpha}\big\rangle_{\rm grav}
=
\frac{c_\alpha}{384\pi^2}\,\frac{\big\langle R\tilde R\big\rangle}{s}
+\Xi_\alpha(T)\,\frac{\partial_t\Upsilon}{s}.
\label{eq:source}
\end{equation}
Both terms are $\mathcal{CP}$-odd and $\mathcal{T}$-odd and can generate an asymmetry in (near) equilibrium when $\Delta L\neq 0$ reactions are active \cite{CohenKaplan1988,Davoudiasl2004,AlexanderYunes2009}. 
A convenient decomposition of the reaction densities is
\begin{equation}
\Gamma_{\alpha\beta}(T)
=
2\,\delta_{\alpha\beta}\,\gamma_{ID,\alpha}
+\gamma_{\Delta L=1,\alpha\beta}
+2\,\gamma_{\Delta L=2,\alpha\beta}
+\gamma_{{\rm fc},\alpha\beta},
\label{eq:Gammaab}
\end{equation}
where $\gamma_{ID,\alpha}$ are inverse-decay densities, $\gamma_{\Delta L=1,\alpha\beta}$ and $\gamma_{\Delta L=2,\alpha\beta}$ are $\Delta L=1,2$ scatterings, and $\gamma_{{\rm fc},\alpha\beta}$ are flavor-changing but $\Delta L=0$ processes driving decoherence in the Yukawa-selected basis \cite{Giudice:2004NPB,Beneke:2011NPB,Drewes:2013IJMPE}. The $\gamma$’s follow from finite-temperature field theory (cutting rules / CTP) via $\gamma=\int d\Pi_{\rm in}\,d\Pi_{\rm out}\,|\mathcal{M}|^2 f_{\rm in}(1\pm f_{\rm out})$ \cite{Giudice:2004NPB,Beneke:2011NPB}. Conventional CP phases enter $|\mathcal{M}|^2$ for decay-driven sources, whereas the geometric source \eqref{eq:source} is linear in the small pseudoscalar background and does not rely on decay asymmetries \cite{DavidsonNardiNir2008}. 
The baryon yield $Y_B\equiv n_B/s$ evolves due to the electroweak anomaly. In the symmetric phase,
\begin{equation}
\frac{dY_B}{dz}
=
-\,\frac{\Gamma_{\rm sph}(T)}{Hz\,s\,T}\,\mathcal{A}_B(T)\,\mu_B
=
-\,\frac{\Gamma_{\rm sph}(T)}{Hz\,s}\,\sum_\alpha \mathcal{K}_\alpha(T)\,Y_{\Delta_\alpha},
\label{eq:BaryonEq}
\end{equation}
where $\Gamma_{\rm sph}(T)$ is the sphaleron rate per unit volume, $\mathcal{A}_B(T)$ is the baryon susceptibility, and $\mu_B/T=\sum_\alpha \mathcal{K}_\alpha Y_{\Delta_\alpha}$ follows from fast chemical equilibrium and hypercharge neutrality \cite{KhlebnikovShaposhnikov:1988NPB,HarveyTurner1990,Weinberg:2008cosmo}. For $\Gamma_{\rm sph}\gg HT^3$ one recovers the usual algebraic relation $B=c_s(B-L)$, while near the electroweak crossover sphalerons switch off and \eqref{eq:BaryonEq} freezes $Y_B$ \cite{Moore:2000PRD,DOnofrio:2014PRL}.
Summing \eqref{eq:Boltzmannz} over flavors gives the evolution of $Y_{B-L}\equiv \sum_\alpha Y_{\Delta_\alpha}$,
\begin{equation}
\frac{d}{dz}\!\left(\sum_\alpha Y_{\Delta_\alpha}\right)
=
\frac{1}{Hzs}\sum_\alpha S^{\,\rm grav}_\alpha
-\frac{1}{Hzs}\sum_{\alpha,\beta}\Gamma_{\alpha\beta}\,Y_{\Delta_\beta}.
\label{eq:BLcons}
\end{equation}
Consistency with anomaly structure requires $\sum_\alpha c_\alpha=0$ (no mixed gravitational anomaly for $B{-}L$), so the Pontryagin piece cancels in $\sum_\alpha S^{\,\rm grav}_\alpha$; therefore only the bias term $\propto \partial_t\Upsilon$ can source $(B{-}L)$ when explicit $(B{-}L)$-violating entries in $\Gamma_{\alpha\beta}$ are present. If $\Delta(B{-}L)=0$ processes are absent, \eqref{eq:BLcons} integrates to constant $Y_{B-L}$, as required \cite{DavidsonNardiNir2008}.
The linear system \eqref{eq:Boltzmannz} admits a closed solution in terms of the flavor-evolution operator. Defining
$\mathsf{G}(z,z')=\mathcal{P}\exp\!\left[-\int_{z'}^{z} d\zeta\,(H\zeta s)^{-1}\mathsf{W}(\zeta)\right]$,
\begin{equation}
\vec{Y}_\Delta(z)
=
\mathsf{G}(z,z_i)\,\vec{Y}_\Delta(z_i)
+\int_{z_i}^{z} dz'\,\mathsf{G}(z,z')\,\frac{\vec{S}^{\,\rm grav}(z')}{Hz's}.
\label{eq:Greensol}
\end{equation}
In the strong-washout regime, where the smallest eigenvalue $\lambda_{\min}(z)$ of $\mathsf{W}(z)$ obeys $\lambda_{\min}\gg Hzs$ during production, the solution adiabatically tracks the source,
$\vec{Y}_\Delta\simeq \mathsf{W}^{-1}\vec{S}^{\,\rm grav}/s$ up to $\mathcal{O}((Hzs)/\lambda_{\min})$ corrections. In the weak-washout regime $\lambda_{\min}\ll Hzs$, one has
$\vec{Y}_\Delta(z)\simeq \int_{z_i}^{z} dz'\,\vec{S}^{\,\rm grav}(z')/(Hz's)$.
Flavor decoherence modifies the projection of $\vec{S}^{\,\rm grav}$ onto the slow eigenmodes of $\mathsf{W}$ and can enhance or suppress transmission depending on alignment and the Yukawa equilibration sequence \cite{Abada:2006NPB,Nardi:2006JHEP,Blanchet:2007JCAP}.
In summary, Eqs.~\eqref{eq:Boltzmannz}-\eqref{eq:BaryonEq} provide a covariant flavored kinetic framework in which a geometric CP-odd source feeds slow charges, washout and flavor transfer control their survival, and sphalerons convert them into a final baryon asymmetry. The dependence on the geometric bias is linear at leading order, while the microphysical dependence is encoded in susceptibilities, reaction densities, and the sphaleron history \cite{Buchmuller:2005Ann,DavidsonNardiNir2008,Drewes:2013IJMPE,AlexanderYunes2009}.

\section{EFT Consistency and Unitarity}\label{sec8}

\begin{figure}[htb]
    \centering
    \includegraphics[width=\linewidth, height=0.8\linewidth]{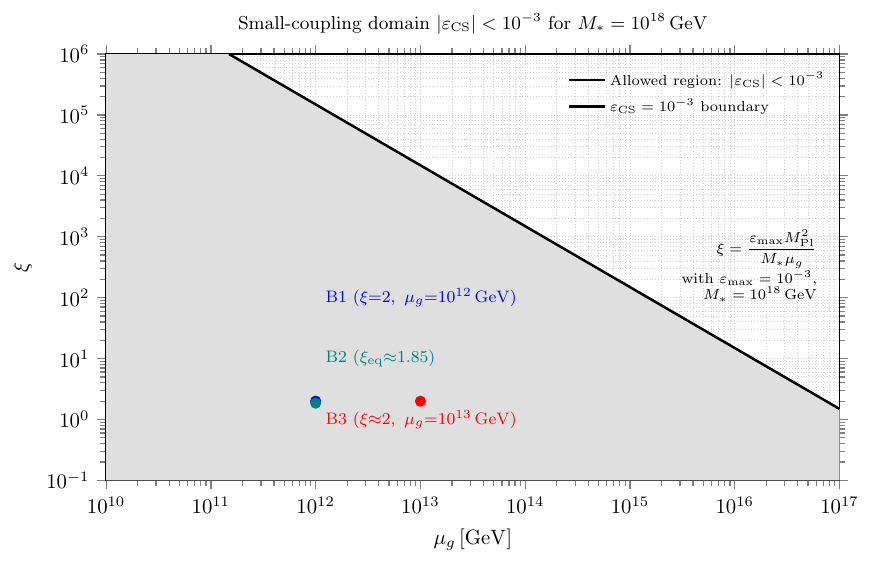}
      \caption{%
    Small-coupling domain for dynamical Chern-Simons parity effects.
    The shaded region satisfies $|\varepsilon_{\rm CS}|<10^{-3}$, consistent with Eq.~\eqref{eq:CSbounds}.
    The boundary is $\xi=\varepsilon_{\max} M_{\rm Pl}^2/(M_\ast \mu_g)$ with $\varepsilon_{\max}=10^{-3}$ and
    $M_\ast=10^{18}\,\mathrm{GeV}$. Benchmarks B1-B3 (Table~\ref{tab:benchmarks}) lie safely inside.
  }
  \label{fig:cs_small_coupling}
\end{figure}

The higher-dimensional curvature couplings, torsion-induced contact interactions, and parity-odd pseudoscalar operators in our construction require an EFT interpretation with a finite domain of validity. We therefore specify the controlled regime in which perturbation theory, gauge invariance, and radiative stability hold, and in which cosmological observables can be computed reliably \cite{Donoghue:1994PRD,Burgess:2004LRR,Weinberg:2008cosmo,Penco:2020EFT,AlexanderYunes2009}. 
At energies below the lightest heavy threshold, the effective theory admits the derivative expansion
\begin{equation}
\mathcal{L}_{\mathrm{eff}}
=
\mathcal{L}_{\mathrm{grav}}+\mathcal{L}_{\mathrm{mat}}
+\frac{1}{M_*}\mathcal{O}_5+\frac{1}{M_*^2}\mathcal{O}_6+\frac{1}{M_*^3}\mathcal{O}_7+\frac{1}{M_*^4}\mathcal{O}_8+\cdots,
\label{eq:Leffhier}
\end{equation}
with the parity-sensitive gravitational sector
\begin{equation}
\mathcal{L}_{\mathrm{grav}}
=
\frac{M_{\mathrm{Pl}}^2}{2}R
+\frac{M_{\mathrm{Pl}}^2}{2\gamma}\,\mathcal{H}
+\frac{\alpha}{4}\,\theta\,R\tilde R
+c_1 R^2+c_2 R_{\mu\nu}R^{\mu\nu}+c_3 R_{\mu\nu\rho\sigma}R^{\mu\nu\rho\sigma}+\cdots,
\label{eq:Lgrav}
\end{equation}
where $\mathcal{H}$ is the Holst density defined in Eq.~\eqref{eq:Holst}, $\gamma$ is the Immirzi parameter, and $c_i/M_*^2$ are radiatively generated or matched curvature-squared Wilson coefficients \cite{Holst1996,Hehl:1976RMP,Donoghue:1994PRD}. The leading dimension-five operators include the derivative portal $(\partial_\mu\Phi)J^\mu_{B-L}$ and the dCS coupling $\theta R\tilde R$, while $\mathcal{O}_6$ contains torsion-induced four-fermion terms and curvature-current operators, etc. The EFT cutoff is the smallest of the relevant heavy/strong-coupling scales,
\begin{equation}
M_* \equiv \min\!\Big\{\,f_\Phi,\; \Lambda_{\mathrm{CS}},\; \Lambda_{\mathrm{tors}},\; \Lambda_{B-L}\,\Big\},
\label{eq:MinCutoff}
\end{equation}
with $\Lambda_{B-L}\sim m_B/g_{BL}$ for the St\"uckelberg $U(1)_{B-L}$ sector. 
For characteristic physical momentum $E$ and curvature scale $\mathcal{R}\sim E^2$, perturbative control requires
\begin{equation}
\epsilon_{\mathrm{curv}}\equiv \frac{\mathcal{R}}{M_*^2}\ll 1,
\qquad
\epsilon_{\mathrm{der}}\equiv \frac{E}{M_*}\ll 1,
\qquad
\epsilon_{\mathrm{CS}}\equiv \frac{\alpha\,E^3}{M_{\mathrm{Pl}}^2\,\Lambda_\theta}\ll 1,
\label{eq:epsilons}
\end{equation}
where $\Lambda_\theta^{-1}\sim|\partial\theta|$ sets the length scale of the slowly varying CS scalar \cite{JackiwPi2003,YunesPretorius:2009PRD,AlexanderYunes2009}. Loop power counting implies $c_i\sim\mathcal{O}(1/16\pi^2)$ at matching; curvature-squared effects remain subleading provided $\epsilon_{\mathrm{curv}}\ll 4\pi$ \cite{Donoghue:1994PRD,Burgess:2004LRR,Penco:2020EFT}. 
Eliminating nondynamical torsion generates the axial--axial contact interaction \cite{Hehl:1976RMP}
\begin{equation}
\mathcal{L}_{AA}=\frac{\lambda_A}{M_{\mathrm{Pl}}^2}\,(\bar{\psi}\gamma_\mu\gamma^5\psi)(\bar{\psi}\gamma^\mu\gamma^5\psi),
\qquad
\lambda_A=\frac{3}{16}\,\frac{1}{1+\gamma^2}.
\label{eq:AAcontact}
\end{equation}
With $a_0\simeq \lambda_A s/(16\pi M_{\mathrm{Pl}}^2)$, $|a_0|<1$ gives
\begin{equation}
E \lesssim \Lambda_{\mathrm{tors}} \equiv \sqrt{\frac{16\pi}{\lambda_A}}\,M_{\mathrm{Pl}}
= \frac{16}{\sqrt{3}}\,\sqrt{\pi(1+\gamma^2)}\,M_{\mathrm{Pl}},
\label{eq:torsionUnit}
\end{equation}
so $\Lambda_{\mathrm{tors}}\gtrsim M_{\mathrm{Pl}}$ and torsion-induced amplitudes are perturbative throughout the sub-Planckian regime. The derivative portal $(\partial_\mu\Phi/M_*)J^\mu$ yields $\mathcal{A}\sim s/M_*^2$, implying
\begin{equation}
E \lesssim \Lambda_{\mathrm{port}}\equiv \sqrt{16\pi}\,M_*,
\label{eq:portalUnit}
\end{equation}
while graviton amplitudes in the dCS sector scale as $|\mathcal{A}_{\mathrm{CS}}|\sim \alpha E^3/M_{\mathrm{Pl}}^2$, giving the intrinsic strong-coupling scale \cite{YunesPretorius:2009PRD,AlexanderYunes2009}
\begin{equation}
E \lesssim \Lambda_{\mathrm{CS}}\equiv \left(\frac{16\pi M_{\mathrm{Pl}}^2}{\alpha}\right)^{1/3}.
\label{eq:CSunit}
\end{equation}
Consistency of \eqref{eq:MinCutoff} requires $M_*\le \Lambda_{\mathrm{CS}}$ whenever parity-odd curvature interactions are retained. 
St\"uckelberg $U(1)_{B-L}$ invariance ($B_\mu\!\to\! B_\mu+\partial_\mu\alpha$, $\sigma\!\to\!\sigma+m_B\alpha$) forces operators to be built from gauge-invariant combinations and conserved currents. Integrating out the heavy vector yields the gauge-invariant portal \cite{Ruegg:2003IMPA,Kors:2004PLB}
\begin{equation}
\mathcal{L}_{\mathrm{portal}}
=
\frac{1}{M_*}\,\partial_\mu \sigma\,J_{B-L}^\mu
+\mathcal{O}\!\left(\frac{\partial^2}{m_B^3}\right),
\qquad
M_*=\frac{m_B}{g_{BL}},
\label{eq:GaugeInvPortal}
\end{equation}
protected against radiative generation of gauge-variant operators. Consistency with gravity further uses the absence of a mixed gravitational anomaly for $(B-L)$ (in the SM+$3\nu_R$), so no net $R\tilde R$ term appears in $\nabla_\mu J^\mu_{B-L}$ \cite{Preskill:1991AnnPhys,Weinberg:2008cosmo}. 
For tensor modes with physical wavenumber $k_{\mathrm{phys}}$, a convenient parameterization of perturbativity in dCS gravity is \cite{YunesPretorius:2009PRD,AlexanderYunes2009}
\begin{equation}
\varepsilon_{\mathrm{CS}}(k)\equiv \frac{\alpha}{M_{\mathrm{Pl}}^2}\,k_{\mathrm{phys}}\,|\dot{\theta}|,
\qquad
\zeta_{\mathrm{CS}}\equiv \frac{\alpha^2}{\kappa \beta}\,\frac{(\partial \theta)^2}{L^4},
\label{eq:CSsmall1}
\end{equation}
with $\kappa=8\pi G$ and $L$ a characteristic curvature radius. The small-coupling EFT domain requires
\begin{equation}
\varepsilon_{\mathrm{CS}}(k)\ll 1,
\qquad
\zeta_{\mathrm{CS}}\ll 1,
\label{eq:CSbounds}
\end{equation}
together with $E\ll \Lambda_{\mathrm{CS}}$ from \eqref{eq:CSunit}. Figure~\ref{fig:cs_small_coupling} illustrates the representative region $|\varepsilon_{\rm CS}|<10^{-3}$ and shows that our benchmarks lie safely within \eqref{eq:CSbounds}.
The derivative portal induces $\mu_{B-L}=u^\mu\partial_\mu\Phi/M_*$, so linear-response susceptibilities require
\begin{equation}
\left|\frac{\mu_{B-L}}{T}\right|=\left|\frac{u^\mu\partial_\mu \Phi}{M_*T}\right|\ll 1.
\label{eq:linresp}
\end{equation}
Spectator dynamics and derivative/curvature expansions require
\begin{equation}
\frac{\rho_\Phi}{3M_{\mathrm{Pl}}^2 H^2}\ll 1,
\qquad
\left|\frac{\mathcal{R}}{M_*^2}\right|\ll 1,
\qquad
\left|\frac{\partial^2\Phi}{M_*^2}\right|\ll 1.
\label{eq:backreact}
\end{equation}
Since $\lambda_A$ in \eqref{eq:AAcontact} decreases with $|\gamma|$, there is no additional sub-Planckian constraint from $\gamma$ beyond $E\ll M_{\mathrm{Pl}}$; in particular,
\begin{equation}
E\ll M_{\mathrm{Pl}}
\ \Longrightarrow\
\text{torsion-induced amplitudes}\ll 1
\ \text{for all}\ \gamma\in\mathbb{R}.
\label{eq:gammaBound}
\end{equation}
For the CS sector, combining \eqref{eq:CSunit} and \eqref{eq:CSbounds} yields the compact requirements
\begin{equation}
\frac{E}{\Lambda_{\mathrm{CS}}}\ll 1,
\qquad
\frac{\alpha\,k_{\mathrm{phys}}\,|\dot{\theta}|}{M_{\mathrm{Pl}}^2}\ll 1,
\label{eq:CScompact}
\end{equation}
ensuring that parity-violating tensor effects are captured at leading order. 
Forward-limit analyticity/positivity constrains parity-even higher-derivative coefficients (e.g.\ $c_1+c_2\gtrsim 0$ in suitable bases) \cite{Adams:2006JHEP,Bellazzini:2020EPJC}; the parity-odd Pontryagin density is not directly constrained in the same way but is bounded by \eqref{eq:CSunit}-\eqref{eq:CScompact}. Axion/Green--Schwarz UV completions naturally generate $\theta R\tilde R$ while preserving Abelian gauge invariance via anomaly inflow \cite{GreenSchwarz:1984PLB,Preskill:1991AnnPhys}. 
Collecting the conditions, the controlled EFT regime for cosmological computations is
\begin{equation}
E\ll M_*,
\qquad
\frac{\mathcal{R}}{M_*^2}\ll 1,
\qquad
\left|\frac{\mu_{B-L}}{T}\right|\ll 1,
\qquad
\varepsilon_{\mathrm{CS}}(k)\ll 1,
\qquad
\frac{E}{\Lambda_{\mathrm{CS}}}\ll 1,
\label{eq:MasterWindow}
\end{equation}
with $M_*$ defined by \eqref{eq:MinCutoff}. Within \eqref{eq:MasterWindow}, the derivative portal, torsion-induced contact sector, and parity-odd gravitational response admit a controlled expansion, ensuring predictive power without trans-Planckian dynamics or loss of unitarity \cite{Donoghue:1994PRD,Burgess:2004LRR,Penco:2020EFT,AlexanderYunes2009}. Existing observational constraints relevant to our parity-odd signatures are quoted in Sec.~\eqref{sec9}, while BBN safety follows from the explicit $\Delta N_{\rm eff}$ bound given in Sec.~\eqref{sec5} and is easily satisfied in the spectator domain adopted here \cite{Weinberg:2008cosmo}.

\section{Parity Tests and Tri-Observable Link}\label{sec9}
The chiral-time equivalence (CTE) mechanism correlates the time orientation selected by a cosmological pseudoscalar background with parity violation in the gravitational and radiation sectors. This correlation is encoded in a single dynamical order parameter whose sign fixes simultaneously the fermionic chirality bias, the handedness of primordial tensor modes, and the rotation of linear polarization. As a consequence, CTE predicts a triplet of linked, independently measurable observables that together render the framework falsifiable: the baryon-to-entropy ratio $\eta_B$, an isotropic cosmic birefringence angle $\Delta\alpha$ rotating CMB linear polarization, and a tensor-mode chirality parameter $\chi_T$ imprinted in the parity-odd CMB spectra. We refer to this predictive structure as the tri-observable relation.
We define the tensor chirality parameter at a pivot wavenumber $k_\star$ by
\begin{equation}
\chi_T(k_\star)\;\equiv\;
\frac{\mathcal{P}_h^{\mathrm{R}}(k_\star)-\mathcal{P}_h^{\mathrm{L}}(k_\star)}
{\mathcal{P}_h^{\mathrm{R}}(k_\star)+\mathcal{P}_h^{\mathrm{L}}(k_\star)},
\qquad -1\le \chi_T \le 1,
\label{eq:chiTdef}
\end{equation}
where $\mathcal{P}_h^{\mathrm{R},\mathrm{L}}$ denote the primordial right/left-handed tensor power spectra; $\chi_T>0$ indicates an excess of right-handed power and $\chi_T<0$ an excess of left-handed power. An isotropic cosmic birefringence angle $\Delta\alpha$ rotates the Stokes parameters $(Q,U)$ and converts E-modes into B-modes, yielding parity-odd CMB correlators $(C_\ell^{TB},C_\ell^{EB}\propto \sin(2\Delta\alpha))$ \cite{LueWangKamionkowski1999,Komatsu:2022AR}. The observed baryon abundance is $\eta_B\equiv n_B/s$, with $s$ the comoving entropy density, fixed by independent cosmological data.
The CTE prediction can be summarized symbolically as
\begin{equation}
\mathcal{R}_{\mathrm{CTE}}:\qquad 
\eta_B \;\longleftrightarrow\; \Delta\alpha \;\longleftrightarrow\; \chi_T,
\label{eq:symbolicTri}
\end{equation}
in the sense that all three are controlled by the same dynamical parity-time order parameter and therefore obey algebraic relations at leading order in small parity violation.
Let $\Phi$ denote the parity-odd cosmological field that selects the time orientation. To leading order in slow, homogeneous evolution, the equilibrium baryon bias is linear in the comoving drift rate $u^\mu \nabla_\mu\Phi$, the tensor chirality inherits the sign of the parity-odd gravitational response, and an isotropic birefringence angle from a parity-odd background inherits the sign of $\Delta\Phi$ over the line-of-sight. Under the assumptions of monotonic slow evolution and weak parity violation one obtains the sign-locking condition
\begin{equation}
\mathrm{sign}\big(\eta_B\big)=\mathrm{sign}\big(\chi_T\big)=\mathrm{sign}\big(\dot{\Phi}\big),
\label{eq:signlocking}
\end{equation}
which is invariant under CPT-preserving transformations because a reversal of time orientation flips the sign of each quantity on the right-hand side while leaving the equality intact. The condition \eqref{eq:signlocking} provides a direct falsifiability test: a statistically significant preference for $\chi_T<0$ together with $\eta_B>0$ would exclude the sign structure of CTE, independently of any amplitude calibration.
The CTE-induced parity violation generates parity-odd CMB spectra through two logically distinct channels: a chiral tensor background and a polarization rotation. Their lowest-order interference yields a parity-odd correlator
\begin{equation}
C_\ell^{TB}\;\simeq\;\mathcal{N}_\ell^{(T)}\,\chi_T\,\mathcal{P}_{\mathcal{R}}\,\sin(2\Delta\alpha),
\label{eq:TBapprox}
\end{equation}
where $\mathcal{P}_{\mathcal{R}}$ is the primordial curvature power spectrum evaluated at the pivot and $\mathcal{N}_\ell^{(T)}$ is a calculable transfer coefficient that captures projection effects and tensor-to-scalar conversion in $\Lambda$CDM \cite{LueWangKamionkowski1999,AlexanderYunes2009,Contaldi:2022JCAP}. Equation \eqref{eq:TBapprox} shows that a joint measurement of $C_\ell^{TB}$ across multipoles fixes the product $\chi_T\sin(2\Delta\alpha)$ up to known transfer factors, providing an anchor for the tri-observable relation below. Current CMB data are consistent with small parity violation; representative constraints include
\begin{equation}
|\Delta\alpha|<0.35^\circ,\qquad |\chi_T|<0.1, 
\eta_B^{\mathrm{obs}}=(8.7\pm0.1)\times 10^{-11},
\label{eq:phenobounds}
\end{equation}
while forthcoming experiments such as {LiteBIRD} and CMB-S4 target sensitivities $\sigma(\Delta\alpha)\sim \mathcal{O}(10^{-2}\text{-}10^{-3})^{\circ}$ and parity-odd spectra at the $r\sim10^{-3}$ frontier \cite{Hazumi:2020LB,Abazajian:2022CMBS4,Planck:2020AandA,Komatsu:2022AR}. Combining \eqref{eq:TBapprox} with \eqref{eq:phenobounds} already yields a conservative lower bound on the CTE calibration factor appearing below.
At leading order in curvature and torsion perturbations and for a single, slowly varying parity-odd background, the three observables admit linear response expansions in a common dimensionless drift parameter $\varepsilon$ that quantifies the amplitude of parity and time-reversal breaking. It is convenient to define $\varepsilon\equiv \mathcal{I}^{-1}u^\mu\nabla_\mu\Phi$, with $\mathcal{I}$ a reference mass scale that drops out of ratios. Linear response then takes the form
\begin{equation}
\eta_B = A_B\,\varepsilon,\qquad
\chi_T = A_T\,\varepsilon,\qquad
\sin(2\Delta\alpha) = A_\alpha\,\varepsilon,
\label{eq:LRexpansions}
\end{equation}
where $A_B,A_T,A_\alpha$ are dimensionless, time-integrated transfer coefficients that depend on the cosmological history and on linear susceptibilities in the radiation phase, on tensor-mode transfer functions, and on the line-of-sight polarization-rotation kernel, respectively \cite{LueWangKamionkowski1999,AlexanderYunes2009,Weinberg:2008cosmo}. The expansions \eqref{eq:LRexpansions} follow from symmetry: each observable is odd under a simultaneous reversal of time orientation and spatial parity, hence only odd powers of $\varepsilon$ can appear at leading order, while higher odd powers are suppressed by additional derivatives of the background.
Eliminating $\varepsilon$ among \eqref{eq:LRexpansions} yields an algebraic relation among the three observables,
\begin{equation}
\eta_B = \kappa_{\mathrm{CTE}}\,\chi_T\,\sin(2\Delta\alpha),
\qquad
\kappa_{\mathrm{CTE}}\equiv \frac{A_B}{A_TA_\alpha},
\label{eq:TriObs}
\end{equation}
which is the tri-observable relation in its model-independent form. The coefficient $\kappa_{\mathrm{CTE}}$ is a calculable, dimensionless number determined by the geometry of the background and by Standard-Model thermodynamics: it factorizes into a baryogenesis susceptibility $(A_B\propto c_s(S/g_*)(H/T)|_{T_D})$, a tensor chirality transfer $A_T$ evaluated at the tensor pivot, and a polarization-rotation line-of-sight weight $A_\alpha$ between recombination and today; here $c_s$ is the sphaleron conversion factor, $S$ the $(B\!-\!L)$ susceptibility count, $g_*$ the relativistic degrees of freedom at decoupling, and $T_D$ the temperature where the slow charge freezes in \cite{Weinberg:2008cosmo,HarveyTurner1990,AlexanderYunes2009}. The sign-locking \eqref{eq:signlocking} is consistent with \eqref{eq:TriObs} since $\kappa_{\mathrm{CTE}}>0$ for the standard cosmological history and linear susceptibilities.
\begin{figure}[htb]
    \centering
    \includegraphics[width=\linewidth, height=170mm]{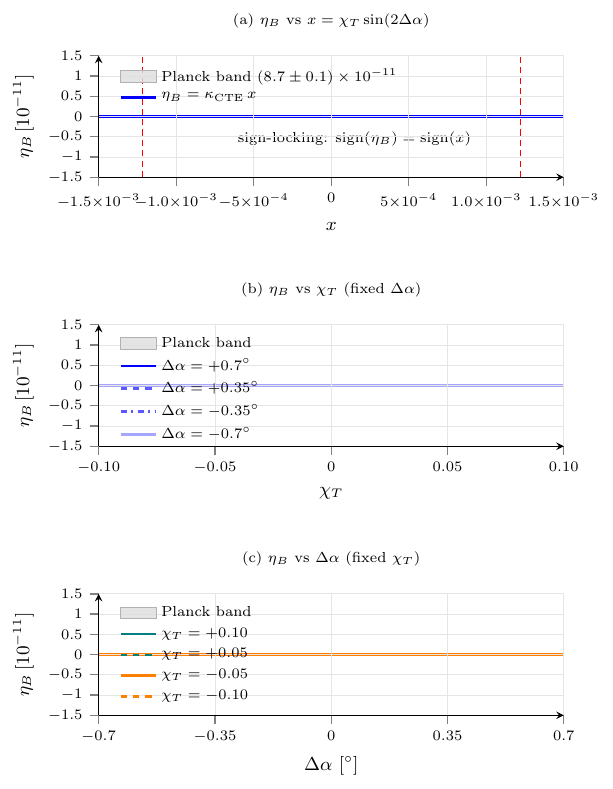}
      \caption{\textbf{Tri-observable relation}
  $\eta_B=\kappa_{\rm CTE}\,\chi_T\sin(2\Delta\alpha)$ with $\eta_B$ shown in units of $10^{-11}$.
  (a) Compressed variable $x=\chi_T\sin(2\Delta\alpha)$. 
  (b) Fixed $\Delta\alpha$: straight lines in $\chi_T$ with slope $\kappa_{\rm CTE}\sin(2\Delta\alpha)$. 
  (c) Fixed $\chi_T$: sinusoidal in $\Delta\alpha$ with amplitude $\kappa_{\rm CTE}\chi_T$.
  Gray bands: Planck; red dashed: bounds $|\chi_T|\le 0.1$, $|\Delta\alpha|\le 0.7^\circ$ (so $|x|\le 0.1\sin 0.7^\circ$).}
  \label{fig:triobservable}
\end{figure}
For phenomenology it is useful to invert \eqref{eq:TriObs} and trade $\kappa_{\mathrm{CTE}}$ for observables. Using the conservative bounds \eqref{eq:phenobounds} one finds
\begin{equation}
\kappa_{\mathrm{CTE}} \gtrsim
\frac{\eta_B^{\mathrm{obs}}}{|\chi_T|\,|\sin(2\Delta\alpha)|}
\gtrsim 7\times 10^{-8},
\label{eq:kappaBound}
\end{equation}
where the numerical estimate uses $|\chi_T|<0.1$ and $|\Delta\alpha|<0.35^\circ$ so that $|\sin(2\Delta\alpha)|<\sin(0.7^\circ)\simeq 1.22\times 10^{-2}$. Any future narrowing of the birefringence or chirality bounds immediately sharpens \eqref{eq:kappaBound}. Conversely, a detection of $\chi_T$ and $\Delta\alpha$ fixes $\kappa_{\mathrm{CTE}}$ from \eqref{eq:TriObs} and predicts $\eta_B$ without recourse to microscopic parameters.
Equation \eqref{eq:TriObs} implies that all CTE predictions for the triple $(\eta_B,\Delta\alpha,\chi_T)$ populate a one-parameter surface in the three-dimensional observable space. See Figure~\ref{fig:triobservable} for a visualization of this linear tri-observable constraint, where the line $\eta_B=\kappa_{\rm CTE}\,\chi_T\sin(2\Delta\alpha)$ is shown together with the current bounds on $|\chi_T|$ and $|\Delta\alpha|$ and the observed $\eta_B$ band. The parity sign is fixed by \eqref{eq:signlocking}. A joint likelihood in $(\eta_B,\chi_T,\Delta\alpha)$ must therefore lie on the curve
\begin{equation}
\mathcal{C}_{\mathrm{CTE}}:\qquad 
\eta_B-\kappa_{\mathrm{CTE}}\,\chi_T\,\sin(2\Delta\alpha)=0,
\label{eq:CTEcurve}
\end{equation}
up to small higher-order corrections suppressed by the square of the parity-violation amplitude. A statistically significant violation of \eqref{eq:CTEcurve} at fixed sign pattern \eqref{eq:signlocking} would falsify the mechanism. The observational strategy is clear: measure or bound $\Delta\alpha$ with delensing-limited E/B rotation estimators, extract $\chi_T$ from parity-odd tensor estimators that compare right/left-handed contributions to $(BB,TB,EB)$ and use the well-measured $\eta_B$ as a normalization. Next-generation surveys targeting $\sigma(\Delta\alpha)\sim 10^{-2}$-$10^{-3}$ degrees and parity-odd tensor signals at the $\chi_T\sim 10^{-2}$ level will test $\kappa_{\mathrm{CTE}}$ over more than an order of magnitude, providing a decisive falsifiability threshold \cite{Hazumi:2020LB,Abazajian:2022CMBS4,Komatsu:2022AR,Contaldi:2022JCAP,Planck:2020AandA}.

\section{Benchmarks and Parameter Inference}\label{sec10}
To confront CTE with data we map a compact parameter set to the triad of observables
\(\mathbf{O}=(\eta_B,\Delta\alpha,\chi_T)\) and define benchmark regimes consistent with the EFT
window of Sec.~\eqref{sec7}. We emphasize that the purpose here is calibration and organization (not a
full global fit): the analysis identifies viable scaling relations, degeneracy directions, and
representative parameter domains compatible with current bounds. 
We take as core parameters
\begin{equation}
\Theta=\{\,M_*,\,\gamma,\,\lambda_\Phi,\,\xi,\,\kappa_{\rm CTE},\,g_{B-L}\,\},
\label{eq:Theta}
\end{equation}
where \(M_*\) is the portal/EFT scale, \(\gamma\) the Immirzi parameter, \(\lambda_\Phi\) the
self-coupling in \(V(\Phi)\), \(\xi\) a dimensionless curvature-response strength (parity-odd
gravitational response), \(\kappa_{\rm CTE}\) a geometric calibration coefficient entering the
tri-observable normalization, and \(g_{B-L}\) the St\"uckelberg gauge coupling (with
\(M_*=m_B/g_{B-L}\) in Sec.~\eqref{sec4}). The forward model is
\begin{equation}
\mathbf{O}_{\rm CTE}(\Theta)=\big(\,\eta_B(\Theta),\,\Delta\alpha(\Theta),\,\chi_T(\Theta)\,\big),
\label{eq:forwardObs}
\end{equation}
evaluated at fixed analysis pivots (tensor chirality at \(k_\star\), birefringence along the CMB line-of-sight).
Near a fiducial point \(\Theta_0\), we linearize
\begin{equation}
\delta\mathbf{O}\equiv \mathbf{O}_{\rm CTE}(\Theta_0+\delta\Theta)-\mathbf{O}_{\rm CTE}(\Theta_0)
=\mathsf{J}(\Theta_0)\,\delta\Theta+\mathcal{O}(\delta\Theta^2),
\qquad
\mathsf{J}_{ai}\equiv \left.\frac{\partial O_a}{\partial \theta_i}\right|_{\Theta_0},
\label{eq:Jacobian}
\end{equation}
so \(\mathsf{J}\) encodes local parameter sensitivities and degeneracy directions. 
It is convenient to factor out a shared small measure of parity-time breaking \(\varepsilon\) and write
\begin{equation}
\eta_B=\mathcal{A}_B(\Theta)\,\varepsilon,
\qquad
\chi_T=\mathcal{A}_T(\Theta)\,\varepsilon,
\qquad
\sin(2\Delta\alpha)=\mathcal{A}_\alpha(\Theta)\,\varepsilon,
\label{eq:LRamplitudes}
\end{equation}
with smooth transfer amplitudes \(\mathcal{A}_B,\mathcal{A}_T,\mathcal{A}_\alpha\). In the slow-drift regime
(\(V(\Phi)=\tfrac{\lambda_\Phi}{4}\Phi^4\) as a representative choice), the baryon channel scales as
\begin{equation}
\mathcal{A}_B \propto \frac{\lambda_\Phi\,\Phi_D^{\,3}}{H_D\,M_*\,T_D},
\label{eq:ABscaling}
\end{equation}
evaluated at the freeze-in/decoupling epoch \(T_D\). For the tensor channel, it is useful to parameterize
\begin{equation}
\mathcal{A}_T=\mathcal{T}_T(\gamma)\,\Xi_T(\xi,M_*),
\qquad
\mathcal{T}_T(\gamma)\sim (1+\gamma^2)^{-p_\gamma}\ \ (p_\gamma=\mathcal{O}(1)),
\label{eq:ATscaling}
\end{equation}
capturing the suppression of torsion-induced parity transfer at large \(|\gamma|\), while \(\Xi_T\) encodes the
dCS response in the small-coupling regime of Sec.~\eqref{sec7}. For birefringence, the portal scale enters through
\(M_*=m_B/g_{B-L}\), and \(\mathcal{A}_\alpha\) is controlled by \(\kappa_{\rm CTE}\) and the same parity-odd
response coefficients that set \(\chi_T\) (explicit kernels are given in Sec.~\eqref{sec9}). 
For forecasting and for organizing parameter scans, it is useful to anchor at representative portal scales:
\begin{table}[t]
\centering
\begin{tabular}{|p{3.4cm}|p{3.2cm}|p{7.2cm}|}
\hline
\textbf{Regime} & \textbf{Scale} & \textbf{Asymptotic response (schematic)} \\
\hline
Strong-gravity geometric & $M_*\simeq 10^{17}\,\mathrm{GeV}$ &
$\eta_B \propto \lambda_\Phi\Phi_D^3/(H_D M_* T_D)$,
\ \ $\chi_T \propto \mathcal{T}_T(\gamma)\Xi_T(\xi,M_*)$,
\ \ $\Delta\alpha \propto \kappa_{\rm CTE}/M_*$
\\ \hline
Intermediate CTE & $M_*\simeq 10^{15}\,\mathrm{GeV}$ &
portal-limited $\eta_B$ with potentially observable $\chi_T$ for $\xi=\mathcal{O}(10^{-1}\!-\!1)$,
and $\Delta\alpha\sim \kappa_{\rm CTE}/M_*$
\\ \hline
Low-energy freeze-out & $M_*\lesssim 10^{12}\,\mathrm{GeV}$ &
$\eta_B$ can saturate early, while $\chi_T$ and $\Delta\alpha$ are diluted/suppressed by the post-inflationary history
\\ \hline
\end{tabular}
\caption{Representative benchmark regimes used as fiducial anchors \(\Theta_0\) for local linearization
\eqref{eq:Jacobian} and for defining informative priors consistent with Sec.~\eqref{sec7}.}
\label{tab:benchmarks}
\end{table} 
Let \(\mathbf{O}_{\rm obs}\) be the data vector with covariance \(\mathbf{C}\). A Gaussian likelihood is
\begin{equation}
\mathcal{L}(\Theta|D)\propto
\exp\!\left[-\frac{1}{2}\big(\mathbf{O}_{\rm CTE}(\Theta)-\mathbf{O}_{\rm obs}\big)^T
\mathbf{C}^{-1}\big(\mathbf{O}_{\rm CTE}(\Theta)-\mathbf{O}_{\rm obs}\big)\right],
\label{eq:likelihood}
\end{equation}
and the posterior \(P(\Theta|D)\propto \mathcal{L}(\Theta|D)\,\pi(\Theta)\) uses priors \(\pi(\Theta)\) enforcing
the EFT window (Sec.~\eqref{sec7}), including \(E\ll M_*\), \(|\mu_{B-L}|/T\ll 1\), and dCS small-coupling bounds.
For near-Gaussian posteriors (or Fisher forecasts) one has
\begin{equation}
\mathbf{\Sigma}=\big(\mathbf{F}+\mathbf{\Lambda}_{\rm prior}\big)^{-1},
\qquad
\mathbf{F}=\mathsf{J}^T\mathbf{C}^{-1}\mathsf{J},
\label{eq:Sigma}
\end{equation}
where \(\mathbf{\Lambda}_{\rm prior}\) is the prior precision (e.g.\ from hard top-hat EFT cuts approximated locally). 
Combining current parity bounds (Sec.~\eqref{sec9}) with EFT consistency (Sec.~\eqref{sec7}) yields a representative,
internally consistent domain
\begin{equation}
\gamma \in [0.1,\,0.3],
\qquad
\kappa_{\rm CTE}\sim 10^{-3},
\qquad
g_{B-L}\lesssim 10^{-2},
\qquad
M_*\gtrsim 10^{15}\,\mathrm{GeV},
\label{eq:reprValues}
\end{equation}
to be interpreted as indicative (the precise region depends on the chosen pivots and covariances). Two robust
degeneracy directions recur in the linearized map \eqref{eq:Jacobian}: (i) a \((\gamma,\kappa_{\rm CTE})\)
degeneracy in parity observables, since both modulate the overall tensor/polarization parity amplitude through
\(\mathcal{T}_T(\gamma)\) and the geometric calibration, and (ii) a \((\lambda_\Phi,\xi)\) degeneracy, since
enhanced drift in \(\Phi\) (raising \(\eta_B\)) can be partially compensated by reduced curvature response in the
parity channel (affecting \(\chi_T\) and \(\Delta\alpha\)) at fixed \(M_*\). These directions are precisely what the
tri-observable program is designed to break: improving polarization sensitivity separates calibration from torsion
suppression, while independent information on the \(\Phi\) potential (or on \(T_D\)) breaks drift-response tradeoffs. 
In practice, \eqref{eq:likelihood}-\eqref{eq:Sigma} can be combined with standard samplers (HMC, nested sampling,
or Metropolis-Hastings) to obtain marginalized posteriors when linearization is insufficient. The benchmarks in
Table~\ref{tab:benchmarks} serve as initialization points (or informative priors) for separate chains; their relative
evidence can be compared when desired. Details of the stiff numerical integration of the coupled system and the construction of forecast bands for
\((\eta_B,\Delta\alpha,\chi_T)\) are given in Appendix~\ref{app:solver}. The outcome is a calibrated mapping from \(\Theta\) to the triad
\((\eta_B,\Delta\alpha,\chi_T)\) consistent with EFT control, providing the quantitative input for the tri-observable
tests developed in Sec.~\eqref{sec9}.

\section{UV Embeddings and Nieh-Yan Option}\label{sec11}

The CTE/ECICS construction is an EFT valid below a cutoff $M_*$ where higher-derivative and higher-curvature corrections remain perturbative. A UV completion must therefore generate (upon integrating out heavy modes) the parity-odd couplings, torsional responses, and Stückelberg structures used here. Natural candidates include string-inspired axion sectors with Green-Schwarz inflow and Kalb-Ramond torsion, loop-quantum-gravity (LQG) scenarios in which $\gamma$ has microscopic meaning, and higher-dimensional gauge-gravity models whose Kaluza-Klein (KK) reductions induce topological couplings \cite{Hehl:1976RMP,Ashtekar:1986PRL,NiehYan1982,Shapiro2002,GreenSchwarz:1984PLB,JackiwPi2003,AlexanderYunes2009,Polchinski:1998v2,KalbRamond:1974PRD,OverduinWesson:1997PhysRept,TaverasYunes2008,Mercuri:2006PRD}.
A generic UV completion may be organized as
\begin{equation}
\mathcal{L}_{\mathrm{UV}}=\mathcal{L}_{\mathrm{EFT}}+\sum_i \frac{c_i}{\Lambda^{n_i-4}}\,\mathcal{O}^{\mathrm{UV}}_i+\cdots,
\label{eq:LUV}
\end{equation}
with heavy scale $\Lambda$, operator dimensions $n_i\equiv \mathrm{dim}(\mathcal{O}^{\mathrm{UV}}_i)$, and Wilson coefficients $c_i=\mathcal{O}(1)$ at matching. A representative UV sector that contains the structures relevant for our EFT is
\begin{multline}
\mathcal{L}_{\mathrm{UV}}\supset
\frac{M_{\mathrm{Pl}}^2}{2}\,R
+\frac{M_{\mathrm{Pl}}^2}{2\gamma_{\mathrm{UV}}}\,\mathcal{H}
+\sum_I \frac{1}{2}(\partial a_I)^2
-\sum_I \frac{a_I}{16\pi^2 f_I}\Big(c_I^{(R)}R\tilde R+c_I^{(G)}\mathrm{Tr}\,G\tilde G+c_I^{(F)}F\tilde F\Big) \\
-\sum_A \frac{1}{4}F_{X_A}^{2}
+\frac{1}{2}\big(m_A X_{A\mu}-\partial_\mu \sigma_A\big)^2
+g_A X_{A\mu}J^\mu_A
+\frac{\kappa_{\mathrm{NY}}}{f_\Phi}\,\Phi\,\mathcal{N}\mathcal{Y}
+\cdots,
\label{eq:UVsector}
\end{multline}
where $\mathcal{H}$ is the Holst density and $\mathcal{N}\mathcal{Y}\equiv T^a\wedge T_a-e^a\wedge e^b\wedge R_{ab}$ is the Nieh-Yan four-form \cite{NiehYan1982,ChandiaZanelli1997}. Matching is defined by integrating out heavy fields at $\Lambda$,
\begin{equation}
e^{\,i W_{\mathrm{IR}}[g,e,\omega,\Phi,\cdots]}=
\int \! \mathcal{D}a_I\,\mathcal{D}X_A\,\mathcal{D}\sigma_A\,
e^{\,i\int d^4x\sqrt{-g}\,\mathcal{L}_{\mathrm{UV}}}.
\label{eq:matchingPath}
\end{equation} 
At tree level, integrating out a heavy St\"uckelberg vector produces the derivative portal. In the low-momentum limit,
\begin{equation}
X_{A\mu}=\frac{1}{m_A}\partial_\mu \sigma_A-\frac{g_A}{m_A^2}J^{\phantom{A}}_{A\mu}
+\mathcal{O}\!\left(\frac{\partial^2}{m_A^3}\right),
\label{eq:solveX}
\end{equation}
which yields
\begin{multline}
\mathcal{L}_{\mathrm{IR}} \supset 
\sum_A\left[\frac{g_A}{m_A}\partial_\mu \sigma_A\,J^\mu_A-\frac{g_A^2}{2m_A^2}J_{A\mu}J_A^\mu\right]
\equiv \frac{1}{M_*}\partial_\mu \sigma\,J^\mu_{B-L}
-\frac{1}{2M_*^2}J_{B-L,\mu}J_{B-L}^\mu+\cdots,\\
M_*=\frac{m_{B-L}}{g_{B-L}},
\label{eq:portalMatch}
\end{multline}
recovering the gauge-invariant portal scale $M_*$ of Sec.~\eqref{sec4}. 
If the light pseudoscalar $\theta$ is a linear combination of UV axions, $\theta=\sum_I U_{\theta I}a_I$, then
\begin{equation}
\frac{\alpha_{\mathrm{IR}}}{4}\,\theta\,R\tilde R
=\sum_I \frac{c_I^{(R)}}{16\pi^2 f_I}\,a_I\,R\tilde R
\;\Longrightarrow\;
\alpha_{\mathrm{IR}}=\sum_I \frac{c_I^{(R)}}{4\pi^2 f_I}\,U_{\theta I},
\label{eq:CSmatch}
\end{equation}
so the parity-odd gravitational response is controlled by axion decay scales and mixing \cite{GreenSchwarz:1984PLB,Polchinski:1998v2,AlexanderYunes2009}. Analogous relations hold for gauge Pontryagin terms. 
In first-order gravity, $\gamma$ weights the Holst density. UV dynamics can renormalize it (e.g.\ radiatively or via integrating out a heavy pseudoscalar coupled to axial torsion), schematically
\begin{equation}
\frac{M_{\mathrm{Pl}}^2}{2\gamma_{\mathrm{IR}}}
=\frac{M_{\mathrm{Pl}}^2}{2\gamma_{\mathrm{bare}}}+\Delta_{\mathrm{Holst}}(\Lambda),
\qquad
\Delta_{\mathrm{Holst}}(\Lambda)=\frac{1}{M^2}\int^{\Lambda}\!\!\frac{d^4p}{(2\pi)^4}\,\Pi_{\mathrm{Holst}}(p^2),
\label{eq:HolstMatch}
\end{equation}
and eliminating nondynamical torsion yields the axial--axial contact operator with coefficient fixed by $\gamma_{\mathrm{IR}}$,
\begin{equation}
\mathcal{L}^{\mathrm{IR}}_{AA}
=\frac{\lambda_A(\gamma_{\mathrm{IR}})}{M_{\mathrm{Pl}}^2}\,
(\bar\psi\gamma_\mu\gamma^5\psi)(\bar\psi\gamma^\mu\gamma^5\psi),
\qquad
\lambda_A(\gamma)=\frac{3}{16}\frac{1}{1+\gamma^2},
\label{eq:AAHolst}
\end{equation}
as used in Sec.~\eqref{sec7} \cite{Hehl:1976RMP,Shapiro2002}. 
The Nieh-Yan density is
\begin{equation}
\mathcal{N}\mathcal{Y}=T^a\wedge T_a-e^a\wedge e^b\wedge R_{ab}=d\!\left(e^a\wedge T_a\right),
\label{eq:NYdef}
\end{equation}
and appears in the chiral anomaly with regulator-dependent coefficient \cite{NiehYan1982,ChandiaZanelli1997,Shapiro2002}. A coupling to $\Phi$ gives a Wess--Zumino-like term
\begin{equation}
S_{\mathrm{NY}}=\frac{\kappa_{\mathrm{NY}}}{f_\Phi}\int_{\mathcal{M}}\Phi\,\mathcal{N}\mathcal{Y}
=\frac{\kappa_{\mathrm{NY}}}{f_\Phi}\int_{\partial\mathcal{M}}\Phi\,e^a\wedge T_a
-\frac{\kappa_{\mathrm{NY}}}{f_\Phi}\int_{\mathcal{M}} d\Phi\wedge e^a\wedge T_a,
\label{eq:SNY}
\end{equation}
which can act as a torsional pump when the bulk term is nonzero (e.g.\ spin-density backgrounds, Kalb--Ramond torsion $H_{\mu\nu\rho}$, or topologically nontrivial tetrads). Then the axial Ward identity receives
\begin{equation}
\big\langle \nabla_\mu J_5^\mu \big\rangle \supset \frac{\kappa_{\mathrm{NY}}}{f_\Phi}\,\big\langle \mathcal{N}\mathcal{Y}\big\rangle,
\label{eq:NYanomaly}
\end{equation}
and a coarse-grained estimate for the induced comoving charge at decoupling $T_D$ is
\begin{equation}
Y_{B-L}^{(\mathrm{NY})}\simeq \frac{\kappa_{\mathrm{NY}}}{f_\Phi}\,\frac{1}{s(T_D)}
\int^{t_D}\!dt\,a^{3}(t)\,\big\langle \mathcal{N}\mathcal{Y}\big\rangle(t),
\label{eq:NYyield}
\end{equation}
typically subdominant because torsion-sourced densities redshift as $a^{-6}$ in Einstein-Cartan cosmology \cite{Hehl:1976RMP,Shapiro2002}. 
UV consistency requires standard bounds
\begin{equation}
\Lambda_{\mathrm{UV}}\gtrsim M_*,
\qquad
c_i(\Lambda)\lesssim \mathcal{O}(1),
\label{eq:UVbounds}
\end{equation}
and cancellation of gauge/gravitational anomalies. In string compactifications, Green--Schwarz inflow from the ten-dimensional two-form $B_{(2)}$ (with $H=dB-\omega_{3L}+\omega_{3Y}$) yields axion-curvature terms matching \eqref{eq:CSmatch}, while RR axions generate St\"uckelberg masses matching \eqref{eq:portalMatch} \cite{GreenSchwarz:1984PLB,Polchinski:1998v2}. KK reductions of higher-curvature theories can likewise induce topological densities and St\"uckelberg couplings with computable coefficients \cite{OverduinWesson:1997PhysRept}. In LQG-motivated settings, $\gamma$ can be promoted to a field and stabilized, producing an IR value $\gamma_{\rm IR}$ as in \eqref{eq:HolstMatch}, while fermions generate \eqref{eq:AAHolst} \cite{TaverasYunes2008,Mercuri:2006PRD}. String backgrounds with Kalb-Ramond torsion $H_{\mu\nu\rho}$ can source torsion and potentially activate \eqref{eq:SNY}-\eqref{eq:NYyield} \cite{KalbRamond:1974PRD,Shapiro2002}.
In summary, UV completions naturally reproduce the EFT backbone: heavy axions generate $\theta R\tilde R$ with normalization fixed by mixing, St\"uckelberg vectors generate the derivative $(B{-}L)$ portal with $M_*=m_B/g_{B-L}$, and first-order gravity fixes torsion-induced contacts through $\gamma_{\rm IR}$. The Nieh-Yan coupling is an optional, model-dependent torsional pump that can supplement (but is not required for) the core CTE mechanism.

\section{Conclusions}
We have developed a geometric mechanism for the origin of the cosmic matter-antimatter asymmetry based on a symmetry principle that ties chiral rotations to the global time orientation of cosmology. The chiral-time equivalence (CTE) framework promotes a diagonal combination of time flow and axial phase to an organizing symmetry of the low-energy theory, thereby fixing a minimal, shift-symmetric operator basis in which a derivative portal to the slow $(B\!-\!L)$ charge and a parity-odd gravitational density provide the leading sources of CP violation. Within this structure, an Einstein-Cartan-Immirzi-Chern-Simons (ECICS) sector yields controlled, parity-violating responses in the tensor channel \cite{Hehl:1976RMP,JackiwPi2003,AlexanderYunes2009}, while the Stückelberg realization of $U(1)_{B-L}$ generates a unique, gauge-invariant portal at dimension five. The resulting geometric baryogenesis scheme is symmetry-driven, economy-minded in its operator content, and anchored in well-defined low-energy dynamics.

A central conceptual outcome is that baryon production can proceed in strict thermal equilibrium once a parity-odd geometric background is present, because the background acts as a spurionic chemical potential for the unique hydrodynamic slow charge, with bias set by susceptibilities of the relativistic plasma \cite{Sakharov1967,Weinberg:2008cosmo,DineKusenko:2003RMP}. In this setting we formulated gravity-assisted leptogenesis and exhibited its flavored nonequilibrium realization through a Boltzmann network that captures washout, decoherence, and flavor projection. The framework reveals two distinguishing signatures. First, a sign-locking principle fixes the common algebraic sign of the baryon excess, the tensor chirality, and the time derivative of the parity-odd background, thereby providing a nontrivial, CPT-respecting correlation among observables. Second, a tri-observable relation links the baryon-to-entropy ratio, the isotropic cosmic birefringence angle, and the tensor-mode chirality parameter into a single algebraic constraint at leading order in parity violation, endowing the mechanism with exceptional predictive power across otherwise independent datasets \cite{LueWangKamionkowski1999,AlexanderYunes2009}.

We established the theoretical coherence of the proposal by deriving EFT consistency bounds that ensure small-coupling control of parity-odd gravitational effects, perturbative unitarity of contact amplitudes including the torsion-induced axial interaction, and gauge invariance in the Stückelberg $U(1)_{B-L}$ sector. Positivity and analyticity considerations for parity-even higher-derivative operators were shown to be compatible with the parity-odd densities present at leading order, while mixed gravitational anomalies are consistently saturated by axionic inflow in ultraviolet realizations. The ECICS-CTE system admits natural embeddings in quantum gravity frameworks: string-inspired axion-torsion couplings generate the gravitational Chern-Simons term through Green-Schwarz anomaly cancellation and furnish Stückelberg masses for anomalous Abelian factors \cite{GreenSchwarz:1984PLB,AlexanderYunes2009}, loop-quantum-gravity interpretations rationalize the appearance and renormalization of the Immirzi parameter \cite{Hehl:1976RMP}, and higher-dimensional reductions provide topological densities with calculable normalizations. The Nieh-Yan four-form offers an optional torsional pump whose contribution to axial charge is subleading in standard cosmologies yet can be dialed in UV scenarios with nontrivial torsion \cite{NiehYan1982,ChandiaZanelli1997,Shapiro2002}, without disturbing the predictive backbone of the CTE mechanism.

On the phenomenological side, we implemented a stiff, adaptive solver for the coupled scalar and flavored kinetic system, constructed parity-sensitive transfer integrals, and propagated parameter covariances into forecast bands for the three key observables. The numerical results delineate a wide, sub-Planckian window in which the EFT remains predictive and the parity-odd signatures are within reach of upcoming surveys. The framework is falsifiable in multiple ways: the sign-locking condition can be confronted with the measured handedness of the tensor spectrum; the tri-observable relation predicts a one-parameter surface in $(\eta_B,\Delta\alpha,\chi_T)$ space that can be tested with joint CMB TB/EB measurements and baryon abundance determinations; and the absolute amplitudes fall within the sensitivity goals of planned polarization and tensor probes such as {LiteBIRD} and CMB-S4 \cite{Planck:2020AandA,Hazumi:2020LB,Abazajian:2022CMBS4}. A decisive outcome is that the same geometric source responsible for the baryon asymmetry necessarily fixes the parity sign of the tensor background and the rotational sense of CMB polarization, thereby tying the microscopic origin of matter to macroscopic parity violation in the sky.

The broader implications of these results extend beyond baryogenesis. A symmetry-based link between time orientation, chirality, and gravitational parity opens a principled avenue for diagnosing parity violation in quantum gravity and for interpreting parity-odd imprints in primordial tensor and polarization data. The EFT-to-UV continuity of the mechanism, complemented by a minimal operator basis, invites systematic extensions that remain under theoretical control, including refined histories of reheating, scale-dependent parity transfer to tensor modes, and correlated signatures in stochastic gravitational-wave backgrounds and large-scale structure parity estimators. From the observational side, a coordinated program that combines high-precision CMB polarization, primordial tensor searches, and independent determinations of the baryon abundance provides the optimal arena for a stringent test of the CTE paradigm. Regardless of the outcome, the exercise sharpens the interface between fundamental symmetry principles and cosmological data, and sets a benchmark for how robust, falsifiable predictions can be distilled from parity-sensitive sectors of quantum gravity \cite{AlexanderYunes2009,LueWangKamionkowski1999,Weinberg:2008cosmo}.

In summary, the CTE construction delivers a unified, symmetry-driven, and testable account of baryogenesis from spacetime geometry. It connects equilibrium bias, flavored nonequilibrium dynamics, parity-violating gravity, and ultraviolet consistency into a coherent narrative that is both calculable and predictive. With forthcoming polarization experiments poised to probe cosmic birefringence and tensor chirality at unprecedented levels, the framework presented here offers a timely and concrete target for discovery or refutation, and a fertile ground for further theoretical developments at the intersection of cosmology, particle physics, and quantum gravity.

\section*{Conflict of Interest}
The authors declare that there are no conflicts of interest regarding this work.

\section*{Data Availability Statement}
This study contains no experimental data. All theoretical results are included in the manuscript.

\section*{Funding Statement}
This work received no external funding.
\bibliographystyle{ytphys.bst}
\bibliography{refs}

\newpage

\appendix

\section{EC Torsion Elimination}\label{appa}

Einstein-Cartan (EC) gravity extends the Levi-Civita geometry by allowing an independent Lorentz connection and a nonvanishing torsion two-form $T^a$. In the presence of spinor matter, torsion couples algebraically to the spin density and, crucially, does not propagate in four dimensions. Eliminating torsion at the action level yields a purely metric-compatible effective field theory in which the connection reduces to the Levi-Civita part and the stress of spin is encoded by local contact interactions. In the baryogenesis framework developed in the main text, this elimination furnishes the torsion-free effective action that underlies the dynamical and phenomenological analyses, while retaining the parity-violating couplings in the gravitational and matter sectors that are required for the chiral-time equivalence (CTE) mechanism.
We work in first-order formalism with coframe $e^a=e^a{}_\mu dx^\mu$ and independent $\mathfrak{so}(1,3)$ connection $\omega^{ab}=\omega^{ab}{}_\mu dx^\mu$ ($\omega^{ab}=-\omega^{ba}$). The curvature and torsion two-forms are
\begin{equation}
R^{ab} = d\omega^{ab} + \omega^a{}_c\wedge \omega^{cb},
\qquad
T^a = De^a = de^a + \omega^a{}_b\wedge e^b,
\label{eq:curvtor}
\end{equation}
and the invariant volume is $e\,d^4x$ with $e\equiv \det(e^a{}_\mu)=\sqrt{-g}$. We consider the EC action augmented by a quadratic torsion term,\footnote{Throughout we set $\kappa^2\equiv 8\pi G$ and adopt the mostly-plus signature.}
\begin{equation}
S_{\mathrm{EC}}[e,\omega;\psi] = \frac{1}{2\kappa^2}\int d^4x\,e\,\big(R(\omega)+\alpha\,\mathfrak{t}\big) + S_{\mathrm{m}}[\psi,e,\omega],
\qquad
\mathfrak{t} \equiv \frac{1}{e}\,\epsilon_{abcd}\,T^a\wedge T^b\wedge e^c\wedge e^d,
\label{eq:SEC}
\end{equation}
where $R(\omega)$ is the Ricci scalar computed from the full connection $\omega$, $\alpha$ is a dimensionless coupling controlling the torsion-squared contribution,\footnote{A torsion-squared term is convenient for bookkeeping; in four dimensions it may be traded against the Nieh-Yan density and a Holst-like piece up to a total derivative \cite{NiehYan1982,ChandiaZanelli1997}.} and $S_{\mathrm{m}}$ is the matter action. For a minimally coupled Dirac field,
\begin{equation}
S_{\mathrm{m}}[\psi,e,\omega] = \int d^4x\,e\!\left[\frac{i}{2}\big(\bar{\psi}\gamma^\mu D_\mu\psi-D_\mu\bar{\psi}\,\gamma^\mu\psi\big)-m\,\bar{\psi}\psi\right],
\qquad
D_\mu\psi \equiv \partial_\mu\psi + \frac{1}{4}\,\omega_{\mu ab}\,\gamma^{ab}\psi,
\label{eq:Sm}
\end{equation}
with $\gamma^\mu=e^\mu{}_a\gamma^a$ and $\gamma^{ab}\equiv \tfrac{1}{2}[\gamma^a,\gamma^b]$. We denote the axial current by the one-form $J_5\equiv J_{5\mu}dx^\mu$ with $J_{5}^\mu\equiv \bar{\psi}\gamma^\mu\gamma^5\psi$.
The field equations for torsion follow from independent variation of the action with respect to the spin connection. Using $\delta R^{ab}=D(\delta\omega^{ab})$ and $\delta T^a=\delta\omega^a{}_b\wedge e^b$, one finds after integration by parts the Cartan equation
\begin{equation}
\frac{1}{\kappa^2}\,\epsilon_{abcd}\,e^c\wedge T^d + \frac{\alpha}{\kappa^2}\,e_{[a}\wedge T_{b]} = \tau_{ab},
\label{eq:CartanEq}
\end{equation}
where $\tau_{ab}\equiv \delta S_{\mathrm m}/\delta \omega^{ab}$ is the spin three-form of matter \cite{Hehl:1976RMP,Shapiro2002}. For a minimally coupled Dirac field the spin density is totally antisymmetric and proportional to the axial current,\footnote{Equivalently, the Dirac spin density reads $\tau_{ab}=\tfrac{1}{4}\,\epsilon_{abcd}\,e^c\wedge e^d\,\star J_5$, where $\star$ is the spacetime Hodge dual \cite{Hehl:1976RMP,Trautman:2006Lect}.}
\begin{equation}
\tau_{ab} = \frac{1}{4}\,\epsilon_{abcd}\,e^c\wedge e^d\wedge J_5.
\label{eq:tauDirac}
\end{equation}
Equation \eqref{eq:CartanEq} is algebraic in torsion. To solve it explicitly, it is convenient to switch to components and decompose torsion $T_{\mu\nu}{}^{\rho}$ into its irreducible pieces under the Lorentz group: a vector trace $T_\mu\equiv T_{\mu\nu}{}^{\nu}$, an axial vector $S^\mu\equiv \epsilon^{\mu\nu\rho\sigma}T_{\nu\rho\sigma}$, and a purely tensor part $q_{\mu\nu\rho}$ with $q_{\mu\nu}{}^{\nu}=0$, $\epsilon^{\mu\nu\rho\sigma}q_{\nu\rho\sigma}=0$ \cite{Hehl:1976RMP,Shapiro2002}. Minimal Dirac matter sources only the totally antisymmetric combination, so $T_\mu=0$, $q_{\mu\nu\rho}=0$, and torsion is entirely encoded in $S^\mu$. In this sector the Cartan equation reduces to the algebraic relation
\begin{equation}
S_\mu = -\,\mathcal{C}(\alpha)\,\kappa^2\,J_{5\mu},
\qquad
\mathcal{C}(\alpha)>0,
\label{eq:Ssolution}
\end{equation}
where $\mathcal{C}(\alpha)$ is a positive coefficient that equals unity in the standard EC theory and is smoothly deformed by the torsion-squared coupling $\alpha$ (see below). In terms of the contorsion one finds the well-known solution
\begin{equation}
K_{abc} = \frac{1}{4}\,\epsilon_{abcd}\,S^d = -\,\frac{\mathcal{C}(\alpha)}{4}\,\kappa^2\,\epsilon_{abcd}\,J_5^d,
\label{eq:contorsionSol}
\end{equation}
which exhibits the purely axial character of Dirac-induced torsion.
To obtain the torsion-free effective action we substitute $\omega=\tilde{\omega}(e)+K(J_5)$ in \eqref{eq:SEC}, where $\tilde{\omega}$ is the torsionless Levi-Civita connection and $K$ is the contorsion \eqref{eq:contorsionSol}. The curvature scalar decomposes as
\begin{equation}
R(\omega) = R(\tilde{\omega}) + \tilde{\nabla}_\mu\big(K_\nu{}^{\nu\mu}-K_{\nu}{}^{\mu\nu}\big) + K_{\mu\nu}{}^{\nu}K_{\rho}{}^{\mu\rho}-K_{\mu\nu\rho}K^{\rho\nu\mu},
\label{eq:Rdecomp}
\end{equation}
so that, up to a total derivative, the torsion dependence of the gravitational Lagrangian is purely quadratic in $K$. Specializing to the purely axial sector one arrives at the standard identity \cite{Hehl:1976RMP,Shapiro2002}
\begin{equation}
e\,R(\omega) = e\,R(\tilde{\omega})-\frac{e}{4}\,S_\mu S^\mu + \text{(total derivative)},
\label{eq:RwithS}
\end{equation}
while the Dirac Lagrangian picks up a linear coupling to $S_\mu$,
\begin{equation}
\mathcal{L}_{\mathrm{m}} = e\!\left[\frac{i}{2}\big(\bar{\psi}\gamma^\mu \tilde{\nabla}_\mu\psi-\tilde{\nabla}_\mu\bar{\psi}\,\gamma^\mu\psi\big)-m\bar{\psi}\psi\right] + \frac{3e}{4}\,S_\mu J_5^\mu,
\label{eq:DiracS}
\end{equation}
where $\tilde{\nabla}_\mu$ is built from $\tilde{\omega}$ alone. The torsion sector of the action reads, after discarding the boundary term,
\begin{equation}
S[S_\mu] = \int d^4x\,\left[-\,\frac{e}{8\kappa^2}\,(2- \hat{\alpha})\,S_\mu S^\mu + \frac{3e}{4}\,S_\mu J_5^\mu\right],
\qquad
\hat{\alpha}\equiv\hat{\alpha}(\alpha),
\label{eq:Ssector}
\end{equation}
where $\hat{\alpha}(\alpha)$ parametrizes the effect of the $T\wedge T$ deformation on the axial quadratic kernel.\footnote{In four dimensions $T^a\wedge T_a$ contributes a quadratic form in the irreducible torsion components; in the purely axial sector it amounts to a multiplicative renormalization of the $S_\mu S^\mu$ coefficient. See \cite{Shapiro2002,Hammond:2002ReptProgPhys} for a detailed decomposition.} Varying \eqref{eq:Ssector} with respect to $S_\mu$ reproduces the algebraic solution \eqref{eq:Ssolution} with
\begin{equation}
\mathcal{C}(\alpha) = \frac{3}{2-\hat{\alpha}(\alpha)}.
\label{eq:Calpha}
\end{equation}
Substituting back yields the effective torsion-free action
\begin{align}
S_{\mathrm{eff}}[e,\psi] &= \frac{1}{2\kappa^2}\int d^4x\,e\,R(\tilde{\omega}) 
+ \int d^4x\,e\!\left[\frac{i}{2}\big(\bar{\psi}\gamma^\mu \tilde{\nabla}_\mu\psi-\tilde{\nabla}_\mu\bar{\psi}\,\gamma^\mu\psi\big)-m\bar{\psi}\psi\right] \nonumber\\
&\quad + \frac{3\,\mathcal{C}(\alpha)\,\kappa^2}{16}\int d^4x\,e\,J_{5\mu}J_5^{\mu},
\label{eq:SeffFinal}
\end{align}
where the last term is the induced axial-axial contact interaction. In the standard EC theory ($\alpha=0\Rightarrow \hat{\alpha}=0\Rightarrow \mathcal{C}=3/2$) one recovers the canonical result \cite{Hehl:1976RMP,Shapiro2002}
\begin{equation}
\mathcal{L}_{4\psi}^{\mathrm{EC}} = \frac{3\kappa^2}{16}\,e\,J_{5\mu}J_5^{\mu}.
\label{eq:AAstandard}
\end{equation}
More general parity-odd deformations of the gravitational sector, such as a Holst density with Immirzi parameter $\gamma$, modify the algebraic map $S_\mu\leftrightarrow J_{5\mu}$ and generate, in addition to \eqref{eq:AAstandard}, a parity-violating vector-axial contact term whose relative weight is controlled by $1/(1+\gamma^2)$ \cite{FreidelMinicTakeuchi:2005PRD,Mercuri:2006PRD,PerezRovelli:2006PRD}. These refinements can be included straightforwardly but are not required for the core elimination presented here.

The elimination procedure preserves local Lorentz invariance, since the torsion equation \eqref{eq:CartanEq} is covariant and the substitution $\omega\mapsto \tilde{\omega}(e)$ yields the unique torsionless connection compatible with the coframe. It is also compatible with the couplings employed in the main manuscript. First, the derivative portal to the CTE scalar $\Phi$ involves the conserved current $J_{B-L}^\mu$ and is independent of the spin connection; torsion elimination therefore leaves it unchanged at leading order. Second, parity-violating gravitational couplings such as the Chern-Simons density $\theta\,R\tilde{R}$ are constructed from the Riemann tensor of $\tilde{\omega}$ in the effective description and retain their role as sources of tensor chirality. Third, the induced contact term \eqref{eq:AAstandard} is parity even and Planck suppressed; it does not obstruct the parity-odd effects needed for geometric baryogenesis, while providing a consistent, UV-insensitive bookkeeping of spin backreaction in the torsionless effective theory.

The inclusion of the $T\wedge T$ term in \eqref{eq:SEC} amounts to a renormalization of the quadratic torsion kernel and therefore to a rescaling of $\mathcal{C}(\alpha)$ in \eqref{eq:Ssolution}-\eqref{eq:SeffFinal}. In a general decomposition into irreducible pieces, $T\wedge T$ mixes the vector and axial sectors; however, for Dirac matter with purely axial spin density only the axial block contributes to the on-shell action, and its effect is captured by $\hat{\alpha}(\alpha)$ in \eqref{eq:Ssector}. From the EFT standpoint this is a higher-dimension operator whose coefficient can be absorbed into the definition of low-energy contact terms and does not alter the structure of the effective, torsion-free action beyond a benign shift of the axial-axial coupling. A detailed component analysis can be found in \cite{Shapiro2002,Hammond:2002ReptProgPhys}.

\section{Tensor Sector in dCS Gravity}\label{appb}

Dynamical Chern-Simons (dCS) gravity augments the Einstein-Hilbert action by a pseudoscalar field $\theta(x)$ that couples linearly to the gravitational Pontryagin density $R\tilde{R}$. The resulting parity-odd interaction modifies the propagation of tensor perturbations on cosmological backgrounds and leads to birefringence of gravitational waves: right- and left-circularly polarized modes acquire different dispersion relations and friction, generating a helicity asymmetry in the tensor power spectrum~\cite{JackiwPi2003,AlexanderYunes2009,LueWangKamionkowski1999,Satoh:2008PRD,YunesPretorius:2009PRD}. In the geometric baryogenesis framework, this helicity asymmetry furnishes a direct, parity-sensitive observable that is correlated with the matter-antimatter asymmetry in the main text, while the technical tensor-sector derivations presented here remain independent of torsion elimination and other appendices.
The dCS action reads
\begin{equation}
S = \frac{M_{\mathrm{Pl}}^2}{2}\int d^4x\sqrt{-g}\,R 
+ \frac{\alpha_{\mathrm{CS}}}{4}\int d^4x\sqrt{-g}\,\theta\,R\tilde{R}
+ S_\theta[\theta,g] + S_{\mathrm{m}}[g,\cdots],
\label{eq:dCSaction}
\end{equation}
where $M_{\mathrm{Pl}}^{-2}=8\pi G$, $\alpha_{\mathrm{CS}}$ is a constant with mass dimension $-1$, $S_\theta$ contains the kinetic and potential terms for $\theta$, and the Pontryagin density is $R\tilde{R}\equiv \tfrac{1}{2}\epsilon^{\mu\nu\rho\sigma}R^\alpha{}_{\beta\mu\nu}R^\beta{}_{\alpha\rho\sigma}$. Varying with respect to the metric yields
\begin{equation}
G_{\mu\nu} + \frac{\alpha_{\mathrm{CS}}}{\sqrt{-g}}\,C_{\mu\nu} 
= M_{\mathrm{Pl}}^{-2}T_{\mu\nu},
\label{eq:dCSeqs}
\end{equation}
where $C_{\mu\nu}$ is the symmetric, traceless Cotton tensor
\begin{align}
C^{\mu\nu} 
&= -(\nabla_\sigma \theta)\,\epsilon^{\sigma\alpha\beta(\mu}\nabla_\alpha R^{\nu)}{}_{\beta}
- \frac{1}{2}(\nabla_\sigma\nabla_\tau \theta)\,{}^{\star}\!R^{\tau(\mu\nu)\sigma}, \nonumber\\
{}^{\star}\!R^{\tau\mu\nu\sigma}&\equiv 
\tfrac{1}{2}\epsilon^{\mu\nu\alpha\beta}R^{\tau}{}_{\alpha\beta}{}^{\sigma}.
\label{eq:Cotton1}
\end{align}
We consider a spatially flat Friedmann-Robertson-Walker (FRW) background in conformal time $\tau$,
\begin{equation}
\bar{g}_{\mu\nu}dx^\mu dx^\nu = a^2(\tau)\Big[-d\tau^2 + d\vec{x}^{\,2}\Big],\qquad 
\mathcal{H}\equiv \frac{a'}{a},
\label{eq:FRWbg}
\end{equation}
and a homogeneous dCS scalar $\theta=\theta(\tau)$. Primes denote $d/d\tau$. The background satisfies the standard FRW equations sourced by $\theta$ and matter, while the Pontryagin density vanishes identically for FRW, so the CS sector modifies only the perturbations to linear order.
We perturb the metric as $g_{\mu\nu}=\bar{g}_{\mu\nu}+h_{\mu\nu}$ and focus on the transverse-traceless (TT) tensor sector. In the synchronous gauge,
\begin{equation}
ds^2 = a^2(\tau)\Big[-d\tau^2 + (\delta_{ij}+h_{ij})dx^i dx^j\Big],\qquad h_{0\mu}=0,
\label{eq:pertmetric}
\end{equation}
with the TT conditions
\begin{equation}
\partial_i h_{ij}=0,\qquad \delta^{ij}h_{ij}=0.
\label{eq:TT}
\end{equation}
We expand $h_{ij}$ in Fourier modes,
\begin{equation}
h_{ij}(\tau,\mathbf{x}) 
= \int\!\frac{d^3k}{(2\pi)^3}\,e^{i\mathbf{k}\cdot \mathbf{x}}\,h_{ij}(\tau,\mathbf{k}),
\qquad k\equiv|\mathbf{k}|.
\label{eq:Fourier}
\end{equation}
Linearizing~\eqref{eq:dCSeqs} around the FRW background with $\theta=\theta(\tau)$ and projecting onto the TT subspace yields the modified wave equation for the tensor modes. The GR piece gives the standard result $h_{ij}''+2\mathcal{H}h_{ij}'+k^2 h_{ij}=0$. The Cotton tensor contributes parity-odd terms proportional to $\theta'(\tau)$ and its derivatives. To leading order in the perturbations and for a homogeneous $\theta$, one finds in Fourier space~\cite{JackiwPi2003,AlexanderYunes2009,YunesPretorius:2009PRD}
\begin{equation}
h_{ij}'' + 2\mathcal{H}h_{ij}' + k^2 h_{ij} 
- 2\,\epsilon_{ikl}^{\ \ m}\,\frac{\alpha_{\mathrm{CS}}}{M_{\mathrm{Pl}}^2}
\,\frac{\theta'}{a^2}\,k_k\,\partial_\tau h_{lj}\Big|_{\mathrm{TT}} = 0,
\label{eq:TensorEqMatrix}
\end{equation}
where $\epsilon^{0ijk}=\epsilon^{ijk}/a^4$ and the spatial Levi-Civita symbol $\epsilon^{ijk}$ is defined with $\epsilon^{123}=+1$. The operator acting on $h_{ij}$ is parity odd and mixes the two helicities with opposite sign.
A compact and physically transparent form arises after decomposing into circular polarization states. Let $e^{R/L}_{ij}(\hat{\mathbf{k}})$ be the circular polarization tensors satisfying
\begin{equation}
i\,\epsilon_{il}^{\ \ mk}\,\hat{k}_k\,e^{R/L}_{lj} = \pm e^{R/L}_{ij},\qquad 
\delta^{ij}e^{R/L}_{ij}=0,\qquad k^i e^{R/L}_{ij}=0,\qquad 
e^{R}_{ij}e^{R,ij}=e^{L}_{ij}e^{L,ij}=2.
\label{eq:polTensors}
\end{equation}
We expand
\begin{equation}
h_{ij}(\tau,\mathbf{k}) = \sum_{\lambda=R,L} h_\lambda(\tau,\mathbf{k})\,e^\lambda_{ij}(\hat{\mathbf{k}}),
\label{eq:helicitydec}
\end{equation}
and define the dimensionless dCS coupling
\begin{equation}
\xi(\tau) \equiv \frac{\alpha_{\mathrm{CS}}}{M_{\mathrm{Pl}}^2}\,\frac{\theta'(\tau)}{a^2(\tau)}.
\label{eq:xiDef}
\end{equation}
Projecting~\eqref{eq:TensorEqMatrix} onto the helicity basis then gives the pair of decoupled equations
\begin{equation}
\big(1-2\lambda k\,\xi\big)\,h_\lambda'' 
+ 2\big(\mathcal{H}-\lambda k\,\xi'\big)\,h_\lambda' + k^2 h_\lambda = 0,
\qquad 
\lambda=\begin{cases}
+1,&\lambda=R,\\[2pt]
-1,&\lambda=L.
\end{cases}
\label{eq:hEqHelicity}
\end{equation}
Equation~\eqref{eq:hEqHelicity} exhibits explicitly the parity-odd nature of the dCS correction: right and left circular modes propagate differently whenever $\xi\neq 0$. The small-coupling regime relevant for cosmology is characterized by
\begin{equation}
\varepsilon_{\mathrm{CS}}(k) \equiv \max_\tau \big|k\,\xi(\tau)\big| \ll 1,
\label{eq:smallcs}
\end{equation}
in which the corrections are perturbative.
Introducing the canonically normalized variables
\begin{equation}
u_\lambda \equiv \frac{a\,M_{\mathrm{Pl}}}{\sqrt{2}}\,\sqrt{1-2\lambda k \xi}\,h_\lambda,
\qquad
z_\lambda \equiv \frac{a\,M_{\mathrm{Pl}}}{\sqrt{2}}\,\sqrt{1-2\lambda k \xi},
\label{eq:canon}
\end{equation}
one obtains the standard Schrödinger-like form
\begin{equation}
u_\lambda'' + \Big[k^2-\frac{z_\lambda''}{z_\lambda}\Big]\,u_\lambda = 0.
\label{eq:uEq}
\end{equation}
The helicity-dependent pump field $z_\lambda$ captures both the usual Hubble friction and the dCS birefringent contribution. In the small-coupling expansion,
\begin{align}
\frac{z_\lambda''}{z_\lambda} 
&= \frac{a''}{a} 
+ \lambda k\Big[ -\,\xi''-2\mathcal{H}\xi' + 2\Big(\frac{a''}{a}-\mathcal{H}^2\Big)\xi \Big] 
+ \mathcal{O}\!\big((k\xi)^2\big).
\label{eq:zpp}
\end{align}
Equation~\eqref{eq:uEq} together with~\eqref{eq:zpp} makes manifest that parity violation enters as a helicity-odd correction to the effective mass term of the tensor modes.
On sub-horizon scales where adiabatic/WKB solutions are valid ($k\gg \mathcal{H},|z_\lambda''/z_\lambda|$), the mode functions approximately obey
\begin{equation}
u_\lambda(\tau,k) \simeq 
\frac{1}{\sqrt{2\omega_\lambda(\tau,k)}}
\exp\!\left[-\,i\!\int^\tau\! d\tilde{\tau}\,\omega_\lambda(\tilde{\tau},k)\right],
\qquad
\omega_\lambda^2(\tau,k) = k^2-\frac{z_\lambda''}{z_\lambda}.
\label{eq:WKB1}
\end{equation}
To leading order in $\varepsilon_{\mathrm{CS}}(k)$ and neglecting slow time variations of $\xi$, the dispersion relation becomes
\begin{equation}
\omega_\lambda(\tau,k) \simeq k\big[1 + \lambda k\,\xi(\tau)\big],
\label{eq:dispersion1}
\end{equation}
so that the phase velocity is helicity dependent,
\begin{equation}
c_{T,\lambda}(\tau,k) \equiv \frac{\omega_\lambda}{k} \simeq 1 + \lambda k\,\xi(\tau),
\label{eq:phasevel}
\end{equation}
while the group velocity obeys the same relation at this order. The parity-odd correction also introduces a helicity-dependent friction through the pump field~\eqref{eq:zpp}; combining~\eqref{eq:hEqHelicity} and~\eqref{eq:canon}, one obtains the amplitude equation
\begin{equation}
\frac{d}{d\tau}\ln|h_\lambda| = -\,\mathcal{H} + \lambda k\big(\xi' + \mathcal{H}\xi\big) 
+ \mathcal{O}\!\big((k\xi)^2\big),
\label{eq:amplitude}
\end{equation}
which shows that a slowly varying $\xi$ produces helicity-dependent damping or amplification, i.e.\ amplitude birefringence~\cite{LueWangKamionkowski1999,AlexanderYunes2009}.
Let $P_h^\lambda(k)$ denote the late-time tensor power spectrum for each helicity, defined by
\begin{equation}
\langle h_\lambda(\mathbf{k})\,h_{\lambda'}(\mathbf{k}')\rangle 
= (2\pi)^3\delta^{(3)}(\mathbf{k}+\mathbf{k}')
\frac{2\pi^2}{k^3}P_h^\lambda(k)\,\delta_{\lambda\lambda'}.
\label{eq:PhDef}
\end{equation}
In the small-coupling regime~\eqref{eq:smallcs} and under slow time evolution of $\xi$, the spectra can be expressed as multiplicative deformations of the GR result,
\begin{equation}
P_h^{R/L}(k) = P_h^{\mathrm{GR}}(k)\big[1 \pm \delta_{\mathrm{CS}}(k)\big] 
+ \mathcal{O}\!\big(\varepsilon_{\mathrm{CS}}^2\big),
\label{eq:PRL}
\end{equation}
with a parity-odd asymmetry
\begin{equation}
\delta_{\mathrm{CS}}(k) = 2k \int_{\eta_{\mathrm{in}}}^{\eta_{\mathrm{fin}}} d\eta\,\xi(\eta)\,\mathcal{W}(k,\eta) 
+ \mathcal{O}\!\big(\partial_\eta\xi/k\big),
\label{eq:deltaCS}
\end{equation}
where $\mathcal{W}(k,\eta)$ is a dimensionless window function peaked around horizon exit or re-entry (depending on whether one considers primordial generation or late-time propagation), normalized so that $\mathcal{W}\to 1$ for slowly varying backgrounds~\cite{Satoh:2008PRD,AlexanderYunes2009}. The leading expression~\eqref{eq:deltaCS} follows either from the adiabatic solution~\eqref{eq:WKB}-\eqref{eq:amplitude} or from the helicity-dependent canonical pump~\eqref{eq:zpp} by matching across the relevant horizon scale. Equation~\eqref{eq:PRL} implies a tensor chirality parameter at a pivot $k_\star$,
\begin{equation}
\chi_T(k_\star) \equiv 
\frac{P_h^{R}(k_\star)-P_h^{L}(k_\star)}{P_h^{R}(k_\star)+P_h^{L}(k_\star)} 
= \delta_{\mathrm{CS}}(k_\star) + \mathcal{O}\!\big(\varepsilon_{\mathrm{CS}}^2\big),
\label{eq:chiT}
\end{equation}
which is the parity diagnostic used in the main text. The small-coupling condition~\eqref{eq:smallcs} ensures $|\delta_{\mathrm{CS}}|\ll 1$ and the perturbative control of the tensor spectra.
The tensor sector of dCS gravity on an FRW background exhibits helicity-dependent propagation governed by~\eqref{eq:hEqHelicity} or, in canonical form,~\eqref{eq:uEq}-\eqref{eq:zpp}. The leading observable consequences are a parity-dependent phase velocity~\eqref{eq:phasevel}, amplitude birefringence~\eqref{eq:amplitude}, and a corresponding helicity asymmetry in the tensor power spectrum~\eqref{eq:PRL} with asymmetry parameter~\eqref{eq:deltaCS}. These formulas feed directly into the definition of the tensor chirality parameter $\chi_T$ without reference to the EC torsion sector or numerical strategies developed elsewhere in the manuscript, and provide the rigorous tensor-level underpinning for the parity-sensitive observables employed in the geometric baryogenesis framework.

\section{Stückelberg Portal Derivation \& Uniqueness}\label{appc}

In a gauge realization of $U(1)_{B-L}$ that is compatible with a heavy vector boson and preserves gauge invariance at all scales below the ultraviolet (UV) completion, the Stückelberg mechanism provides the unique and minimal way to endow the Abelian gauge field with a mass while maintaining a local symmetry. In the geometric baryogenesis framework, this is not merely a kinematical convenience: it fixes the gauge-invariant building blocks out of which all couplings to the chiral-time equivalence (CTE) sector must be assembled and, as a consequence, singles out a unique portal structure between the $U(1)_{B-L}$ sector, the CTE pseudoscalar $\Phi$, and fermionic currents. In particular, the Stückelberg mass consistently coexists with the anomaly-free realization of $(B\!-\!L)$ (e.g.\ in the presence of right-handed neutrinos) and does not introduce additional anomaly inflow beyond those arranged by the CTE construction \cite{Ruegg:2003IMPA,Kors:2004PLB,Kors:2004JHEP,Feldman:2007PRD,Preskill:1991AnnPhys}.
Let $A_\mu$ be the gauge field of $U(1)_{B-L}$ with field strength $F_{\mu\nu}=\partial_\mu A_\nu-\partial_\nu A_\mu$, and let $\sigma$ be the Stückelberg scalar. The Stückelberg Lagrangian, coupled to matter and to the CTE sector through a generic interaction $\mathcal{L}_{\mathrm{int}}$, is
\begin{equation}
\mathcal{L}_{\mathrm{Stuck}} = -\frac{1}{4}F_{\mu\nu}F^{\mu\nu}
-\frac{1}{2}\big(\partial_\mu\sigma-M A_\mu\big)\big(\partial^\mu\sigma-M A^\mu\big)
+\mathcal{L}_{\mathrm{int}}[\Phi,\psi,A_\mu].
\label{eq:Lstueck}
\end{equation}
Under a local $U(1)_{B-L}$ transformation with parameter $\alpha(x)$,
\begin{equation}
A_\mu \to A_\mu + \partial_\mu \alpha,\qquad
\sigma \to \sigma + M \alpha,\qquad
\psi \to e^{i g_{B-L} q_{B-L}\alpha}\psi,
\label{eq:gaugetrf}
\end{equation}
the gauge-invariant combination
\begin{equation}
B_\mu \equiv \partial_\mu \sigma-M A_\mu
\label{eq:Bmu}
\end{equation}
is invariant ($B_\mu\to B_\mu$), and $F_{\mu\nu}$ is invariant as usual. The interaction of the gauge sector with conserved matter current is
\begin{equation}
\mathcal{L}_J = g_{B-L}\,A_\mu J^\mu_{B-L},
\qquad
\partial_\mu J^\mu_{B-L}=0,
\label{eq:LJ}
\end{equation}
with $J^\mu_{B-L}$ the Noether current of the anomaly-free $(B\!-\!L)$ symmetry. It proves convenient to regard $B_\mu$ and $F_{\mu\nu}$ as the only gauge-invariant building blocks involving the Abelian sector in the effective theory.
The most general parity-even, local and Lorentz-invariant interaction between the Stückelberg sector, the CTE scalar $\Phi$, and the matter current at mass-dimension $d\le5$ is spanned by the operators built from $B_\mu$, $F_{\mu\nu}$, $J^\mu_{B-L}$, $\Phi$ and its derivatives \cite{Ruegg:2003IMPA}. Gauge invariance enforces that $A_\mu$ and $\sigma$ appear only through $B_\mu$ and $F_{\mu\nu}$, and renormalizability up to dimension five restricts the operator basis. A general ansatz consistent with these requirements is
\begin{align}
\mathcal{L}_{\mathrm{int}}[\Phi,\psi,A_\mu] 
&= g_{B-L}A_\mu J^\mu_{B-L} + f(\Phi)B_\mu B^\mu 
+ \frac{\beta}{\Lambda}(\partial_\mu\Phi)B^\mu 
+ \frac{\gamma}{\Lambda}J^\mu_{B-L}B_\mu + \cdots,
\label{eq:LintGeneral}
\end{align}
where $f(\Phi)$ is a scalar function, $\Lambda$ is a heavy mass scale and $\beta,\gamma$ are dimensionless Wilson coefficients; the ellipsis denotes operators of dimension $>5$ or containing additional derivatives acting on $F_{\mu\nu}$. The $J^\mu_{B-L}B_\mu$ term is of the same mass-dimension as $A_\mu J^\mu_{B-L}$ and may be absorbed into a field redefinition; we keep it here to exhibit the structure that emerges after eliminating the heavy vector. The term $(\partial\Phi)\cdot B$ is the unique dimension-five, gauge-invariant, shift-symmetric operator that couples the CTE field to the Abelian sector at leading order, while $f(\Phi)B^2$ accounts for kinetic modulation in a background value of $\Phi$. Demanding an approximate shift symmetry for $\Phi$ restricts $f(\Phi)$ to a constant at leading order,\footnote{If $\Phi$ enjoys an approximate shift symmetry ($\Phi\to \Phi+c$), then $f(\Phi)=f_0+\mathcal{O}((\partial\Phi)^2/\Lambda^2)$, and linear terms in $\Phi$ are forbidden in the decoupling limit. A nontrivial $f(\Phi)$ can always be reorganized, via the $\sigma$ equation of motion, into a combination of $B^2$ with constant coefficient plus the allowed derivative operator $(\partial\Phi)\cdot B/\Lambda$.} so we set $f(\Phi)=f_0$ henceforth.
Combining \eqref{eq:Lstueck} and \eqref{eq:LintGeneral} and using $A_\mu = (\partial_\mu\sigma-B_\mu)/M$, one may rewrite the interacting part as
\begin{align}
\mathcal{L}_{\mathrm{int}} 
&= \frac{g_{B-L}}{M}\,\partial_\mu\sigma\,J^\mu_{B-L} 
- \frac{g_{B-L}}{M}\,B_\mu J^\mu_{B-L} 
+ f_0 B_\mu B^\mu 
+ \frac{\beta}{\Lambda}(\partial_\mu\Phi)B^\mu 
+ \frac{\gamma}{\Lambda}J^\mu_{B-L}B_\mu.
\label{eq:LintB}
\end{align}
At energy-momentum scales $E\ll M$, the heavy vector sector may be integrated out by solving its field equation algebraically to leading order in derivatives. Neglecting the Maxwell kinetic term $F_{\mu\nu}F^{\mu\nu}$ at zeroth order in $E/M$ and collecting the terms quadratic and linear in $B_\mu$ yields the algebraic equation
\begin{equation}
(1+2 f_0)\,B_\mu = \frac{g_{B-L}}{M}\,J^\mu_{B-L} 
- \frac{\beta}{2\Lambda}\,\partial^\mu\Phi 
- \frac{\gamma}{2\Lambda}\,J^\mu_{B-L} 
+ \mathcal{O}\!\Big(\frac{\partial}{M}\Big),
\label{eq:Beom}
\end{equation}
whose solution substituted back into \eqref{eq:LintB} gives the low-energy effective portal and contact interactions,
\begin{align}
\mathcal{L}_{\mathrm{eff}} 
&= \frac{g_{B-L}}{M}\,\partial_\mu\sigma\,J^\mu_{B-L}
+ \frac{\beta_{\mathrm{eff}}}{\Lambda}(\partial_\mu\Phi)\partial^\mu\sigma
- \frac{1}{2M_*^2}\,J_{B-L,\mu}J^\mu_{B-L}
- \frac{\eta}{2\Lambda^2}(\partial_\mu\Phi)(\partial^\mu\Phi) + \cdots,
\label{eq:LeffPortal}
\end{align}
with
\begin{equation}
M_* = \frac{M}{g_{B-L}}\sqrt{1+2f_0},\qquad
\beta_{\mathrm{eff}} = -\,\frac{\beta}{2(1+2f_0)},\qquad
\eta = \frac{\beta^2}{4(1+2f_0)}.
\label{eq:params}
\end{equation}
Up to $\mathcal{O}(E/M)$ corrections from the Maxwell term and higher-derivative operators, \eqref{eq:LeffPortal} exhibits the desired Stückelberg portal structure: a derivative coupling between the Stückelberg mode and the conserved $(B\!-\!L)$ current, an optional kinetic mixing between $\Phi$ and $\sigma$ suppressed by $\Lambda^{-1}$, and a current-current contact interaction. In unitary gauge ($\sigma=0$), the derivative portal reappears after integrating out the massive vector as the unique dimension-five operator $(\partial_\mu\Phi\,J^\mu_{B-L}/M_*)$ once the $(\partial\Phi\cdot\partial\sigma)$ mixing in \eqref{eq:LeffPortal} is diagonalized by field redefinitions.

We now show that the portal in \eqref{eq:LeffPortal} is unique under three assumptions: $U(1)_{B-L}$ gauge invariance, renormalizability up to $d\le5$, and absence of additional anomaly contributions beyond those required by the CTE construction. The proof proceeds by operator classification.
First, gauge invariance implies that $A_\mu$ and $\sigma$ enter only through $F_{\mu\nu}$ and $B_\mu$. At $d\le5$, the only independent gauge-invariant contractions with the matter sector and the CTE field are $A_\mu J^\mu_{B-L}$, $B_\mu J^\mu_{B-L}$, $B_\mu B^\mu$, $F_{\mu\nu}F^{\mu\nu}$, and $(\partial_\mu\Phi)B^\mu$. Any term with $\Phi$ without derivatives explicitly breaks at least the approximate shift symmetry of $\Phi$ used in the CTE sector, and is therefore excluded at leading order.
Second, among these operators only $A_\mu J^\mu_{B-L}$ introduces the minimal gauge coupling, while $B_\mu J^\mu_{B-L}$ is redundant at $d=4$ because $B_\mu = \partial_\mu\sigma-M A_\mu$ and the $(\partial_\mu\sigma\,J^\mu_{B-L})$ piece can be shifted into a total derivative in the action when the current is conserved; the remaining $-M A_\mu J^\mu_{B-L}$ renormalizes the gauge coupling. The operator $B_\mu B^\mu$ is already present in the Stückelberg mass term and any modulation $f(\Phi)$ can at most induce, after use of the $\sigma$ equation of motion, a linear combination of $B^2$ with constant coefficient and the allowed dimension-five operator $(\partial\Phi)\cdot B/\Lambda$. The Maxwell kinetic term $F_{\mu\nu}F^{\mu\nu}$ is part of the free Lagrangian and does not generate new portal structures.

Third, potential Chern-Simons-like terms $(A\wedge F)$ or $(\Phi\,F\tilde{F})$ are forbidden by gauge invariance and the requirement that no new anomalies are introduced beyond those already canceled or saturated in the UV completion; in particular, the anomaly-free nature of $(B\!-\!L)$ in the presence of right-handed neutrinos eliminates Wess-Zumino terms that would otherwise couple the Stückelberg scalar linearly to $F\tilde{F}$ \cite{Preskill:1991AnnPhys}. Therefore, up to field redefinitions and the decoupling of higher-derivative suppressed operators, the only independent portal at $d\le5$ is proportional to $(\partial\Phi)\cdot B$, which, after integrating out the heavy vector, is equivalent to the derivative coupling $(\partial\Phi)\cdot J_{B-L}/M_*$ plus contact terms as in \eqref{eq:LeffPortal}. This establishes uniqueness.

The derivation is independent of torsion elimination in Einstein-Cartan gravity (Appendix~\eqref{appa}) because the Stückelberg sector couples to the conserved current $J^\mu_{B-L}$ and to $\Phi$ through gauge-invariant combinations $B_\mu$ and $F_{\mu\nu}$ only; eliminating torsion modifies axial four-fermion operators but leaves $J^\mu_{B-L}$ conserved and the portal unaffected at leading order. It is also orthogonal to the tensor sector of dynamical Chern-Simons gravity (Appendix~\eqref{appb}), since the portal involves neither the Cotton tensor nor the parity-odd gravitational operator $R\tilde{R}$ and contributes to tensor observables only through the overall normalization of the $(B\!-\!L)$ bias after the heavy vector is decoupled. Finally, EFT bounds and UV completions discussed in the main text constrain the parametric domain $(g_{B-L},M,\Lambda)$ but do not alter the operator identity that singles out the portal.
The heavy-vector decoupling controlling the derivative portal requires a hierarchy of scales $E\ll M$, where $E$ is the characteristic energy in the plasma when the slow $(B\!-\!L)$ charge freezes in. A perturbative unitarity estimate obtained from the partial-wave expansion of $\psi\bar{\psi}\to \psi\bar{\psi}$ mediated by the heavy vector yields
\begin{equation}
\sqrt{s} \lesssim \Lambda_{\mathrm{unit}} = \sqrt{16\pi}\,\frac{M}{g_{B-L}},
\label{eq:unitarity}
\end{equation}
so that the EFT cutoff obeys $M_*=\tfrac{M}{g_{B-L}}\sqrt{1+2f_0}\lesssim \Lambda_{\mathrm{unit}}/\sqrt{16\pi}$. In addition, the portal coupling must remain in the linear-response regime for baryogenesis,
\begin{equation}
\left|\frac{\mu_{B-L}}{T}\right| = \left|\frac{u^\mu\partial_\mu \Phi}{M_*T}\right| \ll 1,
\label{eq:linear}
\end{equation}
so as to justify the susceptibility expansion used in the kinetic treatment. Constraints on mixing between $\Phi$ and $\sigma$ from \eqref{eq:LeffPortal} enforce
\begin{equation}
\left|\frac{\beta_{\mathrm{eff}}}{\Lambda}\right|\,\frac{|\partial \Phi|}{M_*} \ll 1,
\label{eq:mixing}
\end{equation}
to ensure that the Stückelberg mode remains nondynamical at freeze-in. These inequalities define a broad, sub-Planckian region where $g_{B-L}\lesssim \mathcal{O}(10^{-2})$, $M\gtrsim \mathcal{O}(10^{15}\,\mathrm{GeV})$, and $\Lambda\gtrsim M$ are compatible with weak coupling and EFT control; the baryogenesis dynamics then depends only on the combination $M_*$ that normalizes the derivative portal in \eqref{eq:LeffPortal}.
The Stückelberg realization of $U(1)_{B-L}$ fixes the gauge-invariant structures available to the low-energy effective theory and, when combined with gauge invariance, renormalizability up to dimension five, and the absence of additional anomalies, uniquely determines the portal between the Abelian sector and the CTE framework. Integrating out the heavy vector produces a derivative coupling $(\partial\Phi)\cdot J_{B-L}/M_*$ and benign contact terms, while maintaining complete consistency with torsion elimination, tensor-sector parity effects, and the EFT/UV structure of the theory. The resulting portal is therefore both structurally unavoidable and phenomenologically adequate for geometric baryogenesis.

\section{Charge Matrix and \texorpdfstring{$S$}{S} Factor}\label{appd}

This appendix formalizes the linear-response map between microscopic particle asymmetries and the slowly evolving conserved (or approximately conserved) charges relevant for baryogenesis. The map is encoded by a \emph{charge matrix} $Q$ assigning the relevant $\mathrm{U}(1)$ quantum numbers (in particular $B\!-\!L$ and/or flavored $B/3-L_\alpha$) to each effectively massless species. Integrating out fast reactions yields an algebraic projector from microscopic chemical potentials to the slow-charge subspace. The resulting contraction with the (inverse) washout operator defines the \emph{$S$ factor}, which controls the equilibrium response of slow charges to spurionic chemical potentials and the efficiency of asymmetry survival \cite{Buchmuller:2005Ann,DavidsonNardiNir2008,Blanchet:2007JCAP,Nardi:2006JHEP,HarveyTurner1990}.
Let $\alpha$ label effectively massless species tracked in the kinetic description (quarks, leptons, Higgs, and optionally equilibrated $\nu_R$). Collect particle--antiparticle asymmetries into $n_\alpha$ and chemical potentials into $\mu_\alpha$. For $|\mu_\alpha|/T\ll 1$,
\begin{equation}
n_\alpha=\chi_\alpha\,\mu_\alpha,
\qquad
\chi_\alpha=\frac{g_\alpha T^2}{6},
\label{eq:linresp1}
\end{equation}
where $g_\alpha$ counts relativistic degrees of freedom \cite{Weinberg:2008cosmo}. Define $\boldsymbol{\chi}\equiv \mathrm{diag}(\chi_\alpha)$ and the $\boldsymbol{\chi}$-weighted inner product $(u,v)_\chi\equiv u^T\boldsymbol{\chi}v$.
Slow charges $X_i$ ($i=1,\dots,N_s$) are linear combinations of species asymmetries,
\begin{equation}
X_i=\sum_\alpha Q_{\alpha i}\,n_\alpha=(Q^T n)_i,
\qquad
n\equiv (n_\alpha),
\qquad
Q\equiv (Q_{\alpha i}).
\label{eq:Xdef}
\end{equation}
For a basis ordered schematically as $(e_L,\mu_L,\tau_L,\ldots; e_R,\mu_R,\tau_R,\ldots; q_L,\ldots; H,\ldots)$, the charge matrix takes the schematic block form
\begin{equation}
Q_{\alpha i}=
\begin{pmatrix}
q_{e_L}^{(i)} & q_{\mu_L}^{(i)} & q_{\tau_L}^{(i)} & \cdots\\
q_{e_R}^{(i)} & q_{\mu_R}^{(i)} & q_{\tau_R}^{(i)} & \cdots\\
\vdots & \vdots & \vdots & \ddots
\end{pmatrix},
\label{eq:Qmatrix}
\end{equation}
with, e.g.\ $q_{\ell_L}^{(B-L)}=-1$, $q_{\nu_R}^{(B-L)}=-1$ (if present), $q_{q_L}^{(B-L)}=+1/3$, $q_H^{(B-L)}=0$. Flavored charges $(B/3-L_\alpha)$ are included by adding columns to $Q$ \cite{Blanchet:2007JCAP,Nardi:2006JHEP}. 
In the linear regime, the fast sector can be formulated directly in terms of chemical potentials as
\begin{equation}
\frac{d\mu}{dt}= -\,\mathsf{K}\,\mu+\mathsf{J}\,\mu_{\mathrm{slow}}+\cdots,
\qquad
\mathsf{K}\equiv \boldsymbol{\chi}^{-1}\boldsymbol{\Gamma},
\qquad
\mathsf{J}\equiv \boldsymbol{\chi}^{-1}Q,
\label{eq:muEOM}
\end{equation}
where $\boldsymbol{\Gamma}$ is the symmetric washout matrix in the species basis and $\mu_{\mathrm{slow}}$ are spurionic chemical potentials conjugate to the slow charges.\footnote{For our purposes, $\mu_{\mathrm{slow}}$ represents a weak driving in the slow sector, e.g.\ a parity-odd bias. The dots denote explicit sources not proportional to $\mu_{\mathrm{slow}}$ and subleading derivative corrections.} In the quasi-stationary limit for fast modes, $d\mu/dt\simeq 0$,
\begin{equation}
\mu=\mathsf{K}^{-1}\mathsf{J}\,\mu_{\mathrm{slow}}
=\boldsymbol{\Gamma}^{-1}Q\,\mu_{\mathrm{slow}},
\label{eq:musteady}
\end{equation}
using $\mathsf{K}^{-1}=\boldsymbol{\Gamma}^{-1}\boldsymbol{\chi}$. The induced slow-charge response is then
\begin{equation}
X\equiv Q^T n=Q^T\boldsymbol{\chi}\,\mu
=Q^T\boldsymbol{\chi}\,\boldsymbol{\Gamma}^{-1}Q\,\mu_{\mathrm{slow}}.
\label{eq:Xresponse}
\end{equation}
This identifies the slow-charge susceptibility matrix in the presence of fast washouts,
\begin{equation}
\mathcal{C}\equiv Q^T\boldsymbol{\chi}\,\boldsymbol{\Gamma}^{-1}Q.
\label{eq:CalC}
\end{equation}
Equivalently, defining $\widehat{Q}\equiv \boldsymbol{\chi}^{1/2}Q$ and $\widehat{\Gamma}\equiv \boldsymbol{\chi}^{-1/2}\boldsymbol{\Gamma}\boldsymbol{\chi}^{-1/2}$,
\begin{equation}
\mathcal{C}=\widehat{Q}^{\,T}\widehat{\Gamma}^{-1}\widehat{Q}.
\label{eq:CalChat}
\end{equation}
The \emph{$S$ factor} is the inverse of $\mathcal{C}$,
\begin{equation}
S\equiv \mathcal{C}^{-1}
=\Big(Q^T\boldsymbol{\chi}\,\boldsymbol{\Gamma}^{-1}Q\Big)^{-1}
=\Big(\widehat{Q}^{\,T}\widehat{\Gamma}^{-1}\widehat{Q}\Big)^{-1},
\label{eq:Sdef}
\end{equation}
so the constitutive relation between slow charges and their spurionic potentials reads
\begin{equation}
\mu_{\mathrm{slow}} = S\,X.
\label{eq:muslow}
\end{equation}
When slow charges are normalized as yields $Y_i\equiv X_i/s$, it is common to write $X_i=(T^2/6)(\widetilde{S}\mu_{\mathrm{slow}})_i$, where $\widetilde{S}$ differs from $S^{-1}$ by the overall factor $T^2/6$ and by the $g_\alpha$ weights absorbed into $\boldsymbol{\chi}$ \cite{Buchmuller:2005Ann,DavidsonNardiNir2008}. 
The matrix $Q$ has rank $r\le N_s$. The symmetric $\widehat{\Gamma}$ is positive semidefinite and typically positive definite on the fast subspace orthogonal to exactly conserved charges. If exact null modes are present, one replaces $\widehat{\Gamma}^{-1}$ by the Moore--Penrose pseudoinverse $\widehat{\Gamma}^{+}$ (equivalently, project out conserved directions before inversion); Eqs.~\eqref{eq:CalChat}-\eqref{eq:Sdef} remain valid with this replacement \cite{Blanchet:2007JCAP,DavidsonNardiNir2008}. Gauge invariance of $\mathrm{U}(1)_{B-L}$ constrains the columns of $Q$ through anomaly cancellation and hypercharge neutrality, but does not alter the linear-algebraic structure of \eqref{eq:Sdef}. CP-violating phases enter the kinetic problem through source terms and the antisymmetric part of the collision operator; to leading order in the linear response relevant for $S$, only the symmetric washout matrix $\boldsymbol{\Gamma}$ contributes, so $S$ is CP-even. 
For a single slow charge, $Q$ is a column vector $q_\alpha$ and
\begin{equation}
S^{-1}=q^T\boldsymbol{\chi}\,\boldsymbol{\Gamma}^{-1}q
=\sum_{\alpha,\beta} q_\alpha\,\chi_\alpha\,(\Gamma^{-1})_{\alpha\beta}\,q_\beta.
\label{eq:ScalarS}
\end{equation}
If $\boldsymbol{\Gamma}$ is diagonal, $\Gamma_{\alpha\beta}=\gamma_\alpha\delta_{\alpha\beta}$,
\begin{equation}
S^{-1}=\sum_\alpha \frac{\chi_\alpha\,q_\alpha^2}{\gamma_\alpha}.
\label{eq:Sdiag}
\end{equation}
In the flavor-democratic limit $\gamma_\alpha=\gamma$,
\begin{equation}
S^{-1}=\frac{1}{\gamma}\sum_\alpha \chi_\alpha q_\alpha^2,
\qquad
S=\frac{\gamma}{\sum_\alpha \chi_\alpha q_\alpha^2}.
\label{eq:Sdem}
\end{equation}
Factoring out the common $T^2/6$ in $\chi_\alpha$, the remaining sum $\sum_\alpha g_\alpha q_\alpha^2$ reproduces the standard susceptibility-counting factors of the electroweak plasma, e.g.\ $S=13$ for the SM without right-handed neutrinos and $S=16$ for the SM with three equilibrated $\nu_R$, in the conventional normalization $X=(T^2/6)\,S\,\mu_{\mathrm{slow}}$ \cite{HarveyTurner1990,Buchmuller:2005Ann}.
In the two-flavor regime with slow-charge basis $X=(B/3-L_\tau,\;B/3-L_{\perp})$ (where $L_\perp$ is orthogonal to $L_\tau$), the susceptibility is a $2\times 2$ matrix and $S=\mathcal{C}^{-1}$. For diagonal $\boldsymbol{\Gamma}$ this reads
\begin{equation}
\mathcal{C}=
\begin{pmatrix}
\sum_\alpha \dfrac{\chi_\alpha\,q^{(\tau)}_\alpha q^{(\tau)}_\alpha}{\gamma_\alpha} &
\sum_\alpha \dfrac{\chi_\alpha\,q^{(\tau)}_\alpha q^{(\perp)}_\alpha}{\gamma_\alpha} \\[10pt]
\sum_\alpha \dfrac{\chi_\alpha\,q^{(\perp)}_\alpha q^{(\tau)}_\alpha}{\gamma_\alpha} &
\sum_\alpha \dfrac{\chi_\alpha\,q^{(\perp)}_\alpha q^{(\perp)}_\alpha}{\gamma_\alpha}
\end{pmatrix},
\qquad
S=\mathcal{C}^{-1},
\label{eq:Stwoflav}
\end{equation}
with off-diagonal entries encoding flavor transfer induced by fast interactions. The generalization to three flavors is immediate by promoting $\mathcal{C}$ and $S$ to $3\times 3$ matrices in the $(B/3-L_e,\;B/3-L_\mu,\;B/3-L_\tau)$ basis. 
For bookkeeping, in $\boldsymbol{\chi}$-orthonormalized variables define the rank-$r$ projector onto the slow-charge subspace,
\begin{equation}
\mathsf{P}_{\mathrm{slow}}\equiv \widehat{Q}\,(\widehat{Q}^{\,T}\widehat{Q})^{-1}\widehat{Q}^{\,T}.
\label{eq:ProjSlow}
\end{equation}
Then $S=(\widehat{Q}^{\,T}\widehat{\Gamma}^{-1}\widehat{Q})^{-1}$ implies that $S$ acts as the inverse metric induced by $\widehat{\Gamma}$ on the slow subspace, and provides a compact way to express how fast washouts project onto slow charges. (This identity is not required in the main text but is occasionally useful in analytic estimates.)
Finally, since $Q$ and $\boldsymbol{\Gamma}$ are purely kinetic/plasma objects, this construction is independent of torsion elimination, St\"uckelberg portal derivations, and tensor-sector parity calculations; those enter only through source terms and the temperature/time dependence of rates.

\section{Slow-Roll and Resonant Yields}\label{appe}

\begin{figure}[htb]
    \centering
    \includegraphics[width=\linewidth, height=0.5\linewidth]{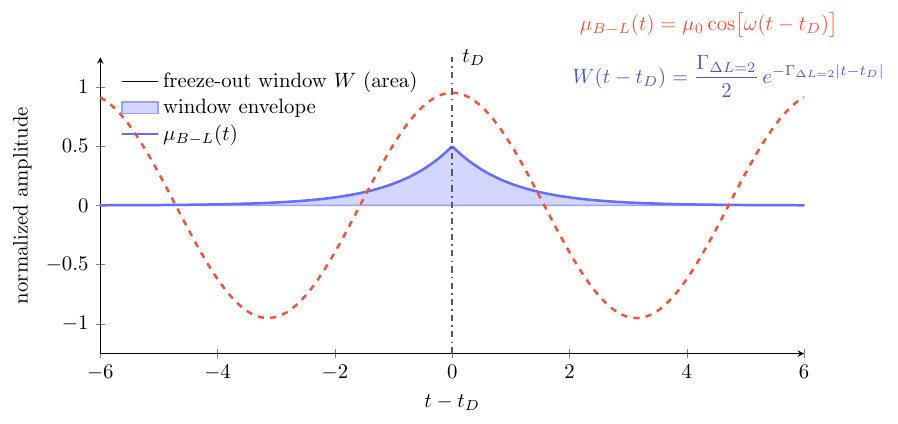}
    \includegraphics[width=\linewidth, height=0.5\linewidth]{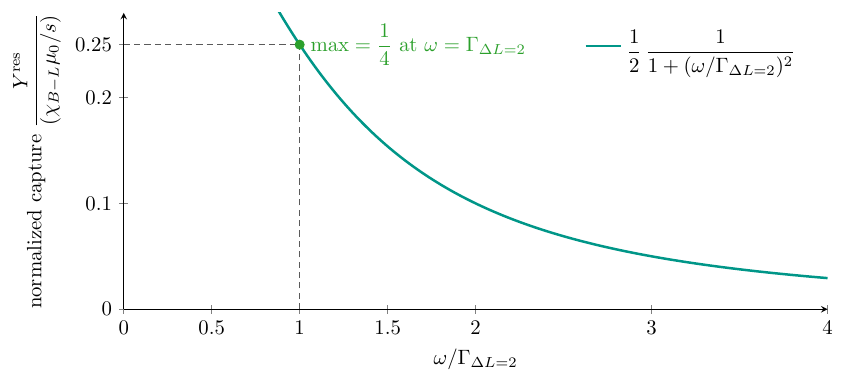}
    \caption{\textbf{Freeze-out kernel and resonant capture.}
  (a) Symmetric exponential window $W(t-t_D)$ weights an oscillatory bias $\mu_{B-L}(t)$.
  (b) The captured fraction is Lorentzian, $(1/2)/[1+(\omega/\Gamma_{\Delta L=2})^2]$, peaking at $1/4$
  for $\omega=\Gamma_{\Delta L=2}$.}
  \label{fig:freezeout-resonant}
\end{figure}

This appendix derives the two parametric regimes used in the main text for generating a slow-charge
asymmetry from a homogeneous, parity-odd background $\Phi(t)$ with a derivative portal. In the
\emph{adiabatic} (slow-drift) regime the plasma tracks an instantaneous equilibrium set by a spurionic
chemical potential $\mu_X(t)=\dot\Phi/M_*$, giving a thermodynamic freeze-out yield at
$T_D$ (defined by $\Gamma_X(T_D)\simeq H(T_D)$). In the \emph{oscillatory} regime, when $\mu_X(t)$ varies
with frequency comparable to the slow relaxation rate, a resonant capture factor appears, yielding a
Lorentzian enhancement peaked at $\omega\simeq \Gamma_X$.
We model the homogeneous background by
\begin{equation}
S_\Phi=\int d^4x\,\sqrt{-g}\,\Big[-\tfrac{1}{2}(\nabla\Phi)^2-V(\Phi)\Big],
\label{eq:SphiE}
\end{equation}
on FLRW with $H=\dot a/a$, and assume a derivative coupling $\mathcal{L}_{\rm int}\supset (\partial_\mu\Phi/M_*)J_X^\mu$,
so that in a homogeneous state
$\mu_X(t)=\dot\Phi(t)/M_*$.  For definiteness we treat $X$ as a slow charge violated by reactions with rate
$\Gamma_X(T)$. 
In the slow-roll regime,
\begin{equation}
\ddot{\Phi}+3H\dot{\Phi}+V'(\Phi)=0,
\qquad
|\ddot{\Phi}|\ll |3H\dot{\Phi}|,\ |V'(\Phi)|,
\label{eq:slowrolleom}
\end{equation}
with slow-roll parameters
\begin{equation}
\epsilon_\Phi\equiv \frac{M_{\mathrm{Pl}}^2}{2}\left(\frac{V'(\Phi)}{V(\Phi)}\right)^2\ll 1,
\qquad
\eta_\Phi\equiv M_{\mathrm{Pl}}^2\frac{V''(\Phi)}{V(\Phi)}\ll 1,
\label{eq:SRparameters}
\end{equation}
one has
\begin{equation}
3H\dot{\Phi}\simeq -V'(\Phi),
\qquad
\mu_X(t)=\frac{\dot{\Phi}}{M_*}\simeq -\,\frac{V'(\Phi)}{3H M_*}.
\label{eq:muSR}
\end{equation}
The kinetic adiabaticity condition is $|\dot\mu_X|\ll \Gamma_X|\mu_X|$; in radiation domination
$|\dot\mu_X|/|\mu_X|\sim\mathcal{O}(H)$, so adiabatic tracking holds for $\Gamma_X\gg H$.
For $|\mu|/T\ll 1$, a relativistic species $\alpha$ has susceptibility $\chi_\alpha=g_\alpha T^2/6$ \cite{LeBellac:2011TF,Weinberg:2008cosmo}.
Summing over species carrying $X$ gives
\begin{equation}
n_X=\sum_\alpha Q_{\alpha X}n_\alpha=\chi_X(T)\,\mu_X,
\qquad
\chi_X(T)\equiv \sum_\alpha Q_{\alpha X}^2\,\chi_\alpha(T),
\label{eq:nXeq}
\end{equation}
and relaxation to equilibrium may be written as
\begin{equation}
\dot n_X=-\,\Gamma_X(T)\big(n_X-\chi_X\mu_X\big).
\label{eq:relax}
\end{equation}
Freeze-out at $T_D$ yields
\begin{equation}
Y_X^{\rm SR}\equiv \frac{n_X}{s}\Big|_{T_D}
=\frac{\chi_X(T_D)}{s(T_D)}\,\mu_X(T_D)
=\frac{15}{4\pi^2}\,\frac{g_{X,{\rm eff}}(T_D)}{g_*(T_D)}\,\frac{\mu_X(T_D)}{T_D},
\qquad
g_{X,{\rm eff}}\equiv \sum_\alpha Q_{\alpha X}^2 g_\alpha,
\label{eq:Yslowroll}
\end{equation}
with $s=(2\pi^2/45)g_*T^3$ \cite{KolbTurner:1990book}. 
The coupling defines a time-dependent interaction Hamiltonian
\begin{equation}
H_{\mathrm{int}}(t)=-\int d^3x\,\mu_X(t)\,J_X^0(t,\mathbf{x}),
\label{eq:Hint}
\end{equation}
so linear response gives
\begin{equation}
\delta\langle J_X^0(t,\mathbf{k}=0)\rangle
=\int_{-\infty}^{t}\!dt'\,G^R_{J^0J^0}(t-t',\mathbf{k}=0)\,\mu_X(t'),
\label{eq:linrespJ}
\end{equation}
with $G^R_{J^0J^0}$ the retarded correlator. In frequency space,
$\delta n_X(\omega)=G^R_{J^0J^0}(\omega,\mathbf{0})\,\mu_X(\omega)$, and the Kubo relation yields
\begin{equation}
\chi_X=\lim_{\omega\to 0}\frac{1}{\omega}\,\mathrm{Im}\,G^R_{J^0J^0}(\omega,\mathbf{0}),
\label{eq:KuboChi}
\end{equation}
reproducing $n_X=\chi_X\mu_X$ in the adiabatic limit. Deviations from equilibrium
$\delta n_X\equiv n_X-\chi_X\mu_X$ relax as
\begin{equation}
\dot{\delta n}_X(t)=-\,\Gamma_X(T)\,\delta n_X(t),
\label{eq:diss}
\end{equation}
with $\Gamma_X$ determined by the low-frequency spectral weight of the violating interactions
(e.g.\ $\Gamma_X=\chi_X^{-1}\lim_{\omega\to 0}\omega^{-1}\mathrm{Im}\,G^R_{\dot J^0\dot J^0}(\omega,\mathbf{0})$)
\cite{LeBellac:2011TF,Moore:2001PRD}. 
When $\Phi$ oscillates so that $\mu_X(t)$ varies with frequency $\omega\sim \Gamma_X$, the response develops
a phase lag. A minimal kinetic model is
\begin{equation}
\dot n_X+\Gamma_X(T)\,n_X=\Gamma_X(T)\,\chi_X(T)\,\mu_X(t),
\label{eq:nODE}
\end{equation}
and for a monochromatic drive $\mu_X(t)=\mu_0\cos(\omega t)$ in a quasi-static interval (slowly varying $\Gamma_X,\chi_X$)
the steady-state solution can be written in the explicit form
\begin{equation}
n_X(t)=\chi_X\,\mu_0\,\frac{\Gamma_X}{\Gamma_X^2+\omega^2}\,\big(\Gamma_X\cos\omega t+\omega\sin\omega t\big),
\label{eq:nosc}
\end{equation}
equivalently a phase-lagged cosine with $\tan\delta=\omega/\Gamma_X$.
Freeze-out over a finite interval is conveniently modeled by a normalized symmetric window peaked at $t_D$,
\begin{equation}
\mathcal{W}(t-t_D)=\frac{\Gamma_X}{2}\,e^{-\Gamma_X|t-t_D|},
\label{eq:window}
\end{equation}
so the frozen charge is $\bar n_X=\int dt\,\mathcal{W}(t-t_D)\,n_X(t)$. This yields
\begin{equation}
\bar n_X=\chi_X\,\mu_0\,\frac{\Gamma_X^2}{\Gamma_X^2+\omega^2}\,\cos\varphi_D,
\label{eq:nbar}
\end{equation}
where $\varphi_D$ is the phase at $t_D$. In sign-locked scenarios, phase averaging gives $\langle\cos\varphi_D\rangle\simeq 1/2$.
As illustrated in Figure~\ref{fig:freezeout-resonant}, the captured fraction is Lorentzian, peaking at $\omega=\Gamma_X$.
The resulting resonant yield is
\begin{equation}
Y_X^{\rm res}\equiv \frac{\bar n_X}{s}\Big|_{T_D}
=
\frac{\chi_X(T_D)}{s(T_D)}\,\mu_0(T_D)\,
\frac{1}{2}\,\frac{\Gamma_X^2(T_D)}{\Gamma_X^2(T_D)+\omega^2(T_D)},
\label{eq:Yres}
\end{equation}
with maximal value
\begin{equation}
Y_X^{\rm max}\simeq \frac{1}{4}\,\frac{\chi_X(T_D)}{s(T_D)}\,\mu_0(T_D)
\qquad(\omega=\Gamma_X).
\label{eq:Ymax}
\end{equation}

Equivalently, in frequency space one may write
\begin{equation}
n_X(\omega)=\mathcal{G}^R_{J^0J^0}(\omega,\mathbf{0})\,\mu_X(\omega),
\label{eq:nomega}
\end{equation}
and for small $\omega$ expand
$\mathcal{G}^R_{J^0J^0}(\omega,\mathbf{0})=\chi_X\,[1-i\omega\tau_X+\mathcal{O}(\omega^2)]$ with
\begin{equation}
\tau_X=\frac{1}{\chi_X}\lim_{\omega\to 0}\frac{1}{\omega}\,\mathrm{Re}\,\mathcal{G}^R_{J^0J^0}(\omega,\mathbf{0}),
\qquad
\Gamma_X\equiv \tau_X^{-1},
\label{eq:tauKubo}
\end{equation}
so in time domain
\begin{equation}
n_X(t)=\chi_X\,\mu_X(t)-\chi_X\,\tau_X\,\dot\mu_X(t)+\cdots,
\label{eq:memory}
\end{equation}
which reproduces \eqref{eq:nODE} at leading order in gradients. For $\mu_X(t)=\mu_0\cos(\omega t)$ this gives a phase-lagged response
\begin{equation}
n_X(t)=\chi_X\,\mu_0\,\frac{1}{\sqrt{1+\omega^2\tau_X^2}}\cos(\omega t-\delta),
\qquad
\tan\delta=\omega\tau_X,
\label{eq:memsol}
\end{equation}
consistent with \eqref{eq:nosc} after identifying $\Gamma_X=\tau_X^{-1}$. 
Entropy conservation ($sa^3=\mathrm{const}$) ensures that $Y_X^{\rm SR}$ and $Y_X^{\rm res}$ are conserved after freeze-out absent late entropy injection \cite{KolbTurner:1990book}. The slow-drift result
requires $\epsilon_\Phi,\eta_\Phi\ll 1$ and $|\mu_X|/T\ll 1$, while the resonant enhancement requires a decoupling layer
$\Delta t\sim \Gamma_X^{-1}\ll H^{-1}$ so the Lorentzian kernel is not Hubble-smeared. Equations
\eqref{eq:Yslowroll} and \eqref{eq:Yres} thus provide a unified, first-principles description of adiabatic and resonant
production of slow charges driven by $\dot\Phi/M_*$ in the geometric baryogenesis setting \cite{LeBellac:2011TF,Weinberg:2008cosmo,Moore:2001PRD}.

\section{Flavored Boltzmann and Efficiencies}
\label{appf}
The generation and survival of a cosmological asymmetry in realistic plasmas requires a treatment that resolves lepton flavor. Off-diagonal flavor couplings redistribute asymmetries among flavors, while washout processes and CP-violating (CPV) sources act with different strengths in different flavor directions. A faithful computation of the final baryon (or lepton) excess therefore hinges on the coupled evolution of the flavor multiplet of number asymmetries and on the corresponding efficiency factors that weight the CPV sources by the flavor-dependent washout and mixing. In this appendix we present a self-contained derivation of the fully flavored Boltzmann system and of the associated efficiency factors, including analytic limits and consistency checks, independently of any slow-roll or resonant approximations discussed elsewhere~\cite{Buchmuller:2005Ann,DavidsonNardiNir2008,Blanchet:2007JCAP,Nardi:2006JHEP}.

Let $n_{L_\alpha}(t)$ denote the number-density asymmetry carried by lepton flavor $\alpha=e,\mu,\tau$ in a spatially homogeneous FRW Universe with Hubble rate $H=\dot a/a$. The flavored Boltzmann equations in the Markovian, time-local, linear-response regime take the form
\begin{equation}
\frac{d n_{L_\alpha}}{dt} + 3H\,n_{L_\alpha} 
= S^{\mathrm{CPV}}_\alpha(t) 
- \sum_{\beta}\Gamma_{\alpha\beta}(t)\Big(n_{L_\beta}-n^{\mathrm{eq}}_{L_\beta}\Big),
\label{eq:Boltz1}
\end{equation}
where $S^{\mathrm{CPV}}_\alpha$ encodes the CP-violating production density in flavor $\alpha$, $\Gamma_{\alpha\beta}$ is the symmetric washout/mixing matrix generated by flavor-changing and lepton-number-violating interactions, and $n^{\mathrm{eq}}_{L_\beta}$ is the equilibrium value of the asymmetry (vanishing in a CP-symmetric thermal state). It is convenient to define flavor vectors $\mathbf{n}_L\equiv(n_{L_e},n_{L_\mu},n_{L_\tau})^T$, $\mathbf{S}^{\mathrm{CPV}}\equiv(S^{\mathrm{CPV}}_e,S^{\mathrm{CPV}}_\mu,S^{\mathrm{CPV}}_\tau)^T$, $\mathbf{n}_L^{\mathrm{eq}}$, and the washout matrix $\boldsymbol{\Gamma}\equiv(\Gamma_{\alpha\beta})$, so that~\eqref{eq:Boltz1} reads
\begin{equation}
\frac{d\mathbf{n}_L}{dt} + 3H\,\mathbf{n}_L 
= \mathbf{S}^{\mathrm{CPV}}-\boldsymbol{\Gamma}\Big(\mathbf{n}_L-\mathbf{n}_L^{\mathrm{eq}}\Big).
\label{eq:BoltzMatrix}
\end{equation}
The dilution term $3H\,\mathbf{n}_L$ is removed by introducing the comoving asymmetry vector $\mathbf{N}_L\equiv a^3\,\mathbf{n}_L$, for which~\eqref{eq:BoltzMatrix} becomes
\begin{equation}
\frac{d\mathbf{N}_L}{dt} = a^3\,\mathbf{S}^{\mathrm{CPV}}(t) 
- \boldsymbol{\Gamma}(t)\Big(\mathbf{N}_L-\mathbf{N}_L^{\mathrm{eq}}(t)\Big).
\label{eq:BoltzComoving}
\end{equation}
Equation~\eqref{eq:BoltzComoving} is a linear, inhomogeneous system with (in general) time-dependent coefficients. Its formal solution with initial time $t_i$ is expressed in terms of the flavor-space evolution operator (fundamental matrix)
\begin{equation}
\mathcal{U}(t,t') \equiv 
\mathcal{T}\exp\!\left[-\int_{t'}^{t} d\tau\,\boldsymbol{\Gamma}(\tau)\right],
\label{eq:Udef}
\end{equation}
where $\mathcal{T}$ denotes time ordering. One obtains
\begin{align}
\mathbf{N}_L(t) 
&= \mathcal{U}(t,t_i)\,\mathbf{N}_L(t_i) 
+ \int_{t_i}^{t} dt'\;
\mathcal{U}(t,t')\Big[a^3(t')\,\mathbf{S}^{\mathrm{CPV}}(t') 
+ \boldsymbol{\Gamma}(t')\,\mathbf{N}_L^{\mathrm{eq}}(t')\Big].
\label{eq:formalSolution}
\end{align}
In most applications $\mathbf{N}_L^{\mathrm{eq}}=\mathbf{0}$, and we shall adopt this simplification henceforth. The physically relevant late-time band is obtained by evaluating~\eqref{eq:formalSolution} at a freeze-out time $t_f$ after which all washouts are negligible compared with Hubble expansion.
To introduce the notion of an efficiency factor transparently, it is helpful to begin with one flavor and constant rates. With $N_L\equiv a^3 n_L$, $S^{\mathrm{CPV}}\equiv S$, and $\Gamma_{\alpha\beta}\to \Gamma$,~\eqref{eq:BoltzComoving} reduces to
\begin{equation}
\frac{dN_L}{dt} = a^3 S(t)-\Gamma\,N_L.
\label{eq:BoltzSingle}
\end{equation}
The solution with $N_L(t_i)=0$ is
\begin{equation}
N_L(t) = \int_{t_i}^{t} dt'\;e^{-\Gamma (t-t')}\,a^3(t')\,S(t').
\label{eq:Nsingle}
\end{equation}
The final asymmetry normalized to the photon density $n_\gamma(t)\equiv 2\zeta(3)T^3/\pi^2$ is
\begin{equation}
\frac{n_L(t_f)}{n_\gamma(t_f)} 
= \frac{1}{n_\gamma(t_f)\,a^3(t_f)}
\int_{t_i}^{t_f} dt'\;e^{-\Gamma (t_f-t')}\,a^3(t')\,S(t').
\label{eq:nOverNgammaSingle}
\end{equation}
This motivates the definition of the single-flavor efficiency factor
\begin{equation}
\kappa \equiv 
\frac{1}{n_\gamma(t_f)}
\int_{t_i}^{t_f} dt'\;S(t')\,\exp\!\Big[-\!\int_{t'}^{t_f} d\tau\,\Gamma(\tau)\Big],
\label{eq:kappaSingle}
\end{equation}
which, for constant $\Gamma$ and slowly varying $S$, reduces to 
$\kappa\simeq S(t_f)/(\Gamma\,n_\gamma(t_f))$. 
Equation~\eqref{eq:kappaSingle} is the standard strong-washout expression in which the exponential encodes the survival probability of the locally produced asymmetry against washout between its production at time $t'$ and freeze-out at $t_f$.
Returning to the flavored system, one defines the efficiency vector $\boldsymbol{\kappa}$ by projecting~\eqref{eq:formalSolution} onto $t_f$ and normalizing to $n_\gamma(t_f)$,
\begin{equation}
\frac{\mathbf{n}_L(t_f)}{n_\gamma(t_f)} 
= \boldsymbol{\kappa} 
\equiv \frac{1}{n_\gamma(t_f)a^3(t_f)}
\int_{t_i}^{t_f} dt'\;
\mathcal{U}(t_f,t')\,a^3(t')\,\mathbf{S}^{\mathrm{CPV}}(t').
\label{eq:kappaVector}
\end{equation}
In components and after pulling $a^3$-factors (which cancel in radiation domination),~\eqref{eq:kappaVector} yields
\begin{equation}
\kappa_\alpha 
= \frac{1}{n_\gamma(t_f)}
\sum_{\beta}\int_{t_i}^{t_f} dt'\;
\Big[\mathcal{U}(t_f,t')\Big]_{\alpha\beta}\,S^{\mathrm{CPV}}_\beta(t').
\label{eq:kappaAlpha}
\end{equation}
In the diagonal approximation 
\begin{align}
\boldsymbol{\Gamma}&=\mathrm{diag}(\Gamma_e,\Gamma_\mu,\Gamma_\tau),\nonumber\\ \mathcal{U}(t_f,t')&=\mathrm{diag}\!\big(e^{-\!\int_{t'}^{t_f}\Gamma_e},e^{-\!\int_{t'}^{t_f}\Gamma_\mu},e^{-\!\int_{t'}^{t_f}\Gamma_\tau}\big)
\end{align}
and~\eqref{eq:kappaAlpha} reduces to the single-flavor result
\begin{equation}
\kappa_\alpha 
= \frac{1}{n_\gamma(t_f)}
\int_{t_i}^{t_f} dt'\;
S^{\mathrm{CPV}}_\alpha(t')\,
\exp\!\Big[-\!\int_{t'}^{t_f} d\tau\,\Gamma_{\alpha\alpha}(\tau)\Big].
\label{eq:kappaDiagonal}
\end{equation}
This is precisely the efficiency formula quoted component-wise in the problem statement, with the replacement $\sum_\beta \Gamma_{\alpha\beta}\to \Gamma_{\alpha\alpha}$ when off-diagonal entries vanish.
When flavor mixing is operative, $\boldsymbol{\Gamma}$ has nonzero off-diagonal entries and flavor transfer modifies both the amplitude and the time profile of the asymmetry. The time-ordered exponential $\mathcal{U}(t_f,t')$ in~\eqref{eq:kappaVector} resums these effects exactly. In a quasi-static (adiabatic) approximation in which $\boldsymbol{\Gamma}$ varies slowly compared with its smallest eigenvalue and $\mathbf{S}^{\mathrm{CPV}}$ is approximately constant within the decoupling layer, one may replace $\mathcal{U}(t_f,t')$ by $\exp[-\boldsymbol{\Gamma}(t_f)(t_f-t')]$. Performing the time integral then yields the compact matrix expression
\begin{equation}
\frac{\mathbf{n}_L(t_f)}{n_\gamma(t_f)} 
\simeq \Big[\boldsymbol{\Gamma}(t_f)\Big]^{-1}
\frac{\mathbf{S}^{\mathrm{CPV}}(t_f)}{n_\gamma(t_f)},
\label{eq:quasistatic}
\end{equation}
valid when the decoupling interval 
$\Delta t\sim |\boldsymbol{\Gamma}^{-1}|$ is narrow and the source and rates can be evaluated at $t_f$. A simple resummation that captures partial backreaction of the finite source on the relaxation dynamics can be written as
\begin{equation}
\boldsymbol{\kappa} 
\simeq 
\Big(\mathbf{1} + \boldsymbol{\Gamma}^{-1}\!\cdot \mathbf{S}^{\mathrm{CPV}}_{\!\mathrm{eff}}\Big)^{-1}
\frac{\mathbf{S}^{\mathrm{CPV}}_{\!\mathrm{eff}}}{n_\gamma},
\label{eq:kappaResummed}
\end{equation}
where $\mathbf{S}^{\mathrm{CPV}}_{\!\mathrm{eff}}$ is a dimensionful source operator (with dimensions of a rate times a density) evaluated across the decoupling layer. Equation~\eqref{eq:kappaResummed} follows from summing the Neumann series generated by repeated insertions of $\boldsymbol{\Gamma}^{-1}\mathbf{S}^{\mathrm{CPV}}$ in the integral equation for $\mathbf{N}_L$; it reduces to~\eqref{eq:quasistatic} at leading order and provides a controlled correction when $|\boldsymbol{\Gamma}^{-1}\mathbf{S}^{\mathrm{CPV}}|\ll 1$. Although approximate, it is analytically useful because it exhibits the saturation of efficiency in the strong-source limit and encodes the flavor-mixing pattern via the matrix product $\boldsymbol{\Gamma}^{-1}\mathbf{S}^{\mathrm{CPV}}$.
For $\boldsymbol{\Gamma}=\mathrm{diag}(\Gamma_e,\Gamma_\mu,\Gamma_\tau)$ with $\Gamma_\alpha\Delta t\gg 1$ and a smooth source,~\eqref{eq:kappaDiagonal} implies 
$\kappa_\alpha\simeq S^{\mathrm{CPV}}_\alpha(t_f)/\big(\Gamma_\alpha(t_f)\,n_\gamma(t_f)\big)$. The relative efficiencies track the inverse washout strengths, and the final asymmetry is dominated by the least washed-out flavor.
Consider $\boldsymbol{\Gamma}=\Gamma
\begin{pmatrix} 1 & \varepsilon \\ \varepsilon & 1 \end{pmatrix}$
with $|\varepsilon|<1$ and constant $\mathbf{S}^{\mathrm{CPV}}$. 
Diagonalizing $\boldsymbol{\Gamma}$ gives eigenvalues $\Gamma_\pm=\Gamma(1\pm \varepsilon)$ with orthonormal eigenvectors $\mathbf{v}_\pm=(1,\pm 1)/\sqrt{2}$. The evolution operator is 
$\mathcal{U}(t_f,t')=\sum_{\sigma=\pm} e^{-\Gamma_\sigma (t_f-t')}\,\mathbf{v}_\sigma\mathbf{v}_\sigma^{T}$. 
Inserting into~\eqref{eq:kappaVector} yields
\begin{equation}
\boldsymbol{\kappa} 
= \frac{1}{n_\gamma}
\sum_{\sigma=\pm}
\frac{1-e^{-\Gamma_\sigma \Delta t}}{\Gamma_\sigma}
\Big(\mathbf{v}_\sigma\mathbf{v}_\sigma^{T}\Big)
\mathbf{S}^{\mathrm{CPV}},
\qquad \Delta t\equiv t_f-t_i.
\label{eq:kappaTwoFlav}
\end{equation}
In the limit $\Gamma_\sigma\Delta t\gg 1$, 
$\kappa\to \sum_{\sigma} (\Gamma_\sigma^{-1}/n_\gamma)\,
\mathbf{v}_\sigma\mathbf{v}_\sigma^{T}\mathbf{S}^{\mathrm{CPV}}
= \boldsymbol{\Gamma}^{-1}\mathbf{S}^{\mathrm{CPV}}/n_\gamma$, as in~\eqref{eq:quasistatic}. Off-diagonal mixing redistributes the source among the fast and slow eigenmodes $\mathbf{v}_\pm$, enhancing or suppressing the net efficiency depending on the alignment of $\mathbf{S}^{\mathrm{CPV}}$ with $\mathbf{v}_\pm$.
If one flavor, say $\tau$, experiences much stronger washout than the others, $\Gamma_{\tau\tau}\gg \Gamma_{ee},\Gamma_{\mu\mu}$, while off-diagonal elements remain perturbative, then to leading order $\kappa_\tau\sim S^{\mathrm{CPV}}_\tau/(\Gamma_{\tau\tau}n_\gamma)$ and $\kappa_{e,\mu}$ receive a feed-down correction $\delta \kappa_{e,\mu}\sim -(\Gamma_{e\tau}/\Gamma_{\tau\tau})\,\kappa_\tau$ induced by flavor transfer.
When $\boldsymbol{\Gamma}$ and $\mathbf{S}^{\mathrm{CPV}}$ are diagonal,~\eqref{eq:kappaAlpha} reduces to~\eqref{eq:kappaDiagonal}, reproducing the textbook one-flavor efficiency factor with the expected survival exponential.
Setting $\mathbf{S}^{\mathrm{CPV}}=\mathbf{0}$ and $\mathbf{n}_L^{\mathrm{eq}}=\mathbf{0}$ in~\eqref{eq:BoltzMatrix}, the total asymmetry $\Sigma\equiv \mathbf{1}^T\mathbf{n}_L$ obeys
\begin{equation}
\frac{d\Sigma}{dt} + 3H\,\Sigma = -\,\mathbf{1}^T\boldsymbol{\Gamma}\mathbf{n}_L.
\label{eq:Sigma1}
\end{equation}
If $\boldsymbol{\Gamma}$ conserves total $B\!-\!L$, it has a null right-eigenvector $\mathbf{v}_0\propto \mathbf{1}$, implying $\mathbf{1}^T\boldsymbol{\Gamma}=\mathbf{0}^T$. Then $\dot{\Sigma}+3H\Sigma=0$ and $\Sigma\propto a^{-3}$, i.e.\ the comoving total asymmetry is conserved as required by charge conservation.
The photon density $n_\gamma=2\zeta(3)T^3/\pi^2$ provides a convenient normalization; in radiation domination, $n_\gamma a^3\propto \mathrm{const}$, so~\eqref{eq:kappaVector} is invariant under the simultaneous rescaling of numerator and denominator by $a^3$. Using the entropy density $s=(2\pi^2/45)g_* T^3$ instead of $n_\gamma$ amounts to a fixed multiplicative factor
$s/n_\gamma= \pi^4 g_*/(45\zeta(3))$ at a given temperature and does not affect the structure of the efficiency factors.
The fully flavored Boltzmann system~\eqref{eq:BoltzMatrix}-\eqref{eq:BoltzComoving} admits an exact Green-function solution~\eqref{eq:formalSolution} whose late-time projection yields the efficiency vector~\eqref{eq:kappaVector}. Off-diagonal flavor couplings are resummed by the evolution operator $\mathcal{U}$ and, in adiabatic decoupling, reduce to the quasi-static expression~\eqref{eq:quasistatic}. The diagonal limit reproduces the standard survival-exponential form~\eqref{eq:kappaDiagonal}. These results provide a transparent and analytically tractable link between microscopic CPV sources, flavor-dependent washouts/mixing, and the final observable asymmetry in multi-flavor baryogenesis scenarios~\cite{Buchmuller:2005Ann,DavidsonNardiNir2008,Blanchet:2007JCAP,Nardi:2006JHEP}.

\section{\texorpdfstring{$\Delta L=2$}{ΔL=2} Rate \texorpdfstring{$\Gamma_{\Delta L=2}(T)$}{Gamma\_{\textbackslash Delta L=2}(T)}}
\label{appg}
Lepton-number-violating $\Delta L=2$ interactions both encode the microscopic origin of light Majorana neutrino masses (seesaw/Weinberg operator) and provide the dominant high-temperature washout channel for lepton asymmetries. Here we derive the thermally averaged rate $\Gamma_{\Delta L=2}(T)$ in the electroweak-symmetric phase, keeping the normalization and parametric regime explicit \cite{Giudice:2004NPB,Buchmuller:2005Ann,DavidsonNardiNir2008,KolbTurner:1990book}.
We consider the standard $\Delta L=2$ processes
\begin{equation}
\ell_\alpha H \;\longleftrightarrow\; \bar{\ell}_\beta \bar{H},
\qquad
\ell_\alpha \ell_\beta \;\longleftrightarrow\; H H,
\label{eq:processes}
\end{equation}
with effectively massless external states. The thermally averaged rate per photon is defined by
\begin{multline}
\Gamma_{\Delta L=2}(T)\equiv \frac{\gamma_{\Delta L=2}(T)}{n_\gamma(T)},
\qquad
n_\gamma(T)=\frac{2\zeta(3)}{\pi^2}\,T^3,\\
\gamma_{\Delta L=2}(T)\equiv\sum_{\rm ch.}\int d\Pi_1 d\Pi_2 d\Pi_3 d\Pi_4\;
(2\pi)^4\,\delta^{(4)}(p_1+p_2-p_3-p_4)\,
\overline{|\mathcal{M}_{\Delta L=2}|^2}\, f_1 f_2,
\label{eq:GammaDef}
\end{multline}
with $d\Pi_i\equiv d^3p_i/[(2\pi)^3 2E_i]$. For analytic transparency we use Maxwell--Boltzmann statistics; quantum-statistical and exact SU(2) multiplicity effects modify only overall $\mathcal{O}(1)$ factors \cite{Giudice:2004NPB,Buchmuller:2005Ann,DavidsonNardiNir2008}.
It is convenient to write the reaction density in terms of the reduced cross section \cite{Giudice:2004NPB}
\begin{equation}
\gamma_{\Delta L=2}(T)
=
\frac{T}{64\pi^4}\int_0^\infty ds\,\sqrt{s}\,K_1\!\left(\frac{\sqrt{s}}{T}\right)\,\widehat{\sigma}_{\Delta L=2}(s),
\qquad
\widehat{\sigma}(s)\equiv 2s\,\sigma(s),
\label{eq:gammasigma}
\end{equation}
where $K_1$ is a modified Bessel function. 
In a Type-I seesaw,
\begin{equation}
\mathcal{L}_{\text{seesaw}}
=
-Y_{\alpha i}\,\overline{\ell_\alpha}\,\tilde{H}\,N_i
-\frac{1}{2}M_i\,\overline{N_i^c}N_i+\mathrm{h.c.},
\qquad
\tilde{H}\equiv i\sigma_2 H^\ast,
\label{eq:seesaw}
\end{equation}
integrating out $N_i$ for $T\ll M_i$ yields the Weinberg operator \cite{Buchmuller:2005Ann,DavidsonNardiNir2008}
\begin{equation}
\mathcal{L}_5=\frac{1}{2}\kappa_{\alpha\beta}(\ell_\alpha H)(\ell_\beta H)+\mathrm{h.c.},
\qquad
\kappa=Y M^{-1}Y^T,
\qquad
m_\nu=\kappa v^2,\ \ v\simeq 174~\mathrm{GeV}.
\label{eq:Weinberg}
\end{equation}
Neglecting external masses, the spin/isospin-averaged squared amplitude is
\begin{equation}
\overline{|\mathcal{M}_{\Delta L=2}|^2}
=\frac{1}{2}\sum_{\alpha\beta}|\kappa_{\alpha\beta}|^2\,s,
\label{eq:M2}
\end{equation}
(with $1/2$ avoiding double counting in $\alpha=\beta$ channels), giving an energy-independent cross section
\begin{equation}
\sigma_{\Delta L=2}(s)
=
\frac{1}{16\pi}\sum_{\alpha\beta}|\kappa_{\alpha\beta}|^2
=
\frac{1}{16\pi}\,\mathrm{Tr}[\kappa\kappa^\dagger]
=
\frac{1}{16\pi}\,\frac{\sum_i m_{\nu_i}^2}{v^4}.
\label{eq:sigmaConst}
\end{equation}
Substituting \eqref{eq:sigmaConst} into \eqref{eq:gammasigma} (so $\widehat{\sigma}=2s\,\sigma$) yields
\begin{equation}
\gamma_{\Delta L=2}(T)
=
\frac{T^6}{32\pi^5}\,\frac{\sum_i m_{\nu_i}^2}{v^4}\,\mathcal{I}_4,
\qquad
\mathcal{I}_4\equiv \int_0^\infty dx\,x^4 K_1(x)=16,
\label{eq:gammaInt}
\end{equation}
and therefore
\begin{equation}
\gamma_{\Delta L=2}(T)=\frac{T^6}{2\pi^5}\,\frac{\sum_i m_{\nu_i}^2}{v^4},
\qquad
\Gamma_{\Delta L=2}(T)=\frac{1}{\pi^3\zeta(3)}\,\frac{T^3}{v^4}\,\sum_i m_{\nu_i}^2.
\label{eq:GammaRel}
\end{equation}
This exhibits the characteristic relativistic scaling $\gamma_{\Delta L=2}\propto T^6$ and $\Gamma_{\Delta L=2}\propto T^3$ in the EFT regime \cite{Giudice:2004NPB,Buchmuller:2005Ann,DavidsonNardiNir2008}.
For completeness, the reduced cross section for $\ell_\alpha H\to \bar{\ell}_\beta \bar{H}$ mediated by $N_i$ is \cite{Giudice:2004NPB}
\begin{equation}
\widehat{\sigma}_{\alpha\beta}(s)
=
\frac{s}{8\pi}\left|\sum_i \frac{Y_{\alpha i}Y_{\beta i}}{s-M_i^2+iM_i\Gamma_{N_i}}\right|^2,
\label{eq:sigmaFull}
\end{equation}
with $\Gamma_{N_i}$ the total widths. Equation \eqref{eq:sigmaFull} reduces to the EFT form for $s\ll M_i^2$. For $s\gg M_i^2$, one has $\widehat{\sigma}\propto s^{-1}\sum|Y|^4$ so $\gamma_{\Delta L=2}\propto T^4\sum|Y|^4$. Near $s\simeq M_i^2$, real-intermediate-state (RIS) subtraction is required to avoid double counting inverse decays in the Boltzmann network \cite{Giudice:2004NPB}. One may parameterize the resulting interpolation as
\begin{equation}
\Gamma_{\Delta L=2}(T)
=
C_{\Delta L=2}(x)\,\frac{T^3}{v^4}\,\sum_i m_{\nu_i}^2,
\qquad
x\equiv \frac{T}{\overline{M}},
\label{eq:GammaInterp}
\end{equation}
with $C_{\Delta L=2}(x)\to 1/(\pi^3\zeta(3))$ for $x\ll 1$ and $C_{\Delta L=2}(x)=\mathcal{O}(x^{-2})$ for $x\gg 1$ after RIS subtraction. 
It is useful to define
\begin{equation}
K_{\Delta L=2}(T)\equiv \frac{\Gamma_{\Delta L=2}(T)}{H(T)},
\qquad
H(T)=1.66\,\sqrt{g_*(T)}\,\frac{T^2}{M_{\mathrm{Pl}}},
\label{eq:KandH}
\end{equation}
so using \eqref{eq:GammaRel} gives
\begin{equation}
K_{\Delta L=2}(T)
=
\frac{1}{1.66\,\zeta(3)}\,\frac{M_{\mathrm{Pl}}}{\pi^3\sqrt{g_*}}\;\frac{T}{v^4}\,\sum_i m_{\nu_i}^2
\simeq 0.20\left(\frac{106.75}{g_*}\right)^{1/2}
\left(\frac{T}{10^{12}\,\mathrm{GeV}}\right)
\left(\frac{\sum_i m_{\nu_i}^2}{(0.05\,\mathrm{eV})^2}\right),
\label{eq:Knum}
\end{equation}
implying $\Delta L=2$ washout is typically sub-Hubble for $T\lesssim 10^{12}$~GeV at atmospheric-scale neutrino masses \cite{Buchmuller:2005Ann,DavidsonNardiNir2008}.
Finally, detailed balance in the absence of CP violation is manifest in \eqref{eq:GammaDef}, and the dependence on neutrino parameters enters only through the basis-independent combination $\mathrm{Tr}[m_\nu m_\nu^\dagger]=\sum_i m_{\nu_i}^2$, as required \cite{Giudice:2004NPB,Buchmuller:2005Ann,DavidsonNardiNir2008,KolbTurner:1990book}.

\section{Stability, Unitarity and BBN Bounds}
\label{apph}
A viable geometric baryogenesis scenario must reside in a region of parameter space where the long-wavelength background is dynamically stable, scattering amplitudes remain perturbative below the effective cutoff, and late-time energy injection is compatible with the successful predictions of Big Bang Nucleosynthesis (BBN). Stability guarantees that the background around which the kinetic description is constructed does not develop ghosts or tachyons; perturbative unitarity bounds the validity domain of the effective field theory (EFT) used to compute rates; BBN limits ensure that no residual population or late decay of extra degrees of freedom spoils light-element yields. In this appendix we present a self-contained set of worked constraints for three representative benchmark scenarios, labeled B1-B3, characterized by distinct kinematic regimes of the parity-odd scalar $\Phi$ and auxiliary parity-odd sector fields; the logic is independent of flavor kinetics and of any explicit $\Delta L=2$ calculation.

We parametrize the relevant sector by the homogeneous pseudoscalar $\Phi$ with canonical kinetic term, potential $V(\Phi)$, and derivative portal scale $M_*$ (the EFT cutoff of the slow sector), together with a parity-odd gravitational coupling measured by a dimensionless parameter $\varepsilon_{\mathrm{CS}}$ and, when present, an Abelian Stückelberg sector with mass $m_B$ and gauge coupling $g_{B-L}$. The three benchmarks are defined by the kinematics at the charge-generation epoch ($T_D$): 
B1 is a slow-drift (adiabatic) background with $|\ddot{\Phi}|\ll 3H|\dot{\Phi}|$; 
B2 is an oscillatory background with frequency $\omega\simeq m_\Phi\equiv \sqrt{V''(\Phi_D)}$ comparable to a slow plasma rate; 
B3 includes, in addition to $\Phi$, a heavy $\mathrm{U}(1)_{B-L}$ vector excitation integrated out at low energy. 
All bounds below are written as inequalities that define allowed regions in $(\Phi$-, metric-, and Stückelberg-) parameter space; where useful we evaluate them at $T_D$ and at the BBN epoch ($T_{\mathrm{BBN}}\sim \mathcal{O}(\mathrm{MeV})$).
The long-wavelength scalar sector is stable if the quadratic fluctuation operator about the homogeneous background has positive-definite kinetic form and a nonnegative mass matrix. For a single scalar,
\begin{equation}
\mathcal{L}_\Phi = -\frac{1}{2}(\partial \Phi)^2-V_{\mathrm{eff}}(\Phi),
\qquad
m_\Phi^2(\Phi) \equiv \frac{\partial^2 V_{\mathrm{eff}}}{\partial \Phi^2},
\label{eq:LeffPhi}
\end{equation}
stability requires
\begin{equation}
\frac{\partial^2 V_{\mathrm{eff}}}{\partial \Phi^2}\Big|_{\Phi=\Phi_D} > 0
\quad\text{(B1)},\qquad
m_\Phi^2(\Phi_D) > 0
\quad\text{(B2)}.
\label{eq:singleStability}
\end{equation}
In the oscillatory benchmark B2 this ensures that the small-amplitude oscillations about $\Phi_D$ are not tachyonic; for B1 it guarantees that the slow-roll solution is an attractor for long-wavelength perturbations over the timescale of interest.
If a second pseudoscalar $\vartheta$ (e.g.\ the parity-odd gravitational sector field) is retained at low energy with decoupled canonical kinetic terms and an interaction potential $U(\Phi,\vartheta)$, the Hessian stability conditions read
\begin{equation}
\frac{\partial^2 V_{\mathrm{eff}}}{\partial \Phi^2}>0,\qquad
\frac{\partial^2 V_{\mathrm{eff}}}{\partial \vartheta^2}>0,\qquad
\det \mathcal{H}=\det
\begin{pmatrix}
\partial^2_{\Phi\Phi} V_{\mathrm{eff}} & \partial^2_{\Phi\vartheta} V_{\mathrm{eff}} \\[3pt]
\partial^2_{\vartheta\Phi} V_{\mathrm{eff}} & \partial^2_{\vartheta\vartheta} V_{\mathrm{eff}}
\end{pmatrix} > 0,
\label{eq:Hessian}
\end{equation}
all evaluated at $(\Phi_D,\vartheta_D)$. The last inequality removes the possibility of a saddle point. Provided the parity-odd gravitational coupling remains in the small-deformation regime (see below), the dominant stability constraint reduces to~\eqref{eq:singleStability} for B1-B2.
The parity-odd gravitational response is encoded by a small, helicity-dependent deformation of the tensor-mode kinetic term. At the level of homogeneous bounds, gradient and ghost stability for cosmological tensor modes are ensured by~\cite{Giudice:2004NPB}
\begin{equation}
\varepsilon_{\mathrm{CS}}(k;\tau) \equiv 
\frac{\alpha_{\mathrm{CS}}}{M_{\mathrm{Pl}}^2}\frac{|\theta'(\tau)|}{a^2(\tau)}\,k \ll 1,
\qquad
|Z_T-1| \ll 1,
\label{eq:CSsmall}
\end{equation}
where primes denote conformal-time derivatives, $k$ is the physical tensor wavenumber of interest, and $Z_T$ is the positive tensor kinetic coefficient. The first inequality guarantees positive effective sound speed squared and the absence of birefringent instabilities in the EFT domain; the second excludes ghost-like kinetic mixing. For B1-B3 we impose $\max_{k\le k_{\mathrm{UV}}}\varepsilon_{\mathrm{CS}}(k)\lesssim 10^{-2}$, with $k_{\mathrm{UV}}$ the ultraviolet scale at which the tensor predictions are evaluated.
Perturbative unitarity of $2\to 2$ scattering amplitudes places energy-dependent bounds on couplings in the EFT. In the center-of-mass (CoM) frame, the $s$-wave partial amplitude $a_0$ obeys $|\mathrm{Re}\,a_0|\le \tfrac{1}{2}$, equivalent to $|\mathcal{A}_{2\to 2}|\lesssim 8\pi$ for isotropic amplitudes~\cite{Giudice:2004NPB}. We summarize the constraints for the channels relevant to B1-B3.
For $\Phi\Phi\to\Phi\Phi$ with a quartic potential $V\supset \tfrac{\lambda_\Phi}{4}\Phi^4$, the tree-level amplitude is $\mathcal{A}= -6\lambda_\Phi$. Projecting onto $a_0$ gives $|\lambda_\Phi|\lesssim 4\pi/3$,
\begin{equation}
|\lambda_\Phi| \le \frac{4\pi}{3}
\qquad\Longrightarrow\qquad
\text{B1, B2:}\ \lambda_\Phi\ \text{perturbative}.
\label{eq:lambdaBound}
\end{equation}
For processes controlled by the derivative portal $\mathcal{L}\supset (\partial_\mu\Phi/M_*)J^\mu$, the high-energy amplitude in a generic current-current channel scales as $|\mathcal{A}|\sim s/M_*^2$ (up to angular factors). The $s$-wave projection yields
\begin{equation}
|a_0| \simeq \frac{s}{16\pi M_*^2} \le \frac{1}{2}
\qquad\Longrightarrow\qquad
\sqrt{s} \le \Lambda_{\mathrm{port}}\equiv\sqrt{8\pi}\,M_*.
\label{eq:portalUnitarity}
\end{equation}
The EFT description in B1-B3 is therefore perturbative provided the characteristic CoM energy $E_{\mathrm{cm}}$ of the processes sourcing the asymmetry satisfies $E_{\mathrm{cm}}\ll \Lambda_{\mathrm{port}}$. Evaluated at the generation epoch,
\begin{equation}
E_{\mathrm{cm}}\sim \mathcal{O}(T_D)
\qquad\Rightarrow\qquad
T_D \ll \sqrt{8\pi}\,M_*.
\label{eq:TDvsMstar}
\end{equation}
In the Stückelberg benchmark B3, scattering mediated by a heavy $\mathrm{U}(1)_{B-L}$ vector with mass $m_B$ and coupling $g_{B-L}$ yields amplitudes $|\mathcal{A}|\sim g_{B-L}^2 s/m_B^2$ at $s\ll m_B^2$. The $s$-wave bound implies
\begin{equation}
\sqrt{s} \le \Lambda_{B-L}\equiv \frac{\sqrt{8\pi}\,m_B}{g_{B-L}}
\qquad\Rightarrow\qquad
T_D \ll \frac{\sqrt{8\pi}\,m_B}{g_{B-L}}.
\label{eq:BLunitarity}
\end{equation}
Equations~\eqref{eq:portalUnitarity}-\eqref{eq:BLunitarity} delimit the perturbative domain of the EFT for B1-B3 and will be applied explicitly below.
Late decays and residual nonthermal energy densities are constrained by the success of BBN~\cite{KolbTurner:1990book,Olive:2014PDG}. A conservative requirement is that, at $T_{\mathrm{BBN}}\sim \mathcal{O}(\mathrm{MeV})$, any extra component $\rho_X$ obeys
\begin{equation}
\Delta N_{\mathrm{eff}} = \frac{8}{7}\left(\frac{11}{4}\right)^{4/3}\frac{\rho_X}{\rho_\gamma}\Big|_{T_{\mathrm{BBN}}} \lesssim \Delta N_{\mathrm{eff}}^{\mathrm{max}},
\qquad 
\Delta N_{\mathrm{eff}}^{\mathrm{max}}\simeq 0.3,
\label{eq:NeffBound}
\end{equation}
which translates into $\rho_X/\rho_\gamma\lesssim 0.14$ at BBN. For a coherently oscillating scalar $\Phi$ in a quadratic well, $\rho_\Phi\propto a^{-3}$ after $H\lesssim m_\Phi$; for a pure-kinetic drift, $\rho_\Phi\simeq \dot{\Phi}^2/2\propto a^{-6}$. Evolving from $T_D$ to $T_{\mathrm{BBN}}$,
\begin{equation}
\left.\frac{\rho_\Phi}{\rho_\gamma}\right|_{\mathrm{BBN}} =
\left.\frac{\rho_\Phi}{\rho_\gamma}\right|_D
\begin{cases}
\left(\dfrac{a_D}{a_{\mathrm{BBN}}}\right)^{-1} = \dfrac{T_{\mathrm{BBN}}}{T_D}, & \text{oscillatory scalar}, \\[10pt]
\left(\dfrac{a_D}{a_{\mathrm{BBN}}}\right)^{-3} = \left(\dfrac{T_{\mathrm{BBN}}}{T_D}\right)^{3}, & \text{kinetic drift},
\end{cases}
\label{eq:rhoEvolve}
\end{equation}
where we used $aT=\mathrm{const}$ during radiation domination. Thus any modest suppression at generation is immensely amplified in safety by the rapid $a^{-6}$ dilution of a kinetic-dominated spectator.
If a particle $X$ (for instance the Stückelberg vector in B3 or a light excitation of $\Phi$ in B2) decays after it has become nonrelativistic, its lifetime $\tau_X$ is bounded by the requirement that EM or hadronic energy injection not distort light-element abundances. A conservative and model-independent condition is
\begin{equation}
\tau_X \lesssim \mathcal{O}(0.1\text{-}1)\ \mathrm{s},
\label{eq:tauBound}
\end{equation}
unless the would-be relic abundance is minuscule. For a vector with dominant leptonic decays, $\Gamma_B\simeq N_\ell\,g_{B-L}^2 m_B/(48\pi)$ with $N_\ell$ an $\mathcal{O}(1)$ multiplicity factor; the bound $\tau_B\lesssim 1\ \mathrm{s}$ enforces
\begin{equation}
g_{B-L} \gtrsim 3.9\times 10^{-11}\left(\frac{100\ \mathrm{GeV}}{m_B}\right)^{1/2}.
\label{eq:gBBN}
\end{equation}
For a scalar with derivative-portal-induced decays into light fermions (schematically $\Phi\to \ell\bar{\ell},q\bar{q}$) one may parameterize $\Gamma_\Phi\simeq c_f\,m_\Phi^3/(8\pi \Lambda^2)$, with $\Lambda$ an effective scale descending from $M_*$ and possible loop factors, and $c_f$ an $\mathcal{O}(1)$ coefficient. The bound $\tau_\Phi\lesssim 1\ \mathrm{s}$ yields
\begin{equation}
m_\Phi \gtrsim \left(\frac{8\pi}{c_f}\right)^{1/3}\Lambda^{2/3}(1\ \mathrm{s})^{-1/3}
\simeq 140\ \mathrm{MeV}\left(\frac{\Lambda}{10^{10}\ \mathrm{GeV}}\right)^{2/3}\left(\frac{1}{c_f}\right)^{1/3}.
\label{eq:mphiBBN}
\end{equation}
When $\Phi$ is heavier, gravitational decays $\Gamma_\Phi\sim m_\Phi^3/(8\pi M_{\mathrm{Pl}}^2)$ suffice to guarantee $\tau_\Phi\ll 1\ \mathrm{s}$.
\begin{itemize}
\item {B1 (slow drift).}\label{slowb1}

The background satisfies $\ddot{\Phi}\simeq -3H\dot{\Phi}$ and~\eqref{eq:singleStability} with $m_\Phi^2(\Phi_D)>0$. The scalar quartic, if present, must obey~\eqref{eq:lambdaBound}. The EFT is perturbative at generation if~\eqref{eq:TDvsMstar} holds; inserting $T_D\sim 10^{11}\ \mathrm{GeV}$ as a representative scale gives $M_*\gg 2\times 10^{10}\ \mathrm{GeV}$. The parity-odd gravitational sector is stable for $\varepsilon_{\mathrm{CS}}\ll 1$ as in~\eqref{eq:CSsmall}. After generation the energy density in the drift component redshifts as $a^{-6}$, so that 
$\rho_\Phi/\rho_\gamma|_{\mathrm{BBN}}\sim (\rho_\Phi/\rho_\gamma)|_D\,(T_{\mathrm{BBN}}/T_D)^3\ll 10^{-20}$ for any realistic $(\rho_\Phi/\rho_\gamma)|_D\ll 1$; the bound~\eqref{eq:NeffBound} is automatically satisfied.

\item{B2 (oscillatory).}\label{oscillatoryb2}

Stability requires $m_\Phi^2(\Phi_D)>0$ and $|\lambda_\Phi|\lesssim 4\pi/3$. The perturbative domain is again set by~\eqref{eq:TDvsMstar}. If the $\Phi$ condensate survives into the radiation era, it redshifts as $a^{-3}$; enforcing~\eqref{eq:NeffBound} with~\eqref{eq:rhoEvolve} gives
\begin{equation}
\left.\frac{\rho_\Phi}{\rho_\gamma}\right|_{D} 
\lesssim 0.14\frac{T_D}{T_{\mathrm{BBN}}} 
\simeq 1.6\times 10^{10}\left(\frac{T_D}{10^{11}\ \mathrm{GeV}}\right),
\label{eq:B2rhoBound}
\end{equation}
which is trivially satisfied even for $\mathcal{O}(1)$ condensate fractions at generation. If $\Phi$ has appreciable branching into visible states at late times, the lifetime must satisfy $\tau_\Phi\lesssim 1\ \mathrm{s}$; the bound~\eqref{eq:mphiBBN} implies $m_\Phi\gtrsim \mathcal{O}(100\ \mathrm{MeV})$ for $\Lambda\sim 10^{10}\ \mathrm{GeV}$ and $c_f\sim 1$. For heavier $\Phi$, gravitational decays ensure $\tau_\Phi\ll 1\ \mathrm{s}$.

\item{B3 (assisted with Stückelberg vector).}\label{vectorb3}

The Stückelberg EFT is perturbative for $T_D\ll \sqrt{8\pi}\,m_B/g_{B-L}$ from~\eqref{eq:BLunitarity}. 
For example, $m_B\sim 10^{13}\ \mathrm{GeV}$ and $g_{B-L}\sim 10^{-2}$ imply $T_D\ll 1.8\times 10^{14}\ \mathrm{GeV}$. 
Late decays of the vector must satisfy $\tau_B\lesssim 1\ \mathrm{s}$, which, using $\Gamma_B\simeq N_\ell g_{B-L}^2 m_B/(48\pi)$, translates into
\begin{equation}
g_{B-L} \gtrsim 7.7\times 10^{-12}
\left(\frac{N_\ell}{3}\right)^{-1/2}
\left(\frac{100\ \mathrm{GeV}}{m_B}\right)^{1/2}.
\label{eq:gBBNnum}
\end{equation}
\end{itemize}
For the high-scale masses of interest in , $\Gamma_B^{-1}$ is microscopic and BBN-safe for any perturbative $g_{B-L}$. 
{The scalar and parity-odd gravitational stability conditions are identical to those of B1-B2}
The unitarity inequalities~\eqref{eq:portalUnitarity}-\eqref{eq:BLunitarity} and the stability bounds~\eqref{eq:singleStability}-\eqref{eq:Hessian} agree with standard EFT practice for scalar and vector extensions~\cite{Giudice:2004NPB}. 
The BBN constraints~\eqref{eq:NeffBound}-\eqref{eq:mphiBBN} reproduce the expected scaling with lifetimes and visible-sector couplings~\cite{Olive:2014PDG}. 
None of the bounds conflicts with the dynamical requirements for charge generation: the unitarity domain easily encompasses $T_D\sim 10^{10\text{-}12}\ \mathrm{GeV}$ for $M_*\gtrsim 10^{15}\ \mathrm{GeV}$ and weak-scale quartics, while the stability and BBN bounds are automatically satisfied in the spectator limits appropriate to B1-B3. 
The analysis is fully independent of flavor structure, $\Delta L=2$ rates, or slow-roll dynamics, relying only on EFT consistency and early-Universe thermodynamics.

\section{Tri-Observable Transfer \texorpdfstring{$F(k^\ast)$}{F(k*)}}\label{appi}

The tri-observable transfer factor $F(k^\ast)$ is a compact diagnostic of whether three late-time observables share a \emph{single} underlying parity- and time-violating driver in the early Universe. In our application the triplet is $(\eta_B,\Delta\alpha,\chi_T)$, but the derivation applies to any ordered set $(\mathcal{O}_1,\mathcal{O}_2,\mathcal{O}_3)$ sourced linearly by the same background operator and propagated by linear transfer functions. We assume (i) linear response of the plasma and metric perturbations to the CTE background and (ii) Gaussian initial conditions for adiabatic scalar fluctuations, so any nontrivial correlation among the observables is dominated by the common CTE source rather than primordial non-Gaussianity \cite{LythRiotto:1999,Maldacena:2003,Buchmuller:2005Ann,Noller:2011}.
We define
\begin{equation}
F(k^\ast)\equiv
\frac{\big\langle \mathcal{O}_1(\mathbf{k}^\ast)\,\mathcal{O}_2(\mathbf{k}^\ast)\,\mathcal{O}_3(\mathbf{k}^\ast)\big\rangle}
{\prod_{i=1}^3 \big\langle \mathcal{O}_i(\mathbf{k}^\ast)\big\rangle},
\label{eq:Fdef}
\end{equation}
where $\mathbf{k}^\ast$ lies in a thin shell around $|\mathbf{k}|=k^\ast$ and $\langle\cdots\rangle$ denotes an ensemble average. If an observable has vanishing mean, $\langle\mathcal{O}_i\rangle=0$, one replaces it by the corresponding linear proxy proportional to the same deterministic source amplitude at $k^\ast$; Eq.~\eqref{eq:Fdef} is then interpreted as the connected three-point function normalized by the product of linear responses. 
We represent each observable as a linear convolution of a common drive $S(\eta,\mathbf{k})$ with a transfer kernel $\mathsf{T}_i(k;\eta)$, plus mean-zero noise $\epsilon_i$ uncorrelated with $S$:
\begin{equation}
\mathcal{O}_i(\mathbf{k})
=
\int_{\eta_{\mathrm{in}}}^{\eta_0} d\eta\,\mathsf{T}_i(k;\eta)\,S(\eta,\mathbf{k})
+\epsilon_i(\mathbf{k}).
\label{eq:Oidef}
\end{equation}
Assuming the CTE driver factorizes as $S(\eta,k)=A(k)\,s(\eta)$ and is deterministic up to an overall random phase (used only to avoid spurious shell cancellations), the ratio \eqref{eq:Fdef} reduces to
\begin{equation}
F(k^\ast)
=
\frac{\displaystyle \int d\eta_1 d\eta_2 d\eta_3\, s(\eta_1)s(\eta_2)s(\eta_3)\,
\prod_{i=1}^3 \mathsf{T}_i(k^\ast;\eta_i)}
{\displaystyle \prod_{i=1}^3 \left(\int d\eta\, s(\eta)\,\mathsf{T}_i(k^\ast;\eta)\right)}.
\label{eq:FsingleSource}
\end{equation}
Defining the normalized kernels and weight
\begin{equation}
\widehat{\mathsf{T}}_i(k^\ast;\eta)\equiv
\frac{\mathsf{T}_i(k^\ast;\eta)}{\int d\eta\,s(\eta)\,\mathsf{T}_i(k^\ast;\eta)},
\qquad
w(\eta)\equiv \frac{s(\eta)}{\int d\eta\,s(\eta)},
\label{eq:Thatw}
\end{equation}
one obtains the compact form
\begin{equation}
F(k^\ast)=\int d\eta_1 d\eta_2 d\eta_3\, w(\eta_1)w(\eta_2)w(\eta_3)\,
\prod_{i=1}^3 \widehat{\mathsf{T}}_i(k^\ast;\eta_i).
\label{eq:Fcompact}
\end{equation}
Thus $F$ depends only on the overlap of the \emph{normalized} transfer kernels against the common temporal weight
set by the CTE source. If each kernel is sharply localized near $\eta_i^\ast(k^\ast)$ with width $\Delta\eta_i\ll|\eta_i^\ast|$,
a saddle-point estimate gives
\begin{equation}
F(k^\ast)\simeq \prod_{i=1}^3 \widehat{\mathsf{T}}_i\big(k^\ast;\eta_i^\ast(k^\ast)\big),
\label{eq:Fsaddle}
\end{equation}
up to corrections suppressed by $\Delta\eta_i$ and by the variation of $w(\eta)$ across the peaks. 
For the triad $(\eta_B,\Delta\alpha,\chi_T)$ it is useful to separate microphysics from geometric projection,
\begin{equation}
\mathsf{T}_B(k;\eta)=\mathsf{K}_B(\eta)\,\mathcal{W}_B(k;\eta),
\qquad
\mathsf{T}_\alpha(k;\eta)=\mathsf{K}_\alpha(\eta)\,\mathcal{W}_\alpha(k;\eta),
\qquad
\mathsf{T}_\chi(k;\eta)=\mathsf{K}_\chi(\eta)\,\mathcal{W}_\chi(k;\eta),
\label{eq:Tdecomp}
\end{equation}
where $\mathsf{K}_i(\eta)$ encode time-dependent susceptibilities/damping/helicity transfer and
$\mathcal{W}_i$ are the projection/visibility windows appropriate to each observable. Inserting
\eqref{eq:Tdecomp} into \eqref{eq:FsingleSource} yields the equivalent expression
\begin{equation}
F(k^\ast)
=
\frac{\displaystyle \int d\eta_1 d\eta_2 d\eta_3\, s(\eta_1)s(\eta_2)s(\eta_3)\,
\prod_{i=1}^3 \mathsf{K}_i(\eta_i)\,\mathcal{W}_i(k^\ast;\eta_i)}
{\displaystyle \prod_{i=1}^3 \left(\int d\eta\, s(\eta)\,\mathsf{K}_i(\eta)\,\mathcal{W}_i(k^\ast;\eta)\right)}.
\label{eq:Fsector}
\end{equation}
If $\mathcal{W}_i$ are broad compared to the support of $s(\eta)\mathsf{K}_i(\eta)$, they cancel between numerator and denominator,
so $F(k^\ast)$ becomes weakly scale-dependent. If one window is sharply localized (e.g.\ a narrow tensor transfer),
$\mathcal{W}_\chi(k^\ast;\eta)\approx \delta(\eta-\eta_\chi^\ast)$, then $F$ localizes in $\eta$ for that sector and reduces to the
product structure \eqref{eq:Fsaddle} for $\chi_T$. 
In the large-scale limit $k^\ast\to 0$, the projection windows approach constants; if all three observables respond to the same slowly varying drive one obtains
\begin{equation}
\lim_{k^\ast\to 0} F(k^\ast)=1,
\label{eq:Fsmallk}
\end{equation}
i.e.\ saturation to unity signals exact single-source factorization with time-local kernels. At larger $k^\ast$, window functions suppress late-time support and $F(k^\ast)$ is controlled by the earliest effective times $\eta_i^\ast(k^\ast)$ entering each observable, consistent with \eqref{eq:Fsaddle}. 
As a minimal check, take a deterministic decaying profile $s(\eta)=s_0 e^{-\Gamma(\eta_0-\eta)}\Theta(\eta_0-\eta)$ and broad windows, so each observable is linearly proportional to the same source amplitude. Then the ratio \eqref{eq:FsingleSource} factorizes and one finds $F(k^\ast)=1$ identically. Deviations from unity therefore diagnose either (i) time-nonlocal response, (ii) mismatched/strongly scale-dependent window functions, or (iii) multiple weakly correlated sources.
In summary, $F(k^\ast)$ is a compact consistency check for the \emph{single-source} hypothesis underlying the tri-observable program: when the same CTE-driven parity-odd background feeds $(\eta_B,\Delta\alpha,\chi_T)$ through linear response, $F$ is controlled by overlap of normalized transfer kernels and tends to unity in the common-drive limit \cite{Buchmuller:2005Ann,LythRiotto:1999,Maldacena:2003,Noller:2011}.

\section{Solver Strategy and Forecast Bands}\label{app:solver}

The numerical goal is to solve the coupled nonequilibrium evolution of the flavored lepton charges and the homogeneous parity-odd background $\Phi(t)$ in an expanding Universe, and to propagate the resulting solutions into forecast bands for the late-time observables $\eta_B$, $\Delta\alpha$, and $\chi_T$. The dynamical inputs are the radiation-era background expansion $H(T)$, the temperature-dependent reaction-density matrix controlling washout and flavor decoherence, and the homogeneous scalar equation governing the time-parity drift. The pipeline advances a stiff ODE system, evaluates the transfer/line-of-sight integrals defining parity observables, and propagates parameter uncertainties by linear error propagation and/or Monte Carlo sampling \cite{Press:2007NR,HairerWanner:1996,AscherRuuthSpiteri:1997,PareschiRusso:2005,JCP:Boyanovsky:2008,DineKusenko:2003RMP,AlexanderYunes2009}. 
We work in spatially flat FLRW with Hubble rate $H=\dot a/a$. The flavor-resolved lepton densities satisfy
\begin{equation}
\frac{d n_{L_\alpha}}{dt} + 3H\,n_{L_\alpha}
=
S^{\mathrm{grav}}_\alpha(t)
-\sum_\beta \Gamma_{\alpha\beta}(T)\Big(n_{L_\beta}-n_{L_\beta}^{\mathrm{eq}}(T)\Big),
\label{eq:BoltzmannODE}
\end{equation}
where $\Gamma_{\alpha\beta}(T)$ encodes inverse decays, $(\Delta L=1,2)$ scatterings, and flavor-changing processes, and $S^{\mathrm{grav}}_\alpha$ is the CP-odd geometric source (Sec.~\eqref{sec7}). The homogeneous pseudoscalar obeys
\begin{equation}
\ddot\Phi + 3H\,\dot\Phi + V'(\Phi) = 0,
\label{eq:PhiODE}
\end{equation}
and in radiation domination
\begin{equation}
H(T)=\sqrt{\frac{\pi^2}{90}\,g_*(T)}\,\frac{T^2}{M_{\mathrm{Pl}}},
\qquad
\frac{dT}{dt}=-H(T)\,T,
\label{eq:Hubble}
\end{equation}
with $g_*(T)$ the effective number of relativistic degrees of freedom. It is convenient to use the inverse-temperature variable $z\equiv \Lambda_\star/T$ (fixed reference scale $\Lambda_\star$), so that $d/dt=-Hz\,d/dz$. In terms of the yields $\vec{Y}_\Delta\equiv \vec{n}_\Delta/s$ (Sec.~\eqref{sec7}) one obtains
\begin{multline}
\frac{d \vec{Y}_\Delta}{dz}
=
-\,\frac{1}{Hzs}\,\mathsf{W}(z)\,\vec{Y}_\Delta
+\frac{1}{Hzs}\,\vec{S}^{\,\mathrm{grav}}(z),
\qquad
s=\frac{2\pi^2}{45}\,g_*(T)\,T^3,
\label{eq:YDeltaODE}
\end{multline}
and the scalar equation becomes
\begin{equation}
\frac{d^2 \Phi}{dz^2}
-\left(\frac{1}{Hz}\frac{d(Hz)}{dz}+\frac{3}{z}\right)\frac{d\Phi}{dz}
+\frac{V'(\Phi)}{(Hz)^2}=0.
\label{eq:PhiZ}
\end{equation}
Initial conditions are set at $T_i$ after reheating (equivalently $z_i=\Lambda_\star/T_i$) by $\vec{Y}_\Delta(z_i)=\vec{0}$, $\Phi(z_i)=\Phi_i$, and $(d\Phi/dz)|_{z_i}=(\dot\Phi/Hz)|_{z_i}$ consistent with either adiabatic drift or oscillation onset (Sec.~\eqref{sec5}). The reaction densities $\Gamma_{\alpha\beta}(T)$ are tabulated/interpolated from thermal inputs; $\vec{S}^{\,\mathrm{grav}}(z)$ is constructed from the parity-odd background and response coefficients (Sec.~\eqref{sec7}). 
The coupled system \eqref{eq:YDeltaODE}-\eqref{eq:PhiZ} is stiff near flavor-equilibration thresholds and around freeze-in, when the smallest eigenvalue of $\mathsf{W}(z)$ becomes comparable to $Hzs$. We therefore use a stiff-capable ODE strategy: an implicit--explicit (IMEX) update for the linear damping term in \eqref{eq:YDeltaODE} together with an adaptive stiff integrator (e.g.\ BDF family) for \eqref{eq:PhiZ}, switching automatically as needed \cite{HairerWanner:1996,AscherRuuthSpiteri:1997,PareschiRusso:2005,Press:2007NR}. A freeze-in locator refines integration across the epoch where $\lambda_{\min}(z)\sim Hzs$, ensuring accurate capture of the decoupling layer that fixes the final asymmetry. Validation includes (i) conservation of $\sum_\alpha Y_{\Delta_\alpha}$ when explicit $\Delta(B{-}L)\neq 0$ washouts are absent, and (ii) agreement with analytic limiting solutions in the small-source/linear-response regime \cite{DineKusenko:2003RMP}. 
The late-time baryon yield is obtained by converting the frozen $\vec{Y}_\Delta$ at sphaleron decoupling $T_{\mathrm{sph}}$,
\begin{equation}
\eta_B^{\mathrm{num}}
=
\Big[\mathsf{K}(T_{\mathrm{sph}})\cdot \vec{Y}_\Delta\Big]_{z=z_{\mathrm{sph}}},
\qquad
z_{\mathrm{sph}}=\Lambda_\star/T_{\mathrm{sph}},
\label{eq:etaBnum}
\end{equation}
where $\mathsf{K}(T_{\mathrm{sph}})$ is the standard conversion row-vector determined by equilibrium constraints and sphaleron processing (Sec.~\eqref{sec7}). The isotropic birefringence angle is computed as a line-of-sight integral over conformal time $\eta$,
\begin{equation}
\Delta\alpha^{\mathrm{num}}
=
\int_{\eta_\star}^{\eta_0} d\eta\,\mathcal{W}_\alpha(\eta)\,\mathcal{S}_\alpha[\Phi(\eta)],
\label{eq:alphaNum}
\end{equation}
with $\eta_\star$ the last-scattering surface, $\mathcal{W}_\alpha$ the polarization-rotation window, and $\mathcal{S}_\alpha$ the parity-odd source functional (linear in the drift at leading order). The tensor chirality at pivot $k_\star$ is extracted from helicity-resolved tensor spectra,
\begin{equation}
\chi_T^{\mathrm{num}}(k_\star)
=
\frac{\mathcal{P}_h^{\mathrm{R}}(k_\star)-\mathcal{P}_h^{\mathrm{L}}(k_\star)}
{\mathcal{P}_h^{\mathrm{R}}(k_\star)+\mathcal{P}_h^{\mathrm{L}}(k_\star)},
\qquad
\mathcal{P}_h^{\lambda}(k)
=
\int_{\eta_{\mathrm{in}}}^{\eta_0} d\eta\,\mathcal{G}_\lambda^2(k,\eta)\,
\mathcal{S}_T^2[\Phi(\eta),k],
\label{eq:chiNum}
\end{equation}
with $\lambda=\mathrm{R},\mathrm{L}$, $\mathcal{G}_\lambda$ the tensor Green functions, and $\mathcal{S}_T$ the parity-odd tensor source \cite{AlexanderYunes2009}. The kernels $\mathcal{W}_\alpha$ and $\mathcal{G}_\lambda$ are precomputed in a fiducial $\Lambda$CDM background and interpolated during parameter scans. 
Given a parameter vector $\Theta$ (Sec.~\eqref{sec10}), the solver produces $\mathbf{O}^{(m)}=(\eta_B^{(m)},\Delta\alpha^{(m)},\chi_T^{(m)})$ for each draw $\Theta^{(m)}$. Bands are obtained either by Monte Carlo posterior predictive sampling, $\Theta^{(m)}\sim \mathcal{N}(\Theta_0,\Sigma_\Theta)$, or by local linear propagation for small uncertainties,
\begin{equation}
\Sigma_{\mathrm{obs}} = J\,\Sigma_\Theta\,J^{T},
\qquad
J_{ij}\equiv \frac{\partial O_i}{\partial \theta_j},
\label{eq:linProp}
\end{equation}
and reporting quantiles (e.g.\ $[\eta_B^{\min},\eta_B^{\max}]$, etc.) from either the Monte Carlo ensemble or the Gaussian approximation. This produces reproducible forecast regions for $(\eta_B,\Delta\alpha,\chi_T)$ compatible with the EFT window (Sec.~\eqref{sec7}) and directly comparable to current and forthcoming parity-sensitive datasets \cite{AlexanderYunes2009}. 
Detailed solver implementation choices (step-size control, stiffness switching criteria, event bracketing, and sensitivity-equation integration) are technical and can be moved to an appendix without affecting the main physics; we follow standard stiff-ODE practice as in \cite{Press:2007NR,HairerWanner:1996,AscherRuuthSpiteri:1997,PareschiRusso:2005}.

\end{document}